\documentclass[twocolumn, twocolappendix]{aastex701}
\shorttitle{Lensing in the Blue III: Shape Catalog}
\shortauthors{Saha et. al}
\usepackage{url}
\usepackage{amsmath}
\usepackage{booktabs} 
\usepackage{soul}
\newcommand{\sayan}[1]{\textcolor{black}{{#1}}}

\begin{document}

\title{Lensing in the Blue III: Weak Lensing Shape Catalogs of 30 Merging Galaxy Clusters}

\newcommand\uoftphysics{Department of Physics, University of Toronto}
\newcommand\utias{University of Toronto Institute for Aerospace Studies (UTIAS)}
\newcommand\princeton{Department of Physics, Princeton University}
\newcommand\princetonmae{Department of Mechanical and Aerospace Engineering, Princeton University}
\newcommand\durham{Department of Physics, Durham University}
\newcommand\durhamcfai{Centre for Advanced Instrumentation (CfAI), Durham University}
\newcommand\asiaa{Centre for Advanced Instrumentation (CfAI), Durham University}
\newcommand\bps{Boston Public School}
\newcommand\stewardobs{Department of Astronomy, Steward Observatory}
\newcommand\uoftastro{David A. Dunlap Dept. of Astronomy and Astrophysics, University of Toronto}
\newcommand\dunlap{Dunlap Institute for Astronomy and Astrophysics, University of Toronto}
\newcommand\jpl{Jet Propulsion Laboratory, California Institute of Technology}
\newcommand\durhamcea{Centre for Extragalactic Astronomy, Department of Physics, Durham University}
\newcommand\durhamicc{Institute for Computational Cosmology, Department of Physics, Durham University}
\newcommand\imperial{Imperial College London}
\newcommand\oslo{Institute of Theoretical Astrophysics, University of Oslo}
\newcommand\sheffield{Department of Physics and Astronomy, The University of Sheffield}
\newcommand\washu{Department of Physics, Washington University in St.\ Louis}
\newcommand\mcdonnell{McDonnell Center for the Space Sciences, Washington University in St. Louis}
\newcommand\madrid{Departamento de F\'isica Te\'orica,   Universidad Aut\'onoma de Madrid}
\newcommand\nymctu{Institute of Physics, National Yang Ming Chiao Tung University}
\newcommand\northeastern{Department of Physics, Northeastern University}
\newcommand\epfl{Laboratoire d'Astrophysique, EPFL, Observatoire de Sauverny}
\newcommand\casewestern{Department of Physics, Case Western Reserve University}
\newcommand\pltr{Scale AI, Inc.}
\newcommand\starspec{StarSpec Technologies Inc.}
\newcommand\mitaa{Department of Aeronautics and Astronautics, Massachusetts Institute of Technology}
\newcommand\camioa{Institute of Astronomy, University of Cambridge}
\newcommand\camdamtp{Department of Applied Mathematics and Theoretical Physics, University of Cambridge}
\newcommand\camkavli{Kavli Institute for Cosmology, University of Cambridge}
\newcommand\nrcherzberg{National Research Council Canada Herzberg Astronomy and Astrophysics Research Centre}
\newcommand\nyu{Center for Data Science, New York University}
\newcommand\caltech{Department of Physics, California Institute of Technology}
\newcommand\noirlab{Cerro Tololo Inter-American Observatory / NSF NOIRLab}
\newcommand\steward{Department of Astronomy/Steward Observatory, University of Arizona}
\newcommand\wpi{Department of Physics, Worcester Polytechnic Institute}

\author[0000-0002-6044-2164]{Sayan Saha}
\affiliation{\northeastern}
\email{sa.saha@northeastern.edu}

\author[0000-0002-9883-7460]{Jacqueline E.\ McCleary}
\affiliation{\northeastern}
\email{}

\author[0000-0002-3745-2882]{Spencer W.\ Everett}
\affiliation{\caltech}
\email{}

\author[0009-0002-2898-7022]{Maya Amit}
\affiliation{\nyu}
\affiliation{\northeastern}
\email{}

\author[0009-0006-2684-2961]{Georgios N.\ Vassilakis}
\affiliation{\camioa}
\affiliation{\camdamtp}
\affiliation{\camkavli}
\affiliation{\northeastern}
\email{}

\author[0000-0001-5101-7302]{Emaad Paracha}
\affiliation{\uoftphysics}
\email{}

\author[0000-0002-5899-3936]{Leo W.H. Fung}
\affiliation{\durhamicc}
\affiliation{\durhamcea}
\email{}

\author[0000-0002-4214-9298]{Steven J. Benton}
\affiliation{\princeton}
\email{}

\author[0000-0002-3636-1241]{William C. Jones}
\affiliation{\princeton}
\email{}

\author[0009-0004-2523-4425]{Gavin  Leroy}
\affiliation{\durhamicc}
\affiliation{\durhamcea}
\email{}

\author[0000-0002-9378-3424]{Eric M.\ Huff}
\affiliation{\jpl}
\email{}

\author[0000-0002-6085-3780]{Richard Massey}
\affiliation{\durhamcea}
\affiliation{\durhamicc}
\affiliation{\durhamcfai}
\email{}

\author{Thuy Vy T.\ Luu}
\affiliation{\princeton}
\email{}

\author[0000-0002-3937-4662]{Ajay S. Gill}
\affiliation{\nrcherzberg}
\email{}

\author[0000-0002-7600-3190]{Mohamed M.\ Shaaban}
\affiliation{\pltr}
\affiliation{\uoftphysics}
\email{}

\author[0009-0008-5546-890X]{Philippe Voyer}
\affiliation{\princeton}
\affiliation{\princetonmae}
\email{}

\author[0009-0007-7393-2983]{Adam Field}
\affiliation{\wpi}
\email{}

\author[0000-0003-0259-3148]{Anthony M. Brown}
\affiliation{\durhamcfai}
\affiliation{\durhamcea}
\email{}

\author[0000-0002-5273-4634]{Giulia Cerini}
\affiliation{\jpl}
\email{}

\author[0000-0003-0438-2133]{Paul Clark}
\affiliation{\durhamcfai}
\email{}

\author[0000-0002-3162-2107]{Matthew Craigie}
\affiliation{\jpl}
\email{}

\author[0000-0002-2036-2506]{Christopher J.\ Damaren}
\affiliation{\utias}
\email{}

\author[0000-0002-1894-3301]{Tim Eifler}
\affiliation{\steward}
\email{}

\author[0000-0002-6066-6707]{David Harvey}
\affiliation{\epfl}
\email{}

\author[0009-0003-9547-0952]{Eric Habjan}
\affiliation{\northeastern}
\email{}

\author{John W. Hartley}
\affiliation{\starspec}
\email{}

\author{Bradley Holder}
\affiliation{\utias}
\email{}

\author[0000-0003-1974-8732]{Mathilde Jauzac}
\affiliation{\durhamicc}
\email{}

\author[ 0000-0002-7633-2883]{David Lagattuta}
\affiliation{\durhamicc}
\email{}

\author[0000-0001-7116-3710]{Jason S.-Y.\ Leung}
\affiliation{\imperial}
\email{}

\author[0000-0002-8896-911X]{Lun Li}
\affiliation{\starspec}
\email{}

\author[0000-0002-2036-7008]{Johanna M. Nagy}
\affiliation{\casewestern}
\email{}

\author{C.\ Barth Netterfield}
\affiliation{\uoftastro}
\affiliation{\dunlap}
\affiliation{\uoftphysics}
\email{}

\author[0000-0002-9618-4371]{Susan F.\ Redmond}
\affiliation{\jpl}
\email{}

\author[0000-0002-4485-8549]{Jason D.\ Rhodes}
\affiliation{\jpl}
\email{}

\author[0000-0002-0086-0524]{Andrew Robertson}
\affiliation{\jpl}
\email{}

\author[0000-0002-3884-0983]{L. Javier Romualdez}
\affiliation{\starspec}
\email{}

\author[0000-0001-5612-7535]{J\"urgen Schmoll}
\affiliation{\durhamicc}
\email{}

\author[0000-0002-7542-0355]{Ellen Sirks}
\affiliation{\madrid}
\email{}

\author[0000-0002-6724-833X]{Sut Ieng Tam}
\affiliation{\nymctu}
\email{}

\author[0000-0002-9740-4591]{Andr\'{e} Z.\ Vitorelli}
\affiliation{\jpl}
\email{}

\author[0000-0001-6455-9135]{Alfredo Zenteno}
\affiliation{\noirlab}
\email{}

\collaboration{all}{SuperBIT collaboration}
\correspondingauthor{Sayan Saha}
\email{sa.saha@northeastern.edu}
\begin{abstract}
We present the weak gravitational lensing dataset from the Super-pressure Balloon-Borne Imaging Telescope (SuperBIT), which imaged 30 galaxy clusters during its 45 night flight in April to May 2023. SuperBIT is a first-of-its-kind balloon-borne imaging telescope that achieved near diffraction-limited observations in near-space conditions above 98\% of the Earth's atmosphere. We use the \textsc{metacalibration} algorithm to obtain calibrated galaxy shapes for our target clusters and several calibration fields, enabling unbiased reconstruction of the weak-lensing signal. We employ several diagnostics throughout the pipeline, including assessments of point-spread function (PSF) modeling residuals and their impact on weak-lensing measurements, as well as tests for correlations between galaxy shapes and measured galaxy and PSF properties. To assess the multiplicative shear bias of the pipeline, we analyze a parallel set of simulated images that incorporate the real observing conditions from the flight, including measured SuperBIT PSFs, observed sky backgrounds, and detector noise, yielding a bias of $(1.54 \pm 6.00)$~per~cent.
\end{abstract}

%\keywords{\uat{Galaxies}{573} --- \uat{Cosmology}{343}}


\section{Introduction}

Galaxy clusters are the largest gravitationally bound structures in the Universe and provide a powerful probe of cosmology. The abundance of clusters as a function of their mass and redshift offers direct insight into the growth of structure, influencing our understanding of
matter density ($\Omega_m$), matter fluctuation amplitude ($\sigma_8$), dark energy equation
of state, and sum of neutrino masses~\citep{intro1, intro2, intro3, intro4}. Beyond their cosmological significance, galaxy clusters, particularly merging systems, provide unique laboratories for studying the interplay between dark matter and baryonic components, and can potentially constrain alternative dark-matter models~\citep{Clowe:2003tk, Robertson:2016xjh, Sirks:2024njj}.

While galaxy richness~\citep{richness1, richness2}, X-ray emission from the hot gas of the intracluster medium~\citep{intro1, xray2}, and the Sunyaev-Zeldovich (SZ) effect~\citep{sz1, sz2} probe the baryonic component of galaxy clusters, the most direct and least biased probe of their total mass, dominated by dark matter, is weak gravitational lensing of background galaxies. Gravitational lensing refers to the deflection of light from background sources due to the curvature of space-time induced by intervening mass. Such deflections \sayan{induce two-dimensional distortions in} the surface brightness profiles of background galaxies. When these distortions are at the level of a few percent or less, the effect is termed \emph{weak} lensing (see \citealt{Bartelmann:1999yn} for a review).
These distortions can be decomposed into two components: magnification (convergence) and stretching (shear). \sayan{While magnification can be measured via flux and size magnification or number-count depletion (e.g.,~\cite{SDSS:2005jin, CARS}), shear remains the primary observable for weak lensing, measured statistically by averaging over many galaxies.} Measurements of shear can then be used to map the intervening mass distribution~\citep{1993ApJ...404..441K}

In this work we present the weak-lensing data obtained with the Super-pressure Balloon-borne Imaging Telescope (SuperBIT)~\citep{Gill2024}, which flew from April to May 2023 at an altitude of approximately 33~km, above $\sim98\%$ of the Earth's atmosphere. Observing in the stratosphere enables near-space-quality imaging at a fraction of the cost of a space mission. \sayan{SuperBIT's diffraction-limited optics provide near-perfect transmission between 280~nm and 900~nm, while the stratospheric environment yields exceptionally low sky backgrounds at blue wavelengths~\citep{Gill:2022bej}}. During the flight, SuperBIT obtained deep imaging of 30 galaxy clusters, most of which are in merging states, each covering a field of view (FOV) of approximately $15' \times 23'$.

This work is Paper III in the \emph{Lensing in the Blue} series. Paper I~\citep{shaaban2022weak} describes the observing conditions available in the stratosphere and demonstrates that, unlike ground- and space-based missions that are typically optimized for weak-lensing observations at red wavelengths, \sayan{SuperBIT's unique observing environment makes lensing studies maximally efficient in blue wavelengths.} Paper II~\citep{mccleary2023} introduced the SuperBIT weak-lensing pipeline and the associated simulation framework, and presented forecasts of the achievable lensing signal reconstruction under expected observing conditions. In the present work (Paper III), we analyze the imaging data obtained during the 2023 flight and construct the weak-lensing shape catalog.
 
Precise measurement of individual galaxy shapes (ellipticities) for weak-lensing science~\citep{tyson_wl, Bernstein:2001nz, Hirata:2003cv}, particularly in the presence of observational non-idealities, is a highly non-trivial step. 
Besides, the applied gravitational shear typically induces distortions at only the percent level, whereas intrinsic galaxy ellipticities are an order of magnitude larger and unknown. Consequently, shear can only be inferred through statistical averaging of galaxy shapes under the assumption that the applied shear is locally constant. The intrinsic scatter in galaxy ellipticities produces the so-called \emph{shape noise}, which constitutes the dominant statistical uncertainty in weak-lensing measurements and can only be reduced by increasing the source density.
\begin{figure}[t]
\centering
\hspace*{-1em}
\includegraphics[width=0.48\textwidth]{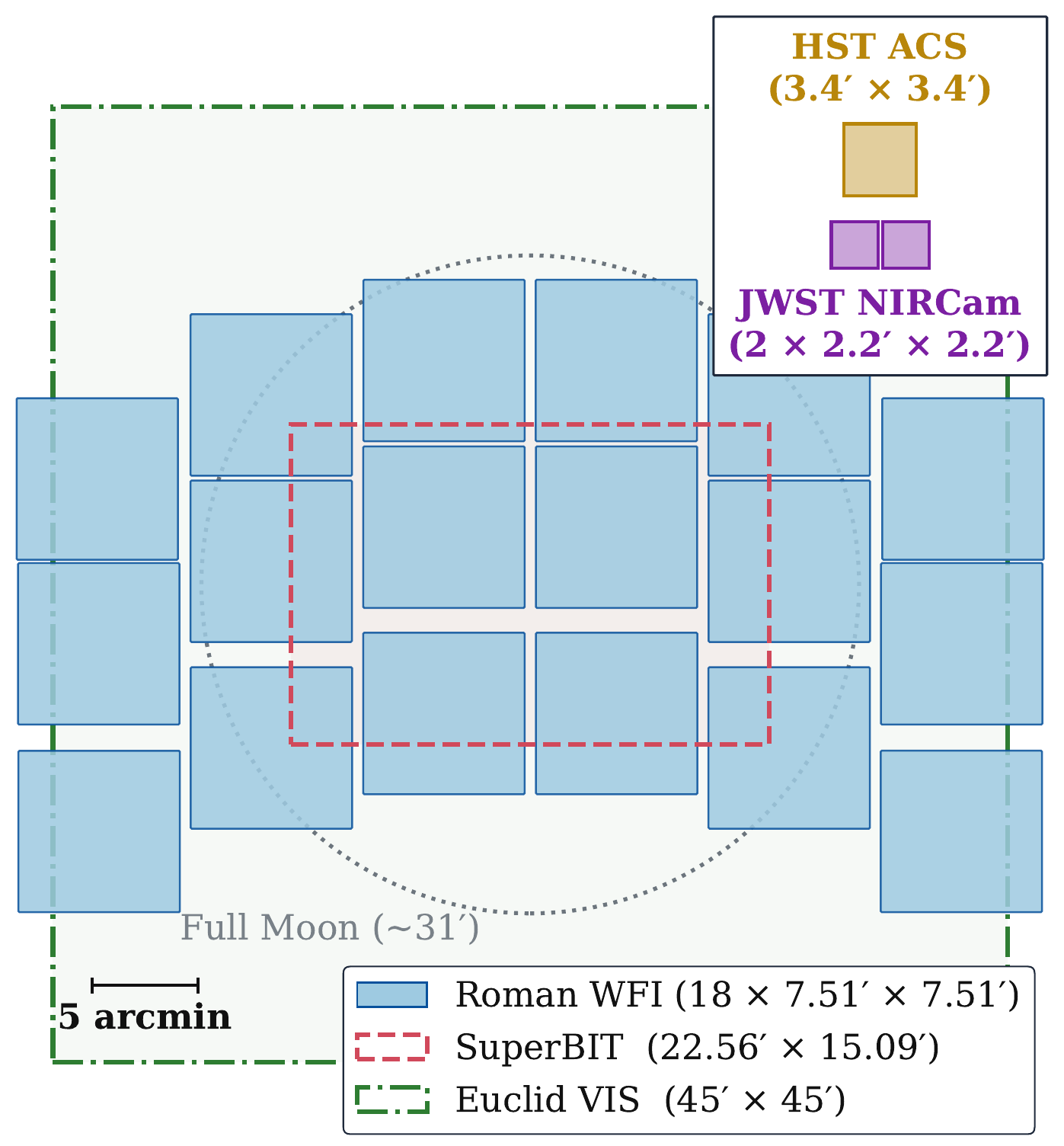}
\caption{Comparison of the field-of-view size of SuperBIT (shown as the red dashed line) with current and Stage~IV space missions: the HST ACS instrument is shown in gold, JWST NIRCam in magenta, Roman Wide Field Instrument (WFI) in sky blue, and Euclid Visible Instrument (VIS)
in green}
\label{fig:size-comp}
\end{figure}
Since its first detections~\citep{Wittman:2000tc, Bacon:2000sy, firstwl3, firstwl4}, weak gravitational lensing has become a cornerstone of precision cosmology, providing a unique probe of the large-scale structure of the Universe and enabling constraints on structure growth and cosmic evolution \citep{Bartelmann:1999yn, Schneider:2005ka}. Technological advances have led to significant improvements in image depth, resolution, and quality, resulting in higher source densities and increased statistical power.
\sayan{This progress has come with an inherent trade-off between ground- and space-based observing strategies.} 

\sayan{Ground-based telescopes benefit from long integration times but are limited by atmospheric seeing; meanwhile, space-based telescopes deliver diffraction-limited image quality but at the cost of heavy oversubscription and comparatively small fields of view such that covering the same sky area demands substantially more observing time. The current state of the art illustrates this contrast. Among wide-field ground-based surveys, the Dark Energy Survey (DES) Year~6 data, spanning 4422~deg$^2$, reach an effective source density $n_{\rm eff} = 8.22$~arcmin$^{-2}$ \citep{DES:2025zvc}, while the Hyper Suprime-Cam (HSC) three-year weak-lensing data attain $n_{\rm eff} = 19.9$~arcmin$^{-2}$ over $\sim$433.48~deg$^2$ \citep{Li:2021mvq}. Space-based missions reach far higher densities over correspondingly smaller areas: the 1000~hr Hubble COSMOS survey ($\sim$1.64~deg$^2$; \citealt{Koekemoer:2007sr}) achieves a source density of 66~arcmin$^{-2}$ \citep{Leauthaud:2007fb}, and the 255~hr JWST COSMOS-Web survey ($\sim$0.54~deg$^2$; \citealt{Casey:2022amu}) reaches 125~arcmin$^{-2}$ \citep{Scognamiglio:2026phv}. The forthcoming Stage-IV surveys are designed to combine wide area with high source density: Euclid's Wide Survey targets $n_{\rm eff} \approx 30$~arcmin$^{-2}$ over 14{,}679~deg$^2$ in six years \citep{euclid2011, Euclid:2024yrr}, the Roman Space Telescope's High Latitude Imaging Survey is forecast to reach $n_{\rm eff} \approx 41.3$~arcmin$^{-2}$ over 2415~deg$^2$ \citep{Zasowski:2025gki, roman2018}, and LSST at the Rubin Observatory is expected to achieve $n_{\rm eff} \approx 27$~arcmin$^{-2}$ over its full 18{,}000~deg$^2$ footprint after ten years of operation \citep{lsst2018, Chang:2013xja}. Against these benchmarks, SuperBIT's 2023 flight had been forecast to reach a source density of $\sim$40~arcmin$^{-2}$ \citep{mccleary2023}; in practice, chronic stability issues during the flight limited this to a best-case $n_{\rm eff} \approx 16.87$~arcmin$^{-2}$ (raw post-selection density 19.71~arcmin$^{-2}$) over 0.0954~deg$^2$ in three hours of integration in its primary detection band, F480W. SuperBIT's FOV, meanwhile, is substantially larger than those of HST and JWST, though smaller than the wide-field imagers on Euclid and Roman (Figure~\ref{fig:size-comp}\footnote{Roman WFI detector assembly has been obtained from \texttt{pysiaf} package~\citep{johannes_sahlmann_2019_3516964}}). With a 1.2~m mirror and the stability issues resolved, SuperBIT's successor mission, GigaBIT~\citep{Voyer2024}, is expected to reach $\sim$50~arcmin$^{-2}$ in 2~hours of integration (Fig.~7.3 in \citealt{ajaythesis}). }

Major efforts over the past decade have focused on improving shape-measurement algorithms to extract accurate shear information from faint and distant sources. Modern approaches rely on calibrating measured shapes to infer the applied shear. While early calibration schemes relied heavily on image simulations \citep{Heymans:2005rv, step2, bridle2010great08, Kitching2012, Mandelbaum:2014fta}, more recent methods use empirical calibration directly on observational data, most notably \textsc{metacalibration}~\citep{Huff:2017qxu, Sheldon:2017szh}, which naturally accounts for many observational non-idealities, \sayan{though it still relies on realistic image simulations for pipeline validation, particularly to determine object selection criteria for unbiased shear estimation}. These methods continue to evolve to mitigate increasingly subtle systematic effects relevant for high-resolution data~\citep{Sheldon:2019uxq, Li:2022, Li:2022qzu}.

 In this work, we first describe the SuperBIT data characteristics and preprocessing steps, including image stacking, source detection, and removal of spurious detections. We then present our modeling of the point-spread function (PSF) for individual exposures and evaluate several PSF diagnostics relevant for weak-lensing analyses. We next describe the shape measurement pipeline based on \textsc{metacalibration} algorithm and perform empirical tests of the resulting shape catalogs, including tests for residual PSF-galaxy correlations. Finally, we validate the pipeline using image simulations, including a suite of unit tests incorporating SuperBIT PSFs and fiducial simulations representative of real data, to estimate shear bias and PSF ellipticity leakage.

The paper is organized as follows. Section~\ref{sec:data} provides an overview of the data and preprocessing steps. Section~\ref{sec:psf-modeling} describes PSF modeling and associated diagnostics. Section~\ref{sec:shape-measure} presents the \textsc{metacalibration} implementation and empirical tests of the resulting shape catalogs. Section~\ref{sec:pipe-val} describes the simulation framework used to validate the pipeline. Finally, Section~\ref{sec:conclusion} summarizes the results and provides an outlook for future work.

\section{Data} \label{sec:data}
\subsection{Instrument \& Flight Overview}
SuperBIT employed a modified Dall-Kirkham telescope, with a 0.5\,m primary mirror and a three-lens corrector to obtain near-diffraction-limited imaging from the stratosphere, across a $\sim$0.1\,deg$^{2}$ field of view \citep{Gill2024}. Flying at an altitude of nearly 33\,km on NASA's Super-pressure balloon, SuperBIT operated above 98\% of Earth's atmosphere. During its 2023 long-duration flight, launched from W\={a}naka, New Zealand on April 16, 2023, SuperBIT circumnavigated the Southern Hemisphere five times over 45 nights, taking multi-band images of more than 30 galaxy clusters for weak lensing analysis.

SuperBIT's optical assembly was mounted on a lightweight aluminum honeycomb gondola with nested gimbal frames providing three-axis stabilization and control in yaw, pitch, and roll. This design helped isolate the telescope's line-of-sight from balloon-induced compound pendulations while compensating for sky rotation to maintain stable pointing for up to several hours \citep{Romualdez2020}. Telescope stabilization was guided by inertial gyroscopes and two wide-field star cameras that matched observed stars to catalogs using \texttt{astrometry.net} \citep{astrometry.net}. 

Additional tip-tilt image stabilization was achieved by using a high-speed fine-steering mirror guided by two focal-plane star cameras \citep{Voyer2024}. These systems collectively achieved a telescope pointing stability of 0.34$^{\prime\prime}$ and a focal-plane image stability of 0.055$^{\prime\prime}$ over (multiple) 5 minute exposures \citep{Gill2024}.

SuperBIT's images were taken with a QHY600M science camera using a Sony IMX-455 CMOS sensor ($9576\times6388$\,pixels, 3.76\,$\mu$m per pixel) with low read noise, high quantum efficiency, and a pixel scale of 0.144$^{\prime\prime}$/pixel. Observations were primarily conducted in three filters: F400W (307--434\,nm), F480W (366--575\,nm), and F600W (518--702\,nm), corresponding to ultraviolet, blue, and green bands, respectively. These bands were chosen for the small diffraction limit and high throughput of the system in those wavelengths, and consequent ability to measure the shapes of faint galaxies \citep{mccleary2023}.

The SuperBIT gondola had six lithium-iron-phosphate batteries (11\,kWh total) charged by solar panels, along with heaters and insulation to maintain stable temperatures while the gondola went through significant day/night swings (from +50$^{\circ}$C to $-60^{\circ}$C) \citep{Gill2024}. Data and telemetry were downlinked using Iridium, Pilot, NASA's TDRSS links, and SpaceX's Starlink \citep{Paracha2024}, with additional physical droppable 5TB storage systems for redundancy, ensuring all science data could be retrieved \citep{Sirks2020}.

SuperBIT's sub-arcsecond stability allowed it to take near-diffraction-limited images, which was well suited to conduct the weak-lensing shape measurements presented in this paper.

\subsection{Image Pre-processing}
We acquired dedicated dark frames prior to sunrise on every flight day and constructed master darks of each exposure time planned for science observations. SuperBIT's main science camera did not use a shutter, so, to prevent sky and sunlight contamination, we kept only the frames that were acquired at sun elevations $< -6.5^{\circ}$. To construct the master dark for each exposure time, we checked the pixel value distributions for each dark. Because histograms showed a narrow Gaussian core with a pronounced positive tail of hot pixels, a simple $\sigma$-clipping would have produced a biased mean dark current. Instead, we defined lower/upper inclusion thresholds (``haircuts'') at the departure from Gaussianity and classified pixels outside of the boundaries as cold/hot, which were then used to create the master dark images and bad pixel masks. 

We then averaged values \emph{per pixel} over the focal plane to generate master darks, but only included pixel samples in the haircut for each exposure time. To be sensitive to temporal effects, a pixel in the final master was kept if it was in the ``haircut'' in $>55\%$ of the total dark images. The threshold was chosen as an optimal trade-off: big enough to down-weight pixels experiencing intermittently high dark current or long-term deterioration, but small or permissive enough to keep otherwise stable pixels that sometimes scattered or had high dark current due to random cosmic rays. Using this cut, the percentage of pixels lost to haircutting was $<2\%$ across the focal plane with a clean, low-noise master dark. In master dark, pixels with no ``good'' contributors were in-filled with a computed median-stack dark for continuity. 

For scientific frames, we first extracted and smoothed cosmic rays with L.A.Cosmic~\citep{2001PASP..113.1420V} and subtracted the corresponding exposure-time master dark to remove bias structure and dark current before obtaining ``cleaned'' science images for astrometric and photometric processing. Hot pixels were identified in the master dark creation process described above, and were flagged separately for treatment in subsequent processing. With these ``clean images'', we were able to suppress and account for the long-tailed hot-pixel population without distorting the Gaussian core in the distribution, achieving spatially smooth residual backgrounds for stable photometric calibration of SuperBIT's scientific frames.

\subsection{Targets and Observation Footprint}
% Exposure time per filter
% 
The primary scientific goal of the SuperBIT mission was to constrain the self-interaction cross-section of dark matter using relative positions of the dark matter, intracluster medium (ICM) and stars, during mergers between galaxy clusters. The technique works as follows. As two galaxy clusters collide, the stars in member galaxies are effectively collisionless, experiencing only gravitational interactions as they pass by one another. By contrast, the ICM is highly collisional, with ram pressure slowing the two clouds of (mainly hydrogen) gas, which lag behind and become spatially offset from the galaxies \citep{Clowe:2003tk}. The location and spatial distribution of the dark matter---which can be probed via gravitational lensing---relative to the ICM and galaxies will depend on the degree of its self-interaction: in the cold dark matter paradigm, dark matter particles are collisionless and so should be entirely coincident with the stellar component.
If the dark matter is self-interacting (SIDM), the merging dark matter halos will experience a mild net deceleration, leading to small spatial offsets from the galaxy distribution \citep{Randall2008,Robertson:2016xjh}. Under certain assumptions, the relative spatial offsets of the dark matter, ICM, and galaxies can then be translated to a measurement of the cross-section for non-gravitational interactions between two dark matter particles, $\sigma/m$ \citep{2015Sci...347.1462H, 2017MNRAS.464.3991H, 2018ApJ...869..104W, Sirks:2024njj}. 
It is easiest to resolve physical separations if the cluster is nearby, but the nearest clusters are inefficient gravitational lenses, and their shear signal is diluted over a large area. For observations with as large a field of view as SuperBIT, the sensitivity of this experiment is optimal at redshift $0.05\lesssim z\lesssim 0.3$ \citep{bulleticity}.

\begin{figure}[t]
\centering
\includegraphics[width=0.48\textwidth]{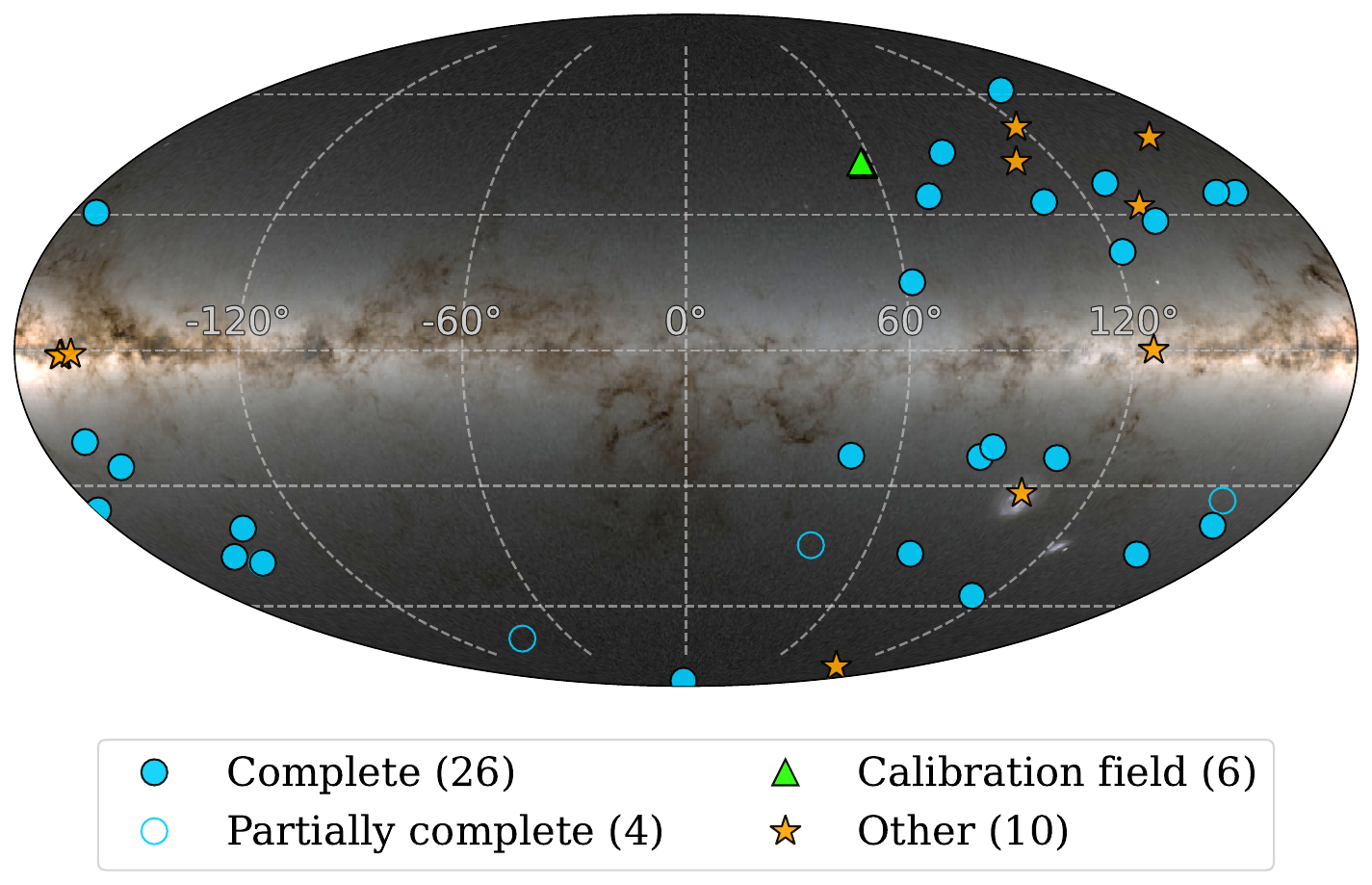}
\caption{On-sky distribution of SuperBIT observations. The filled and unfilled blue points represent the 30 science targets processed in this work. Filled blue points show completed targets, for which we achieved 100\% data acquisition of the planned exposures in the three science bands F400W, F480W, and F600W. Unfilled blue circles represent partially completed targets, for which at least 60\% of planned exposures were acquired. The green triangle marks the calibration fields, and the orange stars (``Other'') denote non-lensing targets, such as nearby galaxies and nebulae.}
\label{fig:general}
\end{figure}

Accordingly, SuperBIT's primary targets comprised a set of massive galaxy clusters, observable from Southern mid-latitudes. \sayan{These targets were selected from catalogs of known merging clusters, including the Merging Cluster Collaboration \citep{Golovich:2017tke}, as well as clusters exhibiting high-significance (SNR $>3$) X-ray or thermal Sunyaev--Zel'dovich (tSZ) detections with multiple peaks}, or known to be clusters merging in a direction close to the plane of the sky from e.g.\ radio shock fronts. Target clusters were chosen to have abundant ancillary data: deep tSZ and X-ray imaging for mapping the ICM and, ideally, Hubble Space Telescope imaging for strong lensing studies of the cluster cores. Several targets also appear in the Local Volume Complete Cluster Survey (LoVoCCS, \citealt{2024ApJ...974...69F}), the Dark Energy Spectroscopic Instrument Data Release 1 \citep{2025arXiv250314745D}, and the Dark Energy Survey \citep{Zenteno2020, 2022PhRvD.105h3528P}. \sayan{During flight, the autonomous target-selection algorithm prioritized observations using a weighted scoring system based on these attributes. Details of the in-flight autopilot algorithm are described in Section 2.2 of \citet{susan2024}.}
%Several of the target clusters appear in the RELICS, CLASH, and Hubble Frontier Fields surveys \citep{2018ApJ...859..159C,2017ApJ...837...97L,2012ApJS..199...25P,2020ApJ...889..189S}.  

Informed by the forecasts of \cite{shaaban2022weak} and \cite{mccleary2023}, the survey strategy consisted of three hours per cluster in F400W and F480W plus 90 minutes in F600W. Template galaxy spectra at redshifts $0 < z_{\rm gal} < 1.5$ showed that cuts in this color-color space could separate foreground and background galaxies of most spectral types from those in a fiducial $z = 0.5$ cluster. %, even with the non-negligible overlap of the SuperBIT $b$ and $g$ bandpasses. 

Based on these criteria, we compiled and ranked a list of $\sim$100 viable clusters. An autonomous ``autopilot'' algorithm on the payload then selected targets from among the list, based on scientific rank as well as target visibility, guide star magnitude, the number of exposures per band of the target already taken, and the previous history of image quality \citep{Gill2024}. When the mission was terminated, SuperBIT had obtained complete or near-complete observations for 30 merging galaxy clusters at $0.01 < z < 1$ (Figure~\ref{fig:general}). Partial data exists for another ten clusters. \sayan{Six pointings within the COSMOS field were also obtained, providing deep multiband photometry used for photometric redshift calibration and validation of our foreground/background separation algorithm. Completed cluster observations and calibration fields are presented in Table~\ref{tab:densities}}.

%Given the limited mission duration, SuperBIT's $ubg$ survey strategy was chosen to maximize the number of clusters observed, albeit at the expense of having sufficient colors for photometric redshift estimation. Accordingly, we separate unlensed foreground galaxies from lensed background sources using selections in galaxy color and magnitude.  %The separation of foreground and background galaxies is an essential part of any weak lensing analysis. 
%This approach has its advantages: although photometric redshifts are a historically more popular approach, mismatched galaxy SED templates can lead to biased distance estimates. Color-based sample selections are more effective in selecting background galaxies than photometric redshifts, resulting in a higher lensing signal-to-noise \cite{2017MNRAS.466.2614M, 2018PASJ...70...30M}. Even with the non-negligible overlap of the SuperBIT $b$ and $g$ bandpasses, based on an evolution of galaxy spectral templates in the range $0 < z_{\rm gal} < 1.5$, \cite{mccleary2023lensing} show that cuts in galaxy color-color space would produce adequate separation for galaxies of most spectral types and a fiducial $z = 0.5$ cluster.

\subsection{Astrometry}\label{subsec:astrometry}
\begin{figure*}[]
%\vspace{-0.3em}
\centering
%\hspace{-3em}
\includegraphics[width=0.95\textwidth]{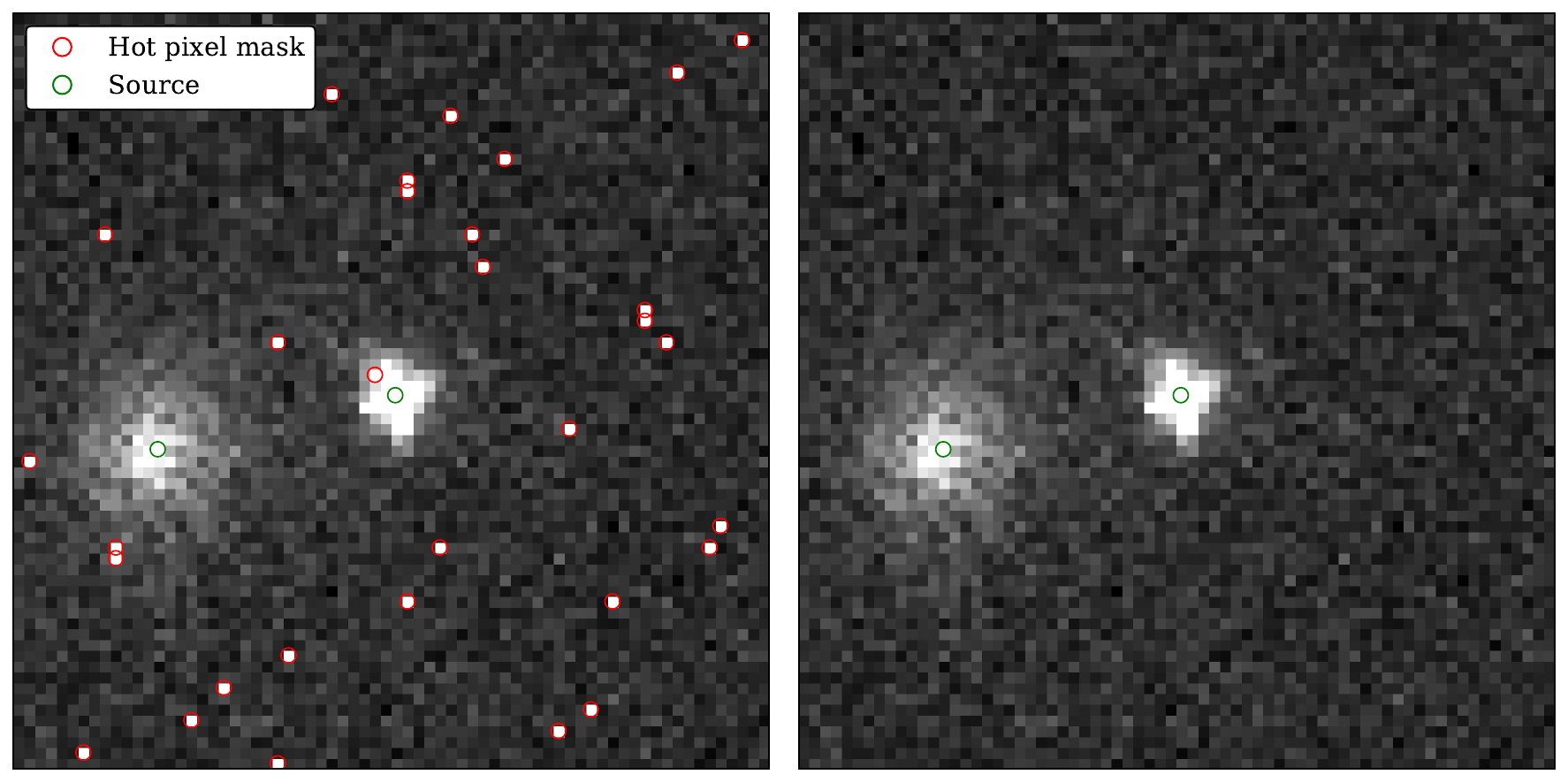}
\caption{Illustration of the hot-pixel mask implementation in our pipeline.
\emph{Left panel:} Example crop of a clean single exposure showing hot pixels (red circles) identified by the pixel mask. Nearby genuine source detections are indicated by green circles.
\emph{Right panel:} Background-subtracted image of the same crop, as output by \textsc{SExtractor} when the hot-pixel mask is provided as an input weight/mask map during source extraction. These background-subtracted images are used in subsequent stages of the pipeline for Bayesian shape measurement with \textsc{ngmix} (detailed in Section~\ref{sec:shape-measure} and Appendix~\ref{appendix:ngmix-workflow}). The hot-pixel masks are also supplied during shape fitting, ensuring that the inference is independent of
those pixels.}
\label{fig:hot-pixels}
\end{figure*}
As is standard practice, SuperBIT exposures were astrometrically calibrated 
%(developing a world coordinate system (WCS) that maps pixel coordinates $(x,y)$ to celestial coordinates $(\theta,\phi)$) 
by matching image sources to a reference catalog and fitting field distortions. To perform these fits, we used the \texttt{solve-field} command from \texttt{astrometry.net} \citep{astrometry.net} after stripping any pre-existing WCS keywords from the FITS headers. Given the frame size, we restricted solutions to the small-field-of-view 5200-series index files built from Tycho-2 and \textit{Gaia}~DR2. In practice, 98\% of fields were solved with index 5202 and the remaining 2\% with index 5200. We removed the larger-scale 5203--5206 indices from \texttt{astrometry.net}'s available dataset because their quad diameters are comparable to a substantial fraction of a SuperBIT frame, reducing the density of fully contained asterisms and yielding less robust WCS matches across the entire field-of-view. In particular, quads 5205--5206 approach the entire frame width, whereas quads 5203--5204 are intermediate and were omitted to maximize contained-quad density, reduce spurious hypotheses, and improve overall WCS solutions across the wide field of view. The resulting TAN-SIP solutions were then converted to TAN-TPV for compatibility with \textsc{SWarp}, preserving the polynomial distortion content. Visual overlays against catalogued sources confirmed accurate source placements across the field. The scripts implementing this workflow are publicly available.\footnote{\url{https://github.com/GeorgeVassilakis/wcs_beheader}}

\subsection{Stacking and Source Detection}
After performing dark subtraction, flat-field correction, and astrometric calibration, the exposures were visually inspected to identify and exclude frames affected by artifacts such as satellite glints, aurora, or other transient anomalies. The resulting ``clean'' exposures constitute the final set used as input for our shape-measurement pipeline.

The first step of the pipeline involves stacking these clean single exposures to produce a deep coadded image, which serves as the basis for source detection. During the inspection of individual exposures, we identified hot pixels distributed across the CCD plane. These pixels are masked using pre-determined hot-pixel maps, which are applied consistently throughout the pipeline. An example of hot pixels in a single exposure, along with their corresponding masks, is shown in the left panel of Figure~\ref{fig:hot-pixels}.

Clean single exposures are stacked using the \textsc{Swarp} algorithm\footnote{\url{https://www.astromatic.net/software/swarp/}}~\citep{swarp}. 
For each exposure, we construct weight maps that assign a value of 1 to valid pixels and 0 to hot pixels, while carefully incorporating the WCS information. These weight maps ensure proper pixel weighting during stacking; because dithering prevents hot pixels from aligning across exposures, the weighting effectively mitigates their impact on the coadded image.
\begin{figure}[]
\centering
\includegraphics[width=0.46\textwidth]{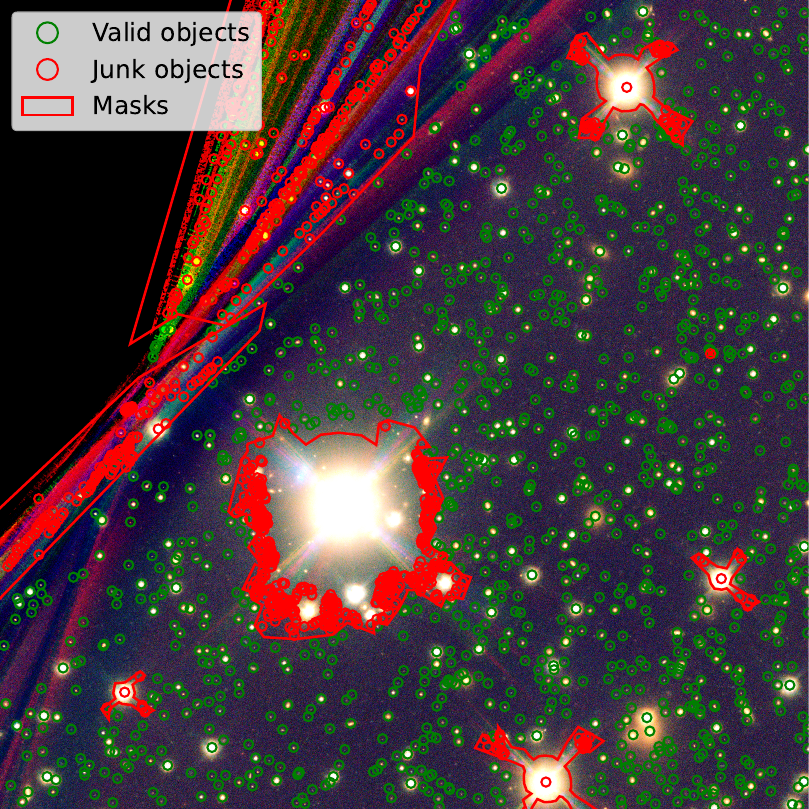}
\caption{Example of a star-mask applied to the detection catalog on the coadded image. Sources within the masked regions (red) are excluded, while the remaining sources (green) are retained and propagated through the pipeline.}
\label{fig:star-masks}
\end{figure}
While the primary science observations were obtained in the F400W, F480W, and F600W filters (corresponding to the $u$, $b$, and $g$ bands), the F480W ($b$-band) images serve as our primary detection band. The header information from the F480W coadd is used to construct the coadds for the F400W and F600W bands, ensuring consistent astrometry across all filters. Sources in the coadds are detected using \textsc{SExtractor} \footnote{\url{https://www.astromatic.net/software/sextractor/}}~\citep{sextractor}. Various configuration settings were tested to optimize source detection. The exposure weight maps are provided to \textsc{SExtractor} to normalize flux measurements, ensuring that photometry is independent of the number of exposures contributing to each object. Source detection is performed in the F480W ($b$) band, while photometric measurements in the F400W ($u$) and F600W ($g$) bands are obtained using the dual-image mode of \textsc{SExtractor} applied to the registered coadds generated with \textsc{Swarp} as described above. We also perform source extraction at the single-exposure level, as we need to identify stars required for PSF modeling
(see Section~\ref{sec:psf-modeling}). In \textsc{SExtractor}, we use the same
binary weight maps (ones and zeros) constructed from our hot-pixel masks as
used in \textsc{SWarp}. This procedure produces background-subtracted images
in which hot pixels are effectively inpainted. An example of such a
background-subtracted image, where the hot pixels have been inpainted, is
shown in the right panel of Figure~\ref{fig:hot-pixels}. We use these background-subtracted images as inputs for Bayesian shape measurement with \textsc{ngmix}, as detailed in Section~\ref{sec:shape-measure} and Appendix~\ref{appendix:ngmix-workflow}. In addition, the hot-pixel mask is supplied to \textsc{ngmix} during model fitting, ensuring that the inferred galaxy properties are independent of the masked pixels.

After obtaining the detection catalog with \textsc{SExtractor}, we identified spurious detections around diffraction spikes of very bright stars in the field of view, as well as near the edges of exposures. Some of our cluster targets contain exceptionally bright brightest cluster galaxies (BCGs), which produce similar artifacts. The most efficient approach, without sacrificing true sources, is to apply masks around bright objects.

For this purpose, we manually create polygonal region files around bright sources using the astronomical visualization software SAOImage DS9\footnote{\url{https://sites.google.com/cfa.harvard.edu/saoimageds9}}~\citep{ds9}. These region files are generated for all valid targets processed through our pipeline. Sources falling within these masked regions are excluded, while the remaining objects are retained for further analysis. \sayan{Figure~\ref{fig:star-masks} shows an example of this masking technique applied to the cluster Abell 3411. Any residual spurious detections not removed by this masking should also be filtered by the selection criteria discussed in Section~\ref{sec:object-selection}.}
\begin{figure}[]
\hspace*{-1.5em}
\centering
\includegraphics[width=0.5\textwidth]{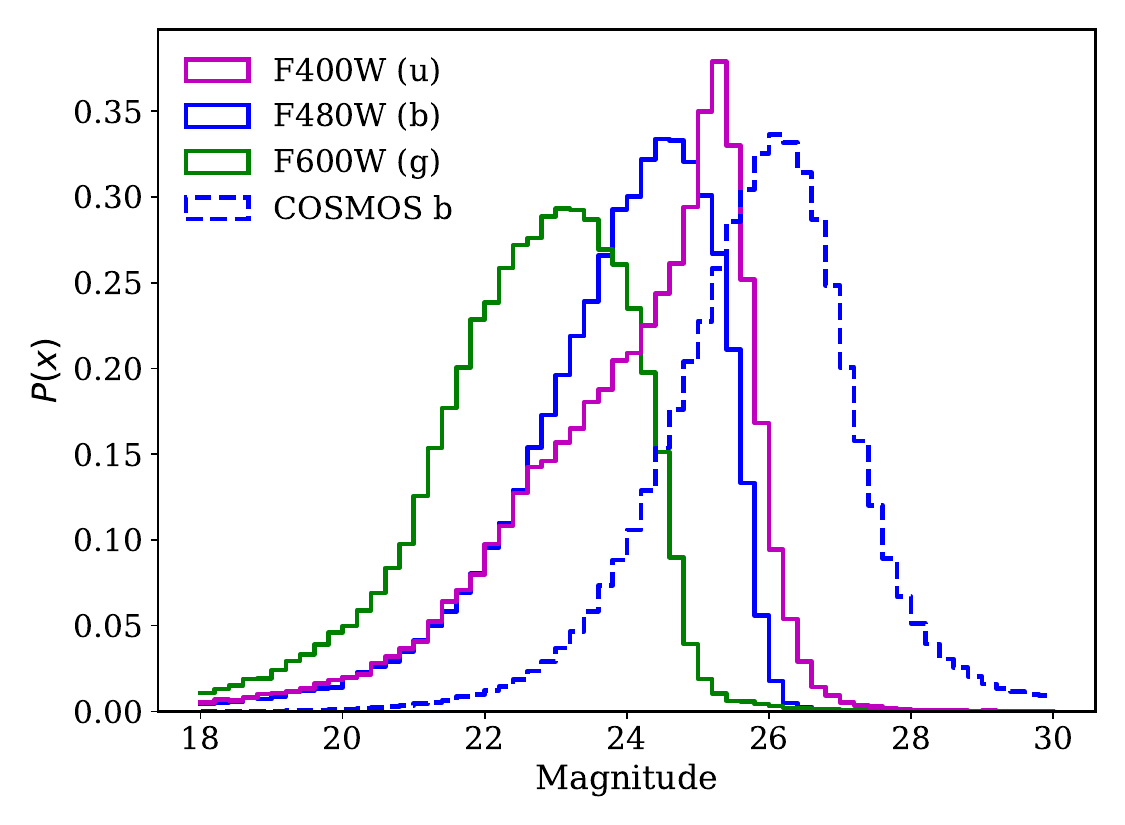}
\caption{Magnitude distributions in the SuperBIT science filters. The F480W ($b$), F600W ($g$), and F400W ($u$) distributions are shown in blue, green, and magenta, respectively. The dashed blue histogram shows the magnitude distribution of all COSMOS catalog objects as they would be observed through the SuperBIT F480W ($b$) filter. The COSMOS catalog is used to sample galaxy properties in the simulations.}
\label{fig:mag-depth}
\end{figure}
\subsection{Photometric Calibration}
\sayan{Photometric calibration is necessary to convert the photometric measurements of detected objects on the coadd, which are in arbitrary instrumental flux units, to an absolute astronomical magnitude scale. We adopt the AB magnitude system,
\begin{equation}\label{eq:m_syn}
    m_{AB} = -2.5\log_{10}\left(\frac{\langle f_\nu \rangle}{3631\ \text{Jy}}\right),
\end{equation}
where $1\,{\rm Jy} = 10^{-26}\,{\rm W\,m^{-2}\,Hz^{-1}}$ (SI). Here $\langle f_\nu \rangle$ is the bandpass-weighted mean flux density per unit frequency:
\begin{equation}
    \langle f_\nu \rangle = \frac{\int f_\lambda(\lambda)\ S(\lambda)\ \lambda\ d\lambda}{\int S(\lambda)\ (c/\lambda)\ d\lambda},
\end{equation}
where $f_\lambda(\lambda)$ is the spectral energy distribution (SED) per unit wavelength of a star, and $S(\lambda)$ is the transmission curve of the corresponding SuperBIT band (see Fig.~2 of \citealt{mccleary2023}).}
\begin{table}[t]
\centering
\caption{Photometric zeropoints (ZP) for each SuperBIT filter.}
\label{tab:calibration}
\begin{tabular}{cc}
\toprule
\textbf{Filter} & \textbf{ZP} \\
\midrule
F400W ($u$) & $28.687 \pm 0.005$ \\
F480W ($b$) & $30.223 \pm 0.004$ \\
F600W ($g$) & $29.610 \pm 0.002$ \\
% F640M ($r$) & $31.440 \pm 0.409$ \\
\bottomrule
\end{tabular}
\end{table}
We compute the synthetic magnitude for all matched stars within the SuperBIT fields of view that have Gaia spectra in Gaia DR3~\citep{gaiadr3, gaiadr3_cv}, using the Gaia XP continuous spectra via \textsc{GaiaXPy}~\citep{gaiaxpy}. This yields $\sim$4500 such stars across 30 SuperBIT targets. For each star we also compute an instrumental magnitude $m_{\rm inst}$ with an unknown zeropoint,
\begin{equation}\label{eq:m_inst}
    m_{\rm inst} = -2.5 \log_{10}(F_{\rm inst}) + \mathrm{ZP},
\end{equation}
where $F_{\rm inst}$ is the \texttt{FLUX\_AUTO} quantity obtained from \textsc{SExtractor}. Equating $m_{\rm inst}$ (Eq.~\ref{eq:m_inst}) with the synthetic AB magnitude $m_{AB}$ (Eq.~\ref{eq:m_syn}) for each star gives a per-star estimate of the zeropoint,
\begin{equation}
    \mathrm{ZP} = m_{AB} + 2.5 \log_{10}(F_{\rm inst}).
\end{equation}
We estimate its uncertainty via jackknife resampling, leaving out one of the 30 targets at a time. The resulting values, with their uncertainties, are reported in Table~\ref{tab:calibration}. The calibrated magnitude distributions of detected objects in all three bands are shown in Figure~\ref{fig:mag-depth}.

\section{PSF Modeling}\label{sec:psf-modeling}

The point-spread function (PSF) describes the response of an imaging system to a point source, i.e., how the image of a distant star appears when recorded by the telescope. Accurate modeling of the PSF is essential for weak-lensing analyses because galaxy shapes must be inferred prior to convolution with the PSF (the \emph{pre-PSF} galaxy properties). In practice, the PSF is characterized using stars within the field of view (FOV) of each exposure, which provide localized measurements of the PSF at their respective positions. Since the PSF can vary significantly across the CCD plane, the aim is to model these spatial variations to ensure accurate characterization of the PSF. \sayan{In this section we discuss the steps involved in PSF modelling and present the assessment results aggregated across all exposures from all targets, while more detailed per-target PSF modelling residuals are discussed in Appendix~\ref{app:rho-stats}.}

\subsection{Selection of Stars}

\begin{figure}[]
\centering
\includegraphics[width=0.46\textwidth]{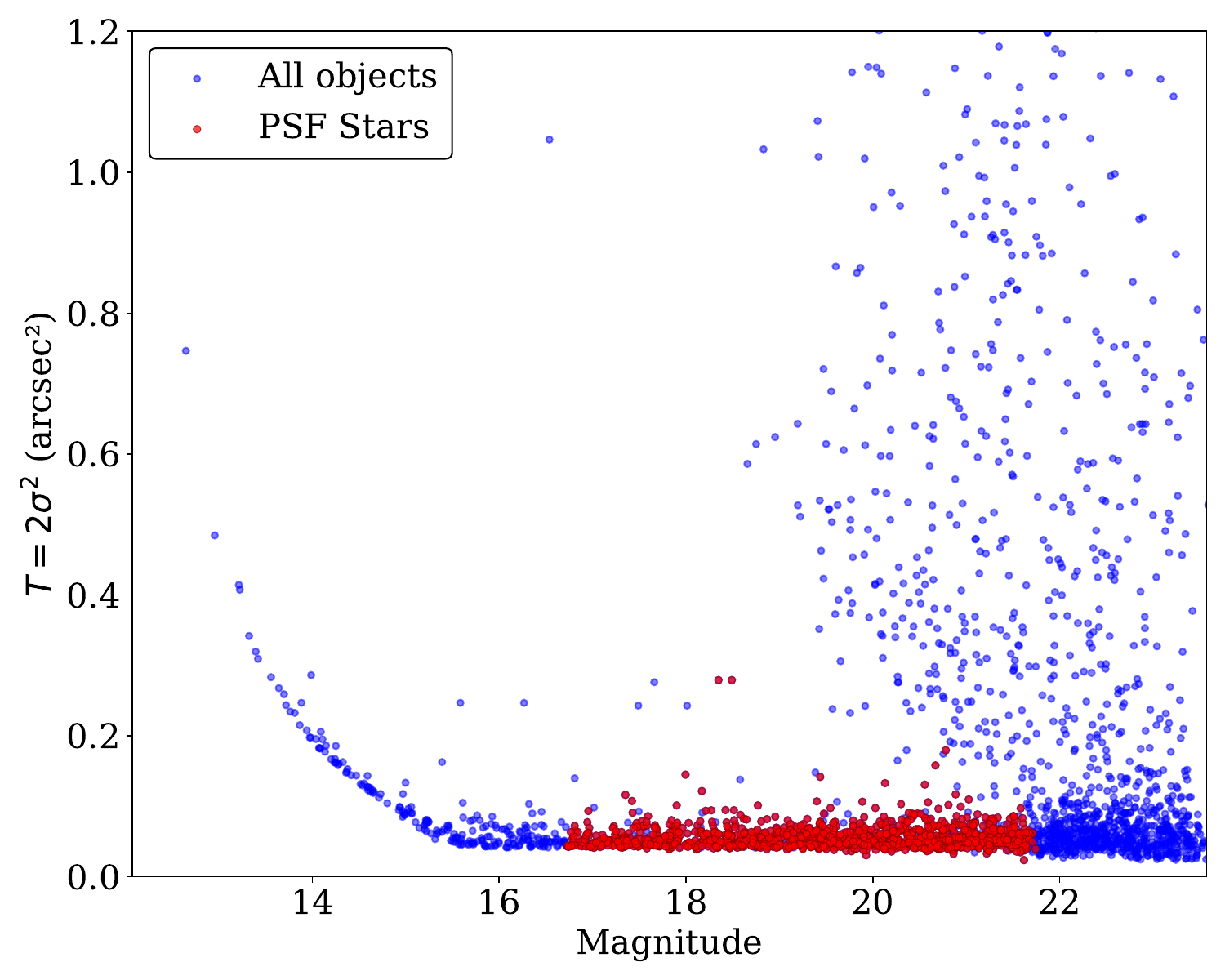}
\caption{Size-magnitude diagram for a single exposure. The stellar locus is visible, allowing the identification of stars suitable for PSF modeling.}
\label{fig:stellar-locus}
\end{figure}

Accurate PSF modeling requires selecting stars that reliably represent the PSF. After detecting objects in an exposure, we identify stars by locating the stellar locus in the size-magnitude diagram (using F480W ($b$)-band magnitudes). This identification is aided by cross-matching the detected sources with the \textit{Gaia} DR3~\citep{gaia2016, gaiadr3} star catalog, an example of which is shown in Figure \ref{fig:stellar-locus}. %plotting the size 
%$T$ versus the \texttt{MAG\_AUTO} parameter from \textsc{SExtractor}.
The size $T$ and ellipticities $e_1$ and $e_2$ are derived from the second-order moments of the light profile, $I(x,\ y)$~\citep{Schneider:1994eq}:
\begin{equation}\label{eq:T_admom}
T = Q_{xx} + Q_{yy},
\end{equation}
\begin{equation}\label{eq:e_admom}
e = e_1 + i e_2 = \frac{Q_{xx} - Q_{yy} + 2i Q_{xy}}{Q_{xx} + Q_{yy} + 2 \sqrt{Q_{xx} Q_{yy} - Q_{xy}^2}},
\end{equation}
where the second-order moments $Q_{\mu\nu}$ are defined as
\begin{equation}
    Q_{\mu\nu} = \frac{\int dx dy\ I(x,\ y)(\mu - \bar{\mu})(\nu - \bar{\nu})}{\int dxdy\ I(x,\ y)}.
\end{equation}
\begin{figure}[]
\centering
\includegraphics[width=0.48\textwidth]{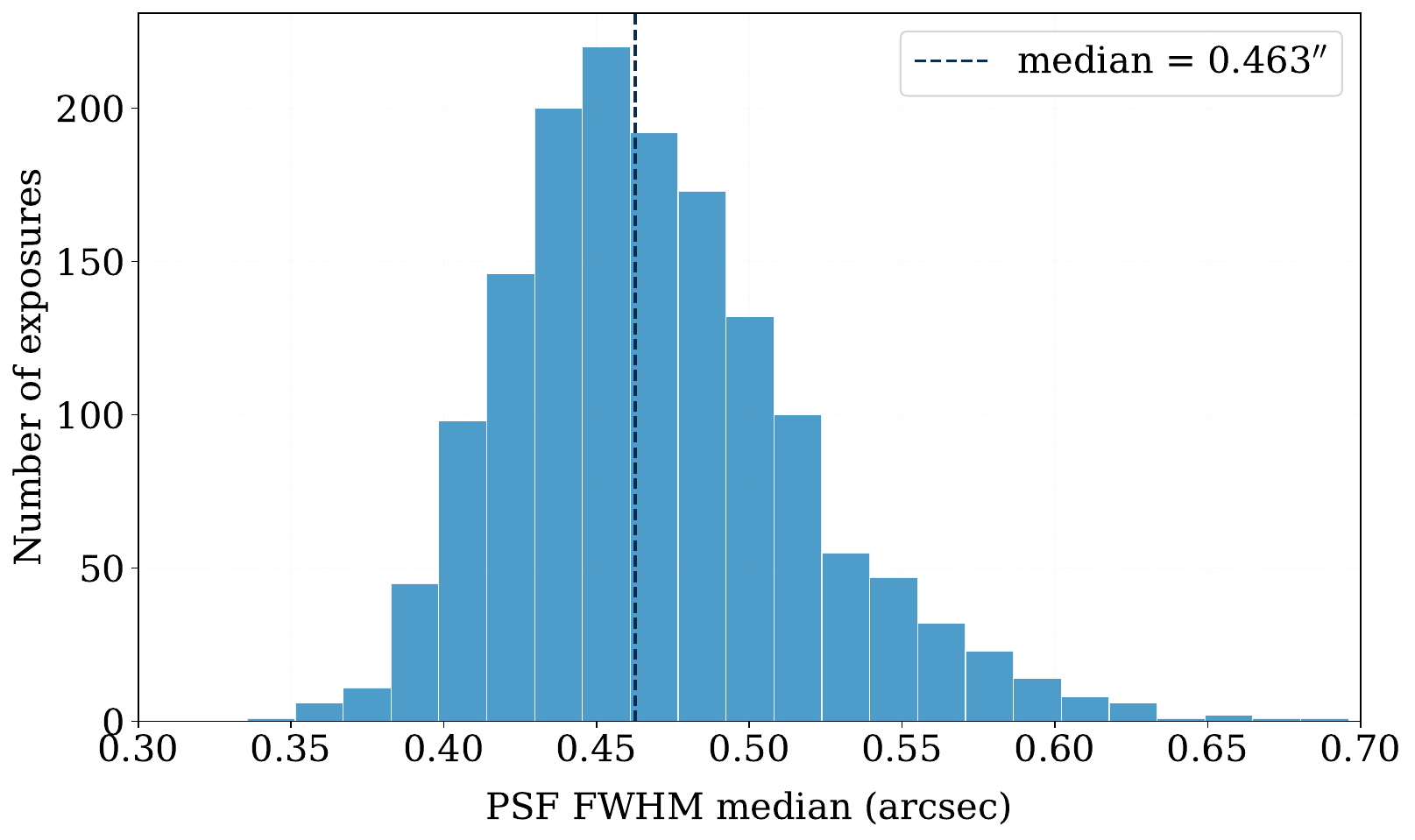}
\caption{Distribution of median PSF FWHM per exposure. This summarizes image quality across the dataset.}
\label{fig:PSF-FWHM}
\end{figure}
The moments $Q_{ij}$ are measured using the \texttt{AdaptiveMom} method implemented in the HSM module~\citep{Mandelbaum:2005wv}. As shown in Figure~\ref{fig:stellar-locus}, the stellar locus is identified by the fact that the size of stars should be independent of their magnitude. However, for very bright stars, CCD saturation occurs, leading to charge overflow. This effect artificially increases the observed size of the stars, producing what are known as saturated stars. On the other hand, galaxies can have a wide range of sizes, typically larger than the PSF, as illustrated in the figure.

To select stars suitable for PSF modeling, we choose those that are at least 1.5 magnitudes fainter than the faintest saturated stars. We further exclude objects in regions where the stellar locus merges with galaxies by imposing a minimum detection SNR (\texttt{SNR\_WIN} from \textsc{SExtractor}) of 20. Consequently, the magnitude range of good stars typically lies between \sayan{$\mathrm{mag}\sim16.7$ and $\mathrm{mag}\sim21.7$}. 
\begin{figure}[]
\centering
\hspace*{-2em}
\includegraphics[width=0.5\textwidth]{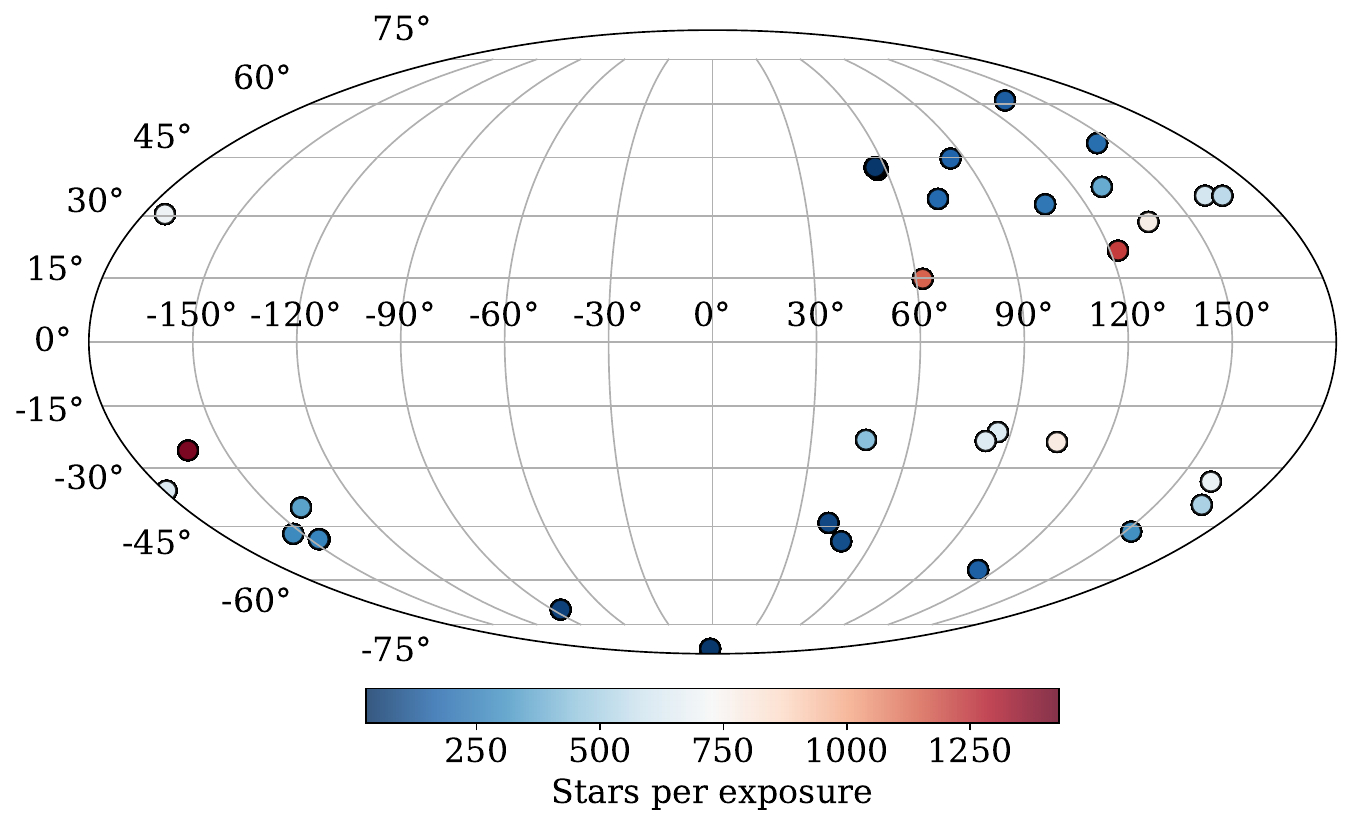}
\caption{Median number of stars per exposure for each target, shown in Galactic coordinates. Targets near the Galactic plane exhibit significantly higher stellar densities.}
\label{fig:star-dist}
\end{figure}
\begin{figure}[]
\centering
\hspace*{-2em}
\includegraphics[width=0.5\textwidth]{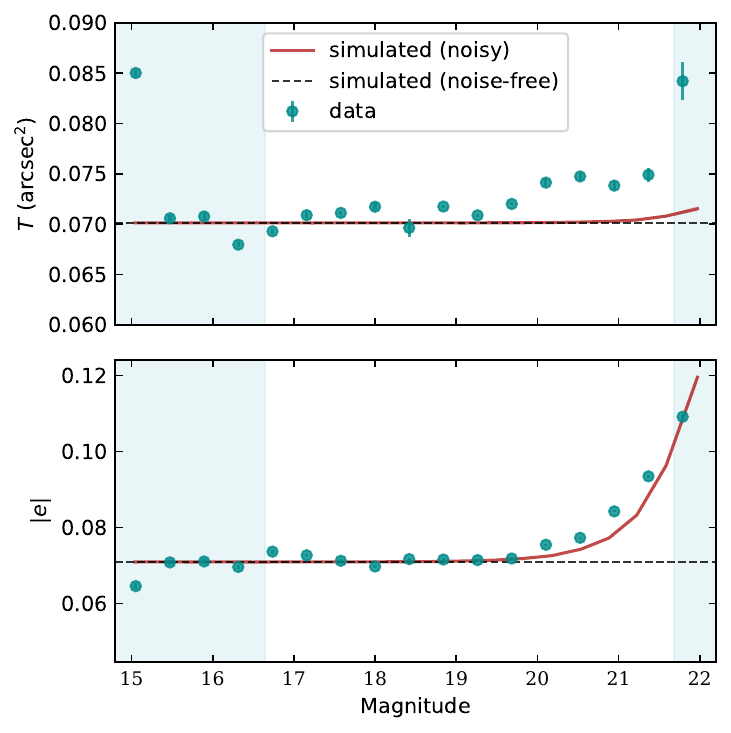}
\caption{Assessment of the brighter-fatter and brighter-rounder effects in SuperBIT CMOS detectors, using stars. The upper panel shows the size ($T$) of detected stars, and the lower panel shows their ellipticity, both as a function of magnitude. Teal points show the measured data, while the black dashed and red solid lines show results from noise-free and noisy simulations. The teal shaded regions correspond to the stars that have been excluded from PSF modelling.}
\label{fig:bf-round-mag}
\end{figure}
\sayan{The temporal stability of the PSF during a single five-minute exposure has been discussed in detail in \citealt{Gill2024} (Sec.~4.1--4.2). In this work, we also note that exposure quality varies significantly across different exposures at the same pointing (details in Appendix~\ref{appendix:exp-qual}). This variability may arise from residual field-dependent image motion when fine guidance relies on a single Focal Star Camera (FSC). A detailed analysis of this effect using SuperBIT flight data, along with implications for GigaBIT, is presented in \citet{Voyer2026}. This instability causes the PSF to become more elliptical and larger near the edges of the field. The scatter in stellar sizes and PSF ellipticity across the CCD therefore serve as useful proxies for exposure quality. Figure~\ref{fig:exp-qual} in Appendix~\ref{appendix:exp-qual} shows an example of how exposure stability evolves across multiple exposures of a single target (Abell 3411 in this case). The upper panel of that figure shows more specifically how the same set of stars appears markedly different between two consecutive exposures, with sizes increasing near the edges; the colorbar represents distance from the exposure centre. For this reason, it is not possible to impose an upper limit on $T$ when selecting stars for PSF modelling: doing so would preferentially select stars from the inner regions of the exposure, resulting in an incorrect PSF model. We therefore rely solely on matching SuperBIT detections to a pure star sample from Gaia for star selection.} 

We adopt a combined scatter-ellipticity exposure-quality criterion to select exposures used in the shape-measurement pipeline. The exposure selection scheme is described in detail in Appendix~\ref{appendix:exp-qual}. \sayan{Since a deeper coadd improves source detection, all exposures are used to construct it, regardless of individual exposure quality; shape measurement, however, is not performed on the coadd. Only the selected exposures are then used to build the Multi-Epoch Data Structure (MEDS)\footnote{\url{https://github.com/esheldon/meds/wiki}}, on which the shape-measurement and metacalibration pipeline (detailed in Section~\ref{sec:metacal}) is run.} The median FWHM of the selected good stars serves as a useful proxy for assessing exposure quality. Figure~\ref{fig:PSF-FWHM} shows the distribution of median PSF FWHM for different exposures. Another important factor is the number of available good stars for PSF fitting. As discussed in the next section, this number can significantly impact the quality of PSF model interpolation. In our observations, which span a wide range of sky positions, the number of stars per exposure varies as expected: fields near the Galactic plane contain more stars, while high-latitude fields contain fewer. The distribution of the number of stars per exposure across different targets is shown in Figure~\ref{fig:star-dist}.
\begin{figure}[]
\centering
\hspace*{-2em}
\includegraphics[width=0.5\textwidth]{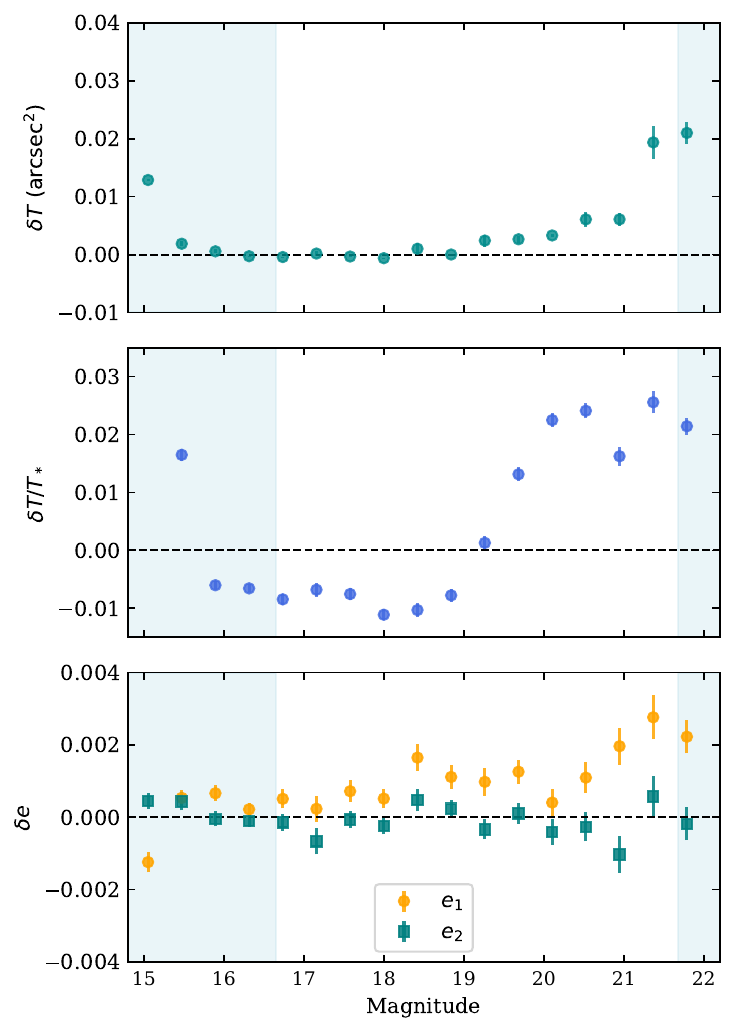}
\caption{PSF model residuals for the reserved test sample as a function of magnitude: size (top), fractional size (middle), and ellipticity components (bottom). The shaded region indicates stars that were excluded from the PSFEx model fitting. The test sample includes the 20\% of good stars reserved for validation, as well as the excluded (bad) stars.}
\label{fig:psf-residual-mag}
\end{figure}
\sayan{\textbf{Brighter-Fatter-Rounder Effect:} Detector response can also depend on the flux of the observed object. Two such phenomena, originally reported for CCDs, are the brighter-fatter (BF) effect \citep{bf1} and the brighter-rounder effect \citep{Finner:2023nwl}. For the SuperBIT CMOS detectors, we investigate these effects in Figure~\ref{fig:bf-round-mag}, where the measured sizes and ellipticities of all detected stars are binned as a function of magnitude. We do not find evidence for a significant BF effect within the PSF star-selection region; instead, we observe a slight fainter-fatter trend.  We do detect a significant brighter-rounder trend. To test whether this trend could simply arise from noise bias in the HSM second-moment based shape measurements of faint, low-S/N objects \citep{noisebias_1, noisebias_2}, we performed the same analysis on simulated star images convolved with a SuperBIT PSF obtained from real data. The test was carried out both without noise (black dashed line in Figure~\ref{fig:bf-round-mag}) and with noise matched to the expected single-exposure noise level (red solid line). For the ellipticity measurements, the noisy simulations reproduce the increase in ellipticity toward fainter magnitudes seen in the real data, indicating that the bias in the HSM moment-based shape measurement is the dominant contributor to the observed brighter-rounder trend. In contrast, the fainter-fatter trend observed in the size measurements is not fully explained by noise bias alone. An important limitation is that \textsc{PSFEx} models the PSF only as a function of detector position and cannot account for flux-dependent variations. If flux-dependent PSF effects are present, \textsc{PSFEx} will be unable to capture them through its spatial model alone. This limitation would be expected to manifest as magnitude-dependent PSF residuals, which we investigate in the next subsection.}
\subsection{PSF Interpolation \& Diagnostics} \label{subsec:psfex}
Stars can be used to trace the spatial variation of the PSF across the field of view. Interpolation is required in regions where no stars are available, particularly at the positions of galaxies, whose intrinsic (pre-PSF) shapes we wish to measure. It is therefore essential to test the PSF model interpolation and to quantify any residuals.

We use the \textsc{PSFEx}\footnote{\url{https://www.astromatic.net/software/psfex/}}~\citep{psfex} software to model the PSF on a pixel basis. We have tested several \textsc{PSFEx} configuration setups and found that the following configuration yields the lowest modeling residuals:
\begin{verbatim}
BASIS_TYPE PIXEL
PSF_SAMPLING 0.5
PSF_SIZE 101,101
PSFVAR_KEYS XWIN_IMAGE,YWIN_IMAGE
PSFVAR_GROUPS 1,1
PSFVAR_DEGREES 5
\end{verbatim}
For each exposure, the good stars selected in the previous subsection are randomly split into training and validation samples in an 80:20 ratio. The PSF model is fitted on the 80\% training set, while the remaining 20\% are reserved to assess the accuracy of the interpolation.
\begin{figure*}[!t]
\centering
\includegraphics[width=\textwidth]{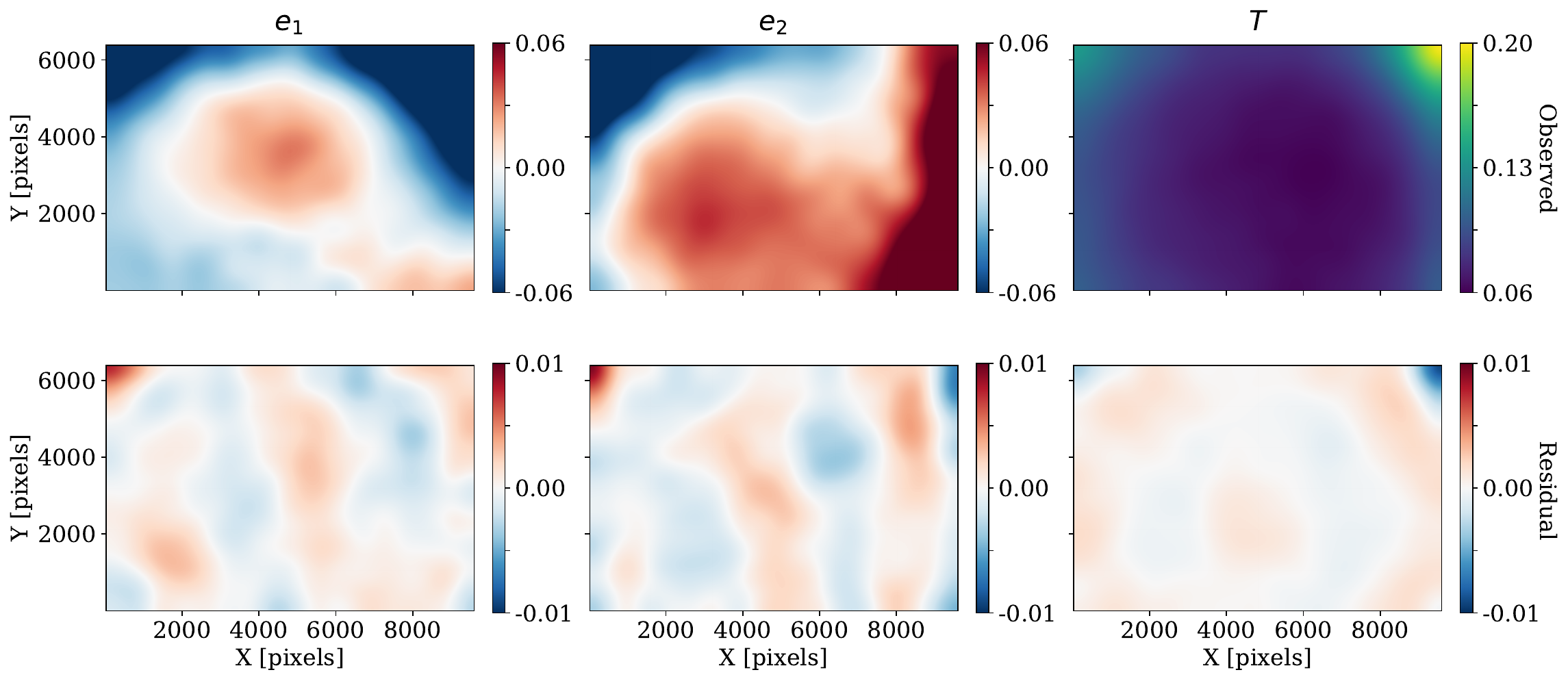}
\caption{Spatial variability of stellar ellipticity and size across CCD coordinates. The top row shows ellipticity and size measurements of the reserved test stars, while the bottom row shows the residuals after subtracting the PSFEx model evaluated at the same positions.}
\label{fig:psf-residual-spatial}
\end{figure*}
\begin{figure*}[!t]
\centering
\includegraphics[width=0.935\textwidth]{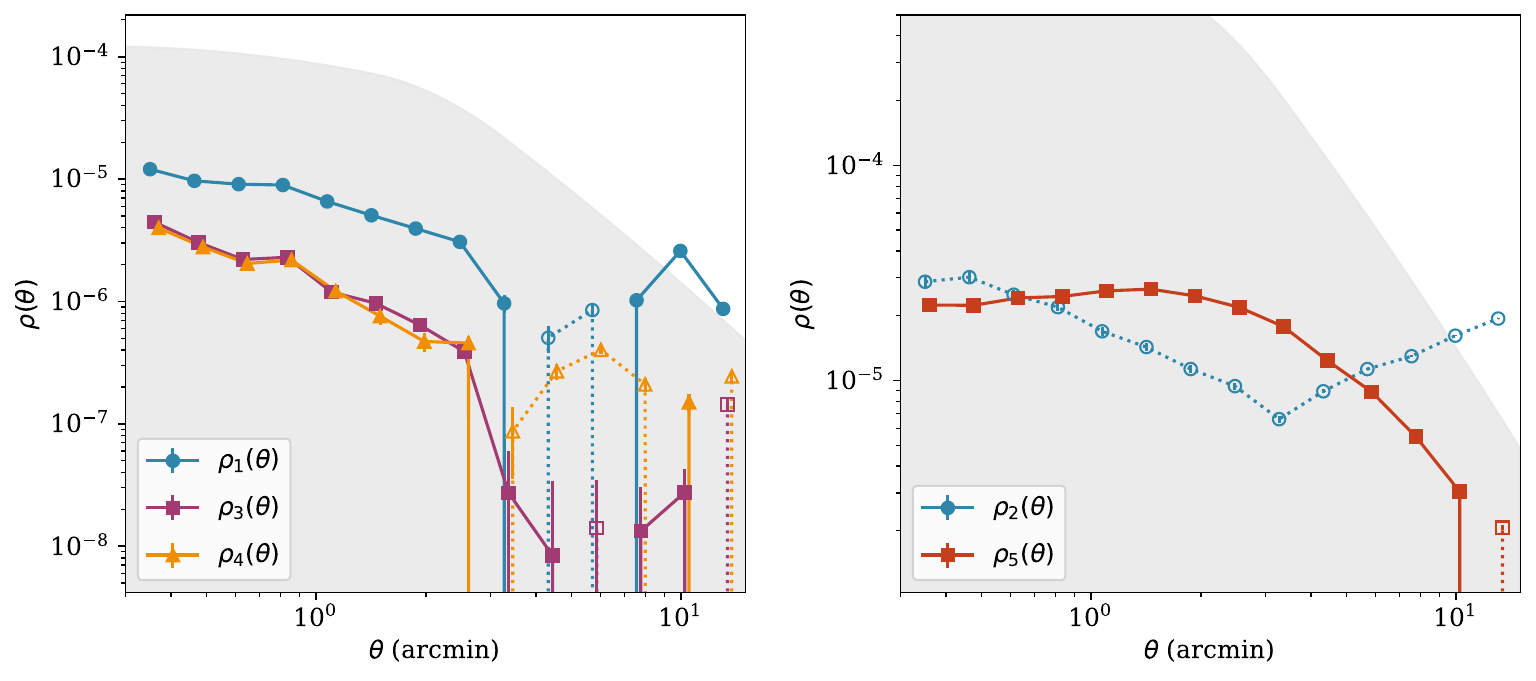}
\caption{
$\rho$ statistics as defined in Eqs.~\ref{eq:rho1}--\ref{eq:rho5}. 
\emph{Left:} $\rho_1$ (blue circles), $\rho_3$ (rose squares), and $\rho_4$ (orange triangles). 
\emph{Right:} $\rho_2$ (blue circles) and $\rho_5$ (red squares). 
In both panels, dotted line segments indicate negative values. 
The statistics are computed using the 20\% reserved test stars that were not included in PSF modeling; the sample includes stars from multiple exposures and targets. 
The shaded region indicates a ``safe zone" corresponding to 5\% of the expected shear two-point correlation function for a representative cluster with mean mass $M_{500}$ and redshift $z$ matching the SuperBIT sample. 
For this estimate, we adopt conservative values of $T_{\mathrm{PSF}}/T_{\mathrm{gal}} = 1$ and leakage strength $\alpha = 0.1$.
}
\label{fig:rho-stats}
\end{figure*}
After fitting the model with the PSFEx settings described above, we evaluate the PSF at the positions of the test stars. For both the observed star images and the PSF model renderings, we measure the sizes and ellipticities using the second-order adaptive moments defined in Eq.~\ref{eq:T_admom} and Eq.~\ref{eq:e_admom}. This procedure is carried out for all exposures across all targets. Figure~\ref{fig:psf-residual-mag} presents the model residuals in size and ellipticity as a function of magnitude. Importantly, the validation sample includes both the 20\% of reserved good stars and the stars excluded by our selection criteria, i.e., those that are either saturated (very bright) or have very low detection SNR. The shaded regions in Figure~\ref{fig:psf-residual-mag} denote the magnitude ranges excluded from the training sample. \sayan{As discussed in the previous subsection, the brighter-rounder trend shows up in the 
$e_1$ residual. We also see a magnitude-dependent size residual, but with a trend opposite to the brighter-fatter effect: fainter objects show higher residuals.}

We further assess the spatial variability of the PSF interpolation quality in CCD coordinates, shown in Figure~\ref{fig:psf-residual-spatial}, where we compare the measured ellipticity and size of the test stars with the \texttt{PSFEx} model predictions and display the corresponding residuals across the detector. 

In principle, PSF modeling residuals can propagate into galaxy shape measurements and subsequently bias the reconstruction of the weak-lensing signal. A standard way to quantify this effect is to inspect the contribution of PSF model residuals to the two-point shear correlation statistics~\citep{Jarvis:2002vs, DES:2015yfn}, defined as
\begin{equation}\label{eq:gg_corr}
\xi_{+}(\theta) = \left\langle e^{}(\mathbf{x}) e(\mathbf{x} + \theta) \right\rangle.
\end{equation}
Although our aim is to measure the true shear from an ensemble of galaxies, all shape measurement algorithms are subject to systematic biases. A common parametrization~\citep{Heymans:2005rv} is
\begin{equation}
\langle e \rangle = (1 + m)g_{\mathrm{true}} + \alpha e_{\mathrm{PSF}} + c,
\end{equation}
where $m$ represents the multiplicative bias, $\alpha$ quantifies the ``leakage" of PSF shape $e_{\mathrm{PSF}}$ into galaxy shapes, and $c$ denotes an additive bias. With this parametrization, the contribution of PSF leakage to the shear two-point function in Eq.\ref{eq:gg_corr} can be expressed as~\citep{DES:2015yfn, DES:2017ibv}:
\begin{eqnarray}
\delta\xi_{+}(\theta) &=& \left\langle \frac{T_{\mathrm{PSF}}}{T_{\mathrm{gal}}} \right\rangle^2 (\rho_1(\theta) + \rho_3(\theta) + \rho_4(\theta)) \nonumber \\
&& - \alpha \left\langle \frac{T_{\mathrm{PSF}}}{T_{\mathrm{gal}}} \right\rangle (\rho_2(\theta) + \rho_5(\theta)),
\end{eqnarray}
where the $\rho$'s are two-point statistics of PSF ellipticity and size residuals, defined as~\citep{rowe2010, DES:2015yfn}:
\begin{eqnarray}\label{eq:rho1}
\hspace{-10pt}\rho_1(\theta) &\equiv& \langle \delta e_{\mathrm{PSF}}^*(\bold{x}) \delta e_{\mathrm{PSF}}(\bold{x} + \bold{\theta}) \rangle, \\
\hspace{-10pt}\rho_2(\theta) &\equiv& \langle e_{\mathrm{PSF}}^*(\bold{x}) \delta e_{\mathrm{PSF}}(\bold{x} + \bold{\theta}) \rangle \label{eq:rho2}\\
\hspace{-10pt}\rho_3(\theta) &\equiv& \left\langle \left( e_{\mathrm{PSF}}^* \frac{\delta T_{\mathrm{PSF}}}{T_{\mathrm{PSF}}} \right)(\bold{x}) \left( e_{\mathrm{PSF}} \frac{\delta T_{\mathrm{PSF}}}{T_{\mathrm{PSF}}} \right)(\bold{x} + \bold{\theta}) \right\rangle \\
\hspace{-10pt}\rho_4(\theta) &\equiv& \left\langle \delta e_{\mathrm{PSF}}^*(\bold{x}) \left( e_{\mathrm{PSF}} \frac{\delta T_{\mathrm{PSF}}}{T_{\mathrm{PSF}}} \right)(\bold{x} + \bold{\theta}) \right\rangle \\
\hspace{-10pt}\rho_5(\theta) &\equiv& \left\langle e_{\mathrm{PSF}}^*(\bold{x}) \left( e_{\mathrm{PSF}} \frac{\delta T_{\mathrm{PSF}}}{T_{\mathrm{PSF}}} \right)(\bold{x} + \bold{\theta}) \right\rangle,
\label{eq:rho5}
\end{eqnarray}
We compute the $\rho$ statistics using the reserved stars in each exposure, including stars from multiple exposures and targets, and present the results in Figure~\ref{fig:rho-stats}. For reference, we also show a shaded ``safe zone" in the figure, defined as 5\% of the expected galaxy two-point correlation signal from a cluster representative of the SuperBIT target sample with mean mass $M_{500} = 6.23 \times 10^{14} M_\odot$ and mean redshift $z = 0.245$. To estimate this safe zone, we adopt a deliberately conservative choice of $T_{\text{PSF}}/T_{\text{gal}}=1$ and PSF leakage strength $\alpha=0.1$. For the background galaxy redshift distribution, we use the COSMOS2015 catalog\footnote{\url{https://irsa.ipac.caltech.edu/cgi-bin/Gator/nph-dd?catalog=cosmos2015}}~\citep{Laigle:2016jxn, cosmos2015},
convert the sources to SuperBIT F480W ($b$) band fluxes using their SEDs and the SuperBIT transmission curve shown in Figure~2 of \citep{mccleary2023}, and then simulate observations with a real SuperBIT PSF (the PSFEx solution derived from observations) and noise conditions. Our fiducial simulations are described in more detail in Section~\ref{subsec:fid-sim}. We run our shape-measurement pipeline on these fiducial simulations and select objects using the selection criteria described in Section~\ref{sec:shape-measure}, yielding the expected redshift distribution of background sources for this analysis. The $\rho$-statistics analysis demonstrates that residual PSF modeling errors are sufficiently controlled to avoid significant contamination of the weak lensing signal within our requirements. All two-point statistics
are computed using the \texttt{TreeCorr} package~\footnote{\url{https://github.com/rmjarvis/TreeCorr}}~\citep{Jarvis:2003wq}.

\sayan{The ellipticity and size residuals in Figures~\ref{fig:psf-residual-mag} and \ref{fig:psf-residual-spatial}, along with the $\rho$-statistics in Figure~\ref{fig:rho-stats}, are aggregated results across all exposures acquired for all science targets. While these results are useful for understanding the average modelling efficiency, we also need to perform these tests separately for each target in order to carry out further science with the weak-lensing signal. In Appendix~\ref{app:rho-stats}, we provide a per-target assessment in terms of mean ellipticity residuals and mean $\rho$-statistics. We also plot the  $\rho$-statistics for each target in Galactic coordinates (Figure~\ref{fig:rho-dist}) to check for any correlation between star availability and modelling quality; we find no such correlation. We would also like to note that prior work in the literature has shown bias contributions from higher-order-moment-based residuals and the corresponding $\rho$-statistics \citep{Zhang:2022aer,DES:2025fvt}, which are of greater concern for cosmic shear analyses. We have not yet investigated such residuals in this work, as we expect cluster weak lensing to be less sensitive to them, but we plan to do so in future work.}

\section{Shape Catalog}\label{sec:shape-measure}
To reconstruct the weak-lensing signal, we require an observable that traces the shear field. 
The standard choice in weak-lensing analyses is the shape of galaxies, which is parameterized by a two-component ellipticity:
\begin{equation}
\bold{e} = e_1 + i e_2;\quad e_1 = e \cos(2\beta),\ e_2 = e \sin(2\beta),
\end{equation}
where $e$ is the ellipticity amplitude and $\beta$ is the position angle of the major axis measured counter-clockwise from the $+x$ image axis. 
Consistent with the moment-based definition in Eq.~\ref{eq:e_admom}, the ellipticity of an object with elliptical isophotes can be expressed as
\begin{equation}
    e = \frac{a-b}{a+b},
\end{equation}
with $a$ and $b$ the semi-major and semi-minor axes, respectively. In this convention, the ensemble average of galaxy ellipticities estimates the reduced shear,
\begin{equation}
    \langle e\rangle \equiv g = \frac{\gamma}{1-\kappa},
\end{equation}
with $\gamma$ and $\kappa$ denoting the shear and convergence, respectively. 
\begin{figure*}[t]
\centering
\includegraphics[width=\textwidth]{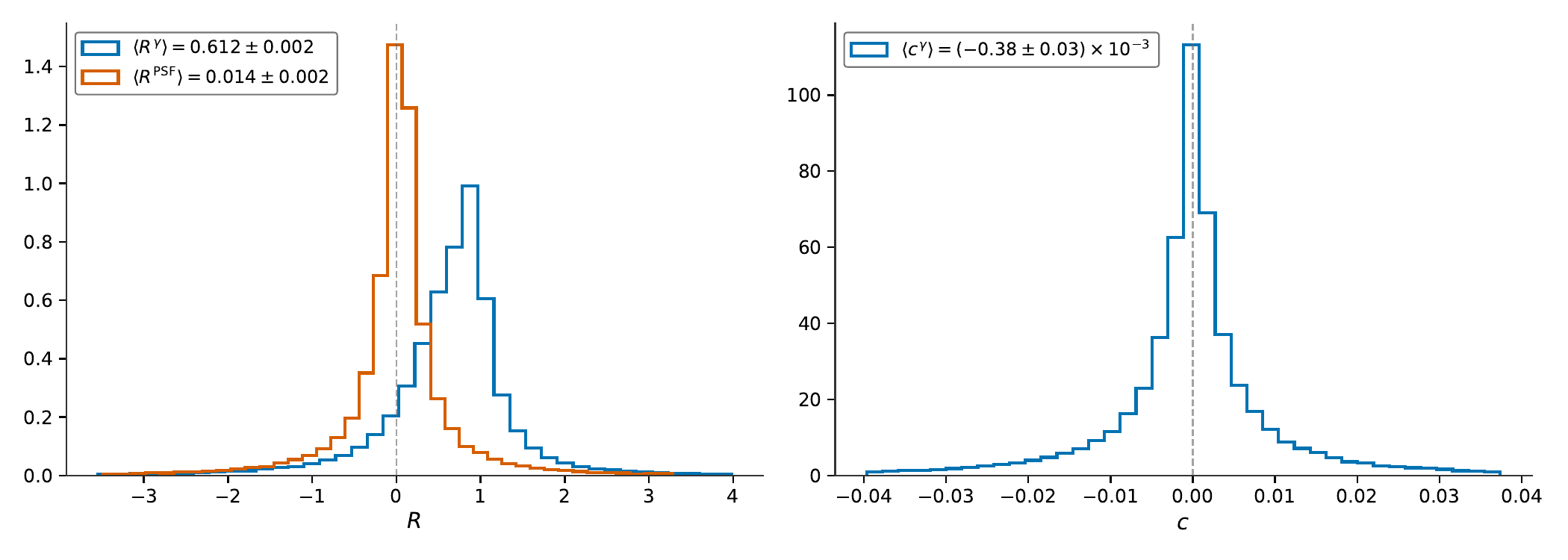}
\caption{Distributions of per-object shear calibration quantities. 
\emph{Left:} The galaxy shear response, $R^{\gamma}$ (Eq.~\ref{eq:shear-response}), and the PSF shear response, $R^{\mathrm{PSF}}$ (Eq.~\ref{eq:PSF-shear-response}), with vertical dashed line marking $R=0$. 
\emph{Right:} The additive correction term, $c^{\gamma}$ (Eq.~\ref{eq:additive-corrections}), with vertical dashed line marking $c=0$. 
Legends report the mean values for each distribution.}
\label{fig:response-distribution}
\end{figure*}

The task of measuring galaxy shapes is highly non-trivial in the presence of observational effects such as the PSF, pixelization, and sky background noise. Since weak lensing is primarily concerned with the applied gravitational shear on galaxies, the problem can be reformulated: rather than requiring perfectly unbiased measurements of intrinsic ellipticity, one can instead quantify how the measured ellipticity of each galaxy responds to an applied external shear $\bold{\gamma} = \gamma_1 + i \gamma_2$. If this shear response, $R$, can be estimated without bias, the weak-lensing shear field can be reconstructed even in the presence of non-ideal observational systematics. Several approaches to estimating this response exist (see \citealt{Mandelbaum:2017jpr} for a detailed review). In this work, we employ the \textsc{Metacalibration} algorithm~\citep{Huff:2017qxu, Sheldon:2017szh}.%, which directly estimates the response from the image data itself, rather than relying on simulations.
\subsection{\textsc{Metacalibration} Overview}\label{sec:metacal}

The fundamental principle behind \textsc{Metacalibration} is to measure the shear response of a shape estimator $\bold{e}$ directly from the data, rather than relying on simulations. The shape estimator can be expanded as a Taylor series in terms of a small applied shear $\bold{\gamma}$:
\begin{equation}\label{eq:metcal1}
\bold{e} = \bold{e}|_{\bold{\gamma}=0} + \left.\frac{\partial \bold{e}}{\partial \bold{\gamma}}\right|_{\bold{\gamma}=0}\bold{\gamma} + \cdots,
\end{equation}
where this relation holds at the per-object level. The per-object response is defined as
\begin{equation}
\bold{R}^{\gamma} = \left.\frac{\partial \bold{e}}{\partial \bold{\gamma}}\right|_{\bold{\gamma}=0}.
\end{equation}
Numerically, the response can be estimated as
\begin{equation}\label{eq:shear-response}
R^{\gamma}_{ij} = \frac{e^+_i - e^-_i}{\Delta \gamma_j},
\end{equation}
where $e^+_i$ is the $i$-th ellipticity component measured on an image sheared by $+\gamma_j$, and $e^-_i$ is the corresponding measurement on an image sheared by $-\gamma_j$. The differential shear in the denominator is $\Delta\gamma_j = 2\gamma_j$; we adopt $\gamma_j = 0.01$.

We estimate these \textsc{Metacalibration} terms using the \textsc{ngmix} package~\footnote{\url{https://github.com/esheldon/ngmix}}~\citep{Sheldon:2014vda}. For each object, \textsc{ngmix} uses the discrete Fourier transform (DFT) of the PSF to deconvolve the galaxy image, applies an artificial shear in four directions (1+, 1$-$, 2+, 2$-$), and reconvolves the result with a slightly dilated version of the PSF using the \texttt{MetacalDilatePSF} class. Since this process does not yield a perfectly round PSF kernel, it is necessary to account for additive shear bias using PSF shear terms produced by \textsc{ngmix}.

Thus, to calculate the shear response, we construct a dictionary of five image versions for each galaxy: $\{\texttt{noshear}, 1+, 1-, 2+, 2-\}$. (Here, \texttt{noshear} refers to deconvolution and reconvolution of the galaxy image with no artificial shear applied.) In addition, we produce four PSF shear terms: $\{\texttt{PSF1+}, \texttt{PSF1-}, \texttt{PSF2+}, \texttt{PSF2-}\}$, where only the reconvolution PSF is sheared (the galaxy image remains unsheared). With these terms, we can similarly estimate the per-object PSF response:
\begin{equation}\label{eq:PSF-shear-response}
R^{\text{PSF}}_{ij} = \frac{e^{\text{PSF},+}_i - e^{\text{PSF},-}_i}{\Delta\gamma_j},
\end{equation}
where $e^{\text{PSF},+}_i$ is the $i$-th ellipticity measured on an image convolved with a PSF sheared by $+\gamma_j$, and $e^{\text{PSF},-}_i$ is the corresponding measurement for $-\gamma_j$. In this case, as well, we adopt $\Delta\gamma_j=2\gamma_j = 0.02$.

Finally, to correct for any surviving additive bias, we calculate additive terms for galaxy shear \citep{Huff:2017qxu}:
\begin{eqnarray}\label{eq:additive-corrections}
\bold{c}^{\gamma} &=& \frac{\bold{e}^+ + \bold{e}^-}{2} - \bold{e}_{\text{noshear}}.
\end{eqnarray}
\begin{figure*}[t]
\centering
\includegraphics[width=\textwidth]{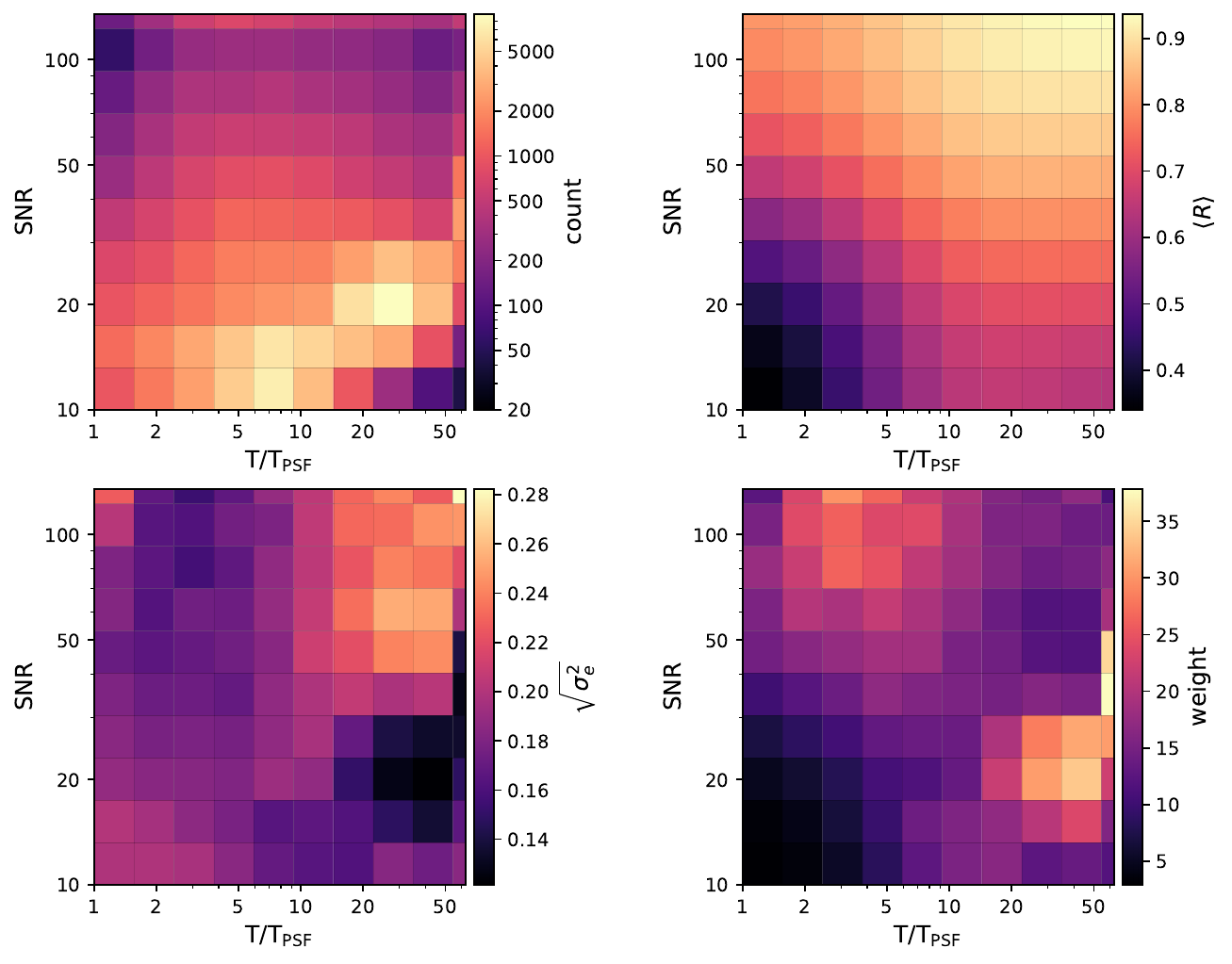}
\caption{Several statistics of the catalog as a function of \textsc{metacalibration} quantities: counts (upper left), $\sqrt{\langle\sigma_{1,2}^2\rangle}$ (lower left), response $R_{\gamma}$ (upper right), and weight as defined in Eq.~\ref{eq:weight} (lower right), shown as a function of gridded SNR and $T/T_{\rm PSF}$ parameter space. The grids have been chosen to include 95\% of the objects in a $10\times10$ logarithmic bin and all objects above 95 percentile have been accumulated in the last bin.}
\label{fig:weights}
\end{figure*}
\sayan{To account for any PSF leakage, we also subtract $\langle R_{\text{PSF}} \rangle e_{\text{PSF}}$, calculated for each individual galaxy. PSF response correction is particularly important in our case, since our reconvolution kernel is a dilated version of the original PSF. We discuss the improvement resulting from this correction using simulations in Appendix~\ref{app:psf-correction}.}

\sayan{We can finally write down the weighted average shear estimator from Eq.~\ref{eq:metcal1} as
\begin{equation}\label{eq:psf_response_corr}
    \langle \bold{\gamma} \rangle \approx \langle \bold{R} \rangle ^{-1} \left \langle (\bold{e}_{\text{noshear}} - \langle \bold{c}_{\gamma}\rangle - \langle \bold{R_{\text{PSF}}}\rangle \bold{e}_{\text{PSF}}) \right\rangle,
\end{equation}}
where
\begin{equation}
    \left<\bold R\right> = \left<\bold R^\gamma\right> + \left<\bold R^S\right>.
\end{equation}
Here, we also account for shear-dependent selection effects described in \citep{Sheldon:2017szh}, using a selection response for an artificially applied shear, defined as
\begin{equation}
    \left<R^S\right>_{ij} = \frac{\left<e_i\right>^{S+}-\left<e_i\right>^{S-}}{\Delta \gamma_j},
\end{equation}
where $\left<e_i\right>^{S+}$ corresponds to the mean of the $i$th ellipticity component measured on the \emph{unsheared} images, but for the sample selected using the images sheared by $+\gamma_j$, and similarly for $\left<e_i\right>^{S-}$. Figure~\ref{fig:response-distribution} shows the distributions of the multiplicative response and the additive correction for the final galaxy sample. 
%In the following subsection, we examine how the response varies with different \textsc{metacalibration} observables, which will guide the choice of selection cuts adopted in this work.
\subsection{Object selection for \textsc{Metacalibration}}\label{sec:object-selection}
Object selection is one of the most critical steps in weak-lensing analyses, as it directly impacts the fidelity of the final cosmological constraints and mass reconstructions. The aim is to maximize statistical power while ensuring that the \textsc{Metacalibration} response remains stable as a function of measurement-noise proxies. We select galaxies based on (i) the SNR of the galaxy model fit and (ii) the ratio of the intrinsic galaxy size to the PSF size at the galaxy's position.

\textsc{ngmix} provides an estimate of the SNR as part of the model-fitting process for each galaxy. This SNR corresponds to a matched-filtering estimate of the fitted model and is defined as
\begin{equation}
    \text{SNR} = \frac{\sum_p \frac{m_p I_p}{\sigma_p^2}}
    {\left(\sum_p \frac{m_p^2}{\sigma_p^2}\right)^{1/2}}, 
\label{eq:snr-ngmix}
\end{equation}
where the sum runs over image pixels $p$, $m_p$ is the profile of the best-fit galaxy model in pixel space, $I_p$ is the observed pixel intensity, and $\sigma_p$ is the pixel noise standard deviation.

The second proxy, $T/T_{\text{PSF}}$, quantifies the relative size of the galaxy with respect to the PSF, where $T$ denotes the pre-PSF galaxy size. By ``pre-PSF" we mean the size parameter of the best-fit galaxy model before convolution with the best-fit PSF model, as inferred from the full Bayesian forward-modeling framework implemented in \textsc{ngmix}. A brief overview of the \textsc{ngmix} workflow is provided in Appendix~\ref{appendix:ngmix-workflow}. 

Based on these measurement quantities, we apply a set of selection cuts to define our \textsc{Metacalibration} catalog. Our choices closely follow the criteria used in the SuperBIT forecasting analysis~\citep{mccleary2023} and the DES Y3 weak lensing analysis~\citep{DES:2020ekd}. Specifically, we impose:
\begin{enumerate}
    \item $10 < \text{SNR} < 1000$,
    \sayan{\item $\frac{T}{T_{\text{PSF}}} > 1.0$,}
    \item $T < 10$,
\end{enumerate}
We adopt a gridding approach to calculate the shear response $R_{\gamma}$ defined over the SNR-$T/T_{\rm PSF}$ parameter space. Each galaxy is assigned an $R_{\gamma}$ value corresponding to its grid cell. As demonstrated in Section \ref{subsec:fid-sim}, this gridded estimation of the response substantially reduces shear bias. \sayan{We discuss the stability of the shear-response at each bin in Appendix~\ref{app:r_stability}.} In addition to the shear response, we include a global selection response term estimated from the full sample rather than from the SNR-$T/T_{\rm PSF}$ grid. The final, response-calibrated shapes in our catalog are obtained by dividing by the total response.
In addition to these cuts, we apply color-based selections and red-sequence matching to separate foreground galaxies from background sources when reconstructing the weak-lensing signal (i.e., convergence maps). The details of this selection will be presented in the \emph{Lensing in the Blue IV: Convergence Maps} paper (Saha et al. 2026, in prep.).
\subsection{Weights}
\begin{deluxetable*}{lccccccc}
\tablecaption{
Summary of source number densities for the SuperBIT targets. 
The target type, sky position, redshift and three estimates of the source number density are shown for each target. 
The ``Raw'' column gives the source density after removing stars and applying masks. 
The ``Selected'' column reports the unweighted source density after applying the selection cuts described in  Section~\ref{sec:metacal}. 
The final column, $n_{\rm eff}$, shows the effective weighted source number density of selected objects as defined in Eq.~\ref{eq:neff}.
\label{tab:densities}
}
\tabletypesize{\footnotesize}
\tablewidth{0pt}
\tablehead{
\colhead{Target} &
\colhead{Target Type} &
\colhead{RA} &
\colhead{Dec} &
\colhead{Redshift} &
\colhead{Raw} &
\colhead{Selected} &
\colhead{$n_{\mathrm{eff}}$ (H12)} \\
\colhead{} &
\colhead{} &
\colhead{(deg)} &
\colhead{(deg)} &
\colhead{} &
\colhead{(arcmin$^{-2}$)} &
\colhead{(arcmin$^{-2}$)} &
\colhead{(arcmin$^{-2}$)}
}
\startdata
1E0657 Bullet & Merging Clusters & 104.61450 & -55.95475 & 0.296 & 22.87 & 12.27 & 9.29 \\
Abell 141 & Merging Clusters &  16.40085 & -24.66079 & 0.230 & 11.36 & 9.84 & 8.24 \\
Abell 1689 & Merging Clusters & 197.87025 &  -1.34168 & 0.183 & 22.17 & 20.35 & 17.03 \\
Abell 2163 & Merging Clusters & 243.94512 &  -6.14784 & 0.203 & 15.00 & 10.92 & 8.71 \\
Abell 2345 & Merging Clusters & 321.78648 & -12.16679 & 0.179 & 18.45 & 15.52 & 12.62 \\
Abell 2384a & Merging Clusters & 328.09425 & -19.56626 & 0.094 & 20.35 & 17.41 & 14.11 \\
Abell 2384b & Merging Clusters & 328.03567 & -19.72170 & 0.094 & 16.45 & 13.52 & 10.58 \\
Abell 3192 & Merging Clusters &  59.71568 & -29.94346 & 0.425 & 15.96 & 13.55 & 10.78 \\
Abell 3365 & Merging Clusters &  87.19233 & -21.92990 & 0.093 & 11.40 &  7.75 & 6.07 \\
Abell 3411 & Merging Clusters & 130.45046 & -17.49227 & 0.162 & 21.40 & 16.27 & 12.85 \\
Abell 3526 & Merging Clusters & 192.19688 & -41.30712 & 0.011 & 20.60 & 11.45 & 7.42 \\
Abell 3571 & Merging Clusters & 206.87965 & -32.86029 & 0.039 & 20.27 & 14.63 & 11.15 \\
Abell 3667 & Merging Clusters & 303.15580 & -56.84331 & 0.056 & 18.45 & 13.12 & 10.39 \\
Abell 3716S & Merging Clusters & 312.98849 & -52.69637 & 0.045 & 17.74 & 14.28 & 11.24 \\
Abell 3827 & Merging Clusters & 330.43705 & -59.95795 & 0.098 & 16.30 & 13.77 & 11.04 \\
Abell S0592 & Merging Clusters &  99.68232 & -53.97385 & 0.222 & 22.55 & 18.81 & 15.49 \\
ACT-CL J0012.8-0855 & Filament &   3.21349 &  -8.91999 & 0.338 & 12.10 & 10.57 & 8.74 \\
ACT-CL J0411.2-4819 & Merging Clusters & 62.81384 & -48.32236 & 0.418 & 20.09 & \_\_ & \_\_ \\
MACS J0416.1-2403 & Merging Clusters &  64.03787 & -24.07053 & 0.397 & 11.40 & 10.15 & 8.35 \\
MACS J1105.7-1014 & Merging Clusters & 166.45958 & -10.25243 & 0.466 & 13.81 & 11.96 & 9.44 \\
MACS J1931.8-2635 & Merging Clusters & 292.95667 & -26.57583 & 0.352 & 28.49 & 7.01 & 4.61 \\
MS1008.1-1224 & Merging Clusters & 152.60835 & -12.65294 & 0.306 & 15.89 & 13.73 & 10.89 \\
MS2137-2353 & Merging Clusters & 325.06132 & -23.66031 & 0.313 & 14.08 & 10.59 & 8.23 \\
PLCK G287.0+32.9 & Merging Clusters & 177.70640 & -28.08444 & 0.383 & 17.80 & 14.94 & 11.86 \\
RXC J1314.4-2515 & Merging Clusters & 198.63567 & -25.26848 & 0.244 & 15.20 & 12.06 & 9.61 \\
RXC J1514.9-1523 & Merging Clusters & 228.76418 & -15.36761 & 0.223 & 18.16 & 13.65 & 10.80 \\
RXC J2003.5-2323 & Merging Clusters & 300.88723 & -23.40022 & 0.317 & 17.30 & 9.23 & 6.81 \\
RX J1347.5-1145 & Merging Clusters & 206.872788 & -11.745483 & 0.451 & 17.18 & 14.53 & 11.67 \\
SMACSJ0723.3-7327 & Merging Clusters & 110.85730 & -73.43672 & 0.390 & 18.83 & 12.66 & 9.78 \\
SMACSJ2031.8-4036 & Exotic Lens & 307.96747 & -40.61824 & 0.331 & 23.56 & 18.60 & 14.74 \\
Z20-SPT-CLJ0135-5904 & Merging Clusters &  23.97592 & -59.10169 & 0.490 & 20.55 & 19.40 & 16.71 \\
COSMOS113 & Calibration field & 149.85274 &   1.75418 & \_\_ & 12.25 & 10.53 & 8.39 \\
COSMOSa & Legacy tile & 150.29180 &   1.90032 & \_\_ & 14.77 & 12.82 & 10.59 \\
COSMOSb & Legacy tile & 150.05422 &   1.86288 & \_\_ & 15.68 & 13.94 & 11.41 \\
COSMOSg & Legacy tile & 150.08010 &   2.56667 & \_\_ & 18.03 & 16.61 & 13.87 \\
COSMOSk & Legacy tile & 150.22175 &   2.23472 & \_\_ & 15.38 & 13.66 & 11.23 \\
COSMOSo & Legacy tile & 150.26667 &   2.45508 & \_\_ & 13.64 & 11.51 & 9.10 \\
\enddata
\end{deluxetable*}
To obtain a high-fidelity shear signal from the source catalog, it is important to assign weights to galaxies based on the precision of their shape measurements. Intuitively, objects with more precise measurements should contribute more strongly, while noisier measurements should be down-weighted.

We follow the weighting scheme adopted in~\cite{DES:2025zvc,DES:2020ekd, Anbajagane:2025mso}, in which weights are derived in a two-dimensional parameter space of SNR and size ratio $T/T_{\text{PSF}}$. Specifically, we bin the galaxies in this space, and within each bin estimate the shape noise variance $\sigma_e^2$ and the mean shear response $\langle R^{\gamma}\rangle$, where $\langle R^{\gamma}\rangle$ is the average of the diagonal components $R^{\gamma}_{11}$ and $R^{\gamma}_{22}$. The weight assigned to galaxies in bin $(i,j)$ is then defined as
\begin{equation}\label{eq:weight}
    w_{ij} = \frac{\langle R^\gamma\rangle_{ij}^2}{(\sigma_e^2)_{ij}},
\end{equation}
where $i$ and $j$ index the size ratio and SNR bins, respectively. 
The shape noise variance in each bin is estimated as
\begin{equation}
    (\sigma_e^2)_{ij} = \frac{1}{N_{ij}} \sum_{k=1}^{N_{ij}} \left( e_{k,1}^2 + e_{k,2}^2 \right),
\end{equation}
where $N_{ij}$ is the number of galaxies in bin $(i,j)$, and the sum runs over all galaxies in that bin.
Figure~\ref{fig:weights} shows the binned distributions of the raw galaxy counts, $\sqrt{\sigma_{e}}$, $\langle R^{\gamma} \rangle$, and the weight $w$ in a $10\times10$ logarithmic grid. The upper bounds of the parameter space are chosen to include 95\% of the sample, with galaxies at higher SNR or $T/T_{\text{PSF}}$ accumulated in the final bins. Galaxies are then assigned weights based on their corresponding bin, or the nearest bin if they lie outside the defined range.
\subsection{Number Density}
For any weak-lensing shape catalog, it is essential to quantify the statistical power available for constraining physical parameters. This is typically characterized by the weighted effective source number density. Following the definition in \cite{H12}, we compute
\begin{eqnarray}\label{eq:neff}
\hspace{-20pt}&n_{\text{eff}} =& \frac{1}{A} \frac{\left(\sum_i w_i\right)^2}{\sum_i w_i^2},
%\hspace{-20pt}&\sigma^2_{e} =& \frac{1}{2} \left[ \frac{\sum_i (w_i e_{1,i})^2}{\left(\sum_i w_i\right)^2} + \frac{\sum_i (w_i e_{2,i})^2}{\left(\sum_i w_i\right)^2} \right] \left[ \frac{\left(\sum_i w_i\right)^2}{\sum_i w_i^2} \right],
\end{eqnarray}
where $A$ is the survey area, and $w_i$ is the weight of galaxy $i$, obtained from the gridded SNR-$T/T_{\text{PSF}}$ space. Effective source densities for all targets are presented in Table~\ref{tab:densities}. %The effective source number density, determines the overall constraining power of the weak-lensing signal for each target. 

% \begin{figure}[]
% \centering
% \hspace*{-0.8cm}
% \includegraphics[width=0.5\textwidth]{selected_density_footprint.pdf}
% \caption{Unweighted source density for each SuperBIT target, after the selection has been applied.}
% \label{fig:density-footprint}
% \end{figure}

\subsection{Empirical Tests of the Shape Catalog}\label{sec:shape-cat-test}

\begin{figure*}[t]
\centering
\includegraphics[width=0.7\textwidth]{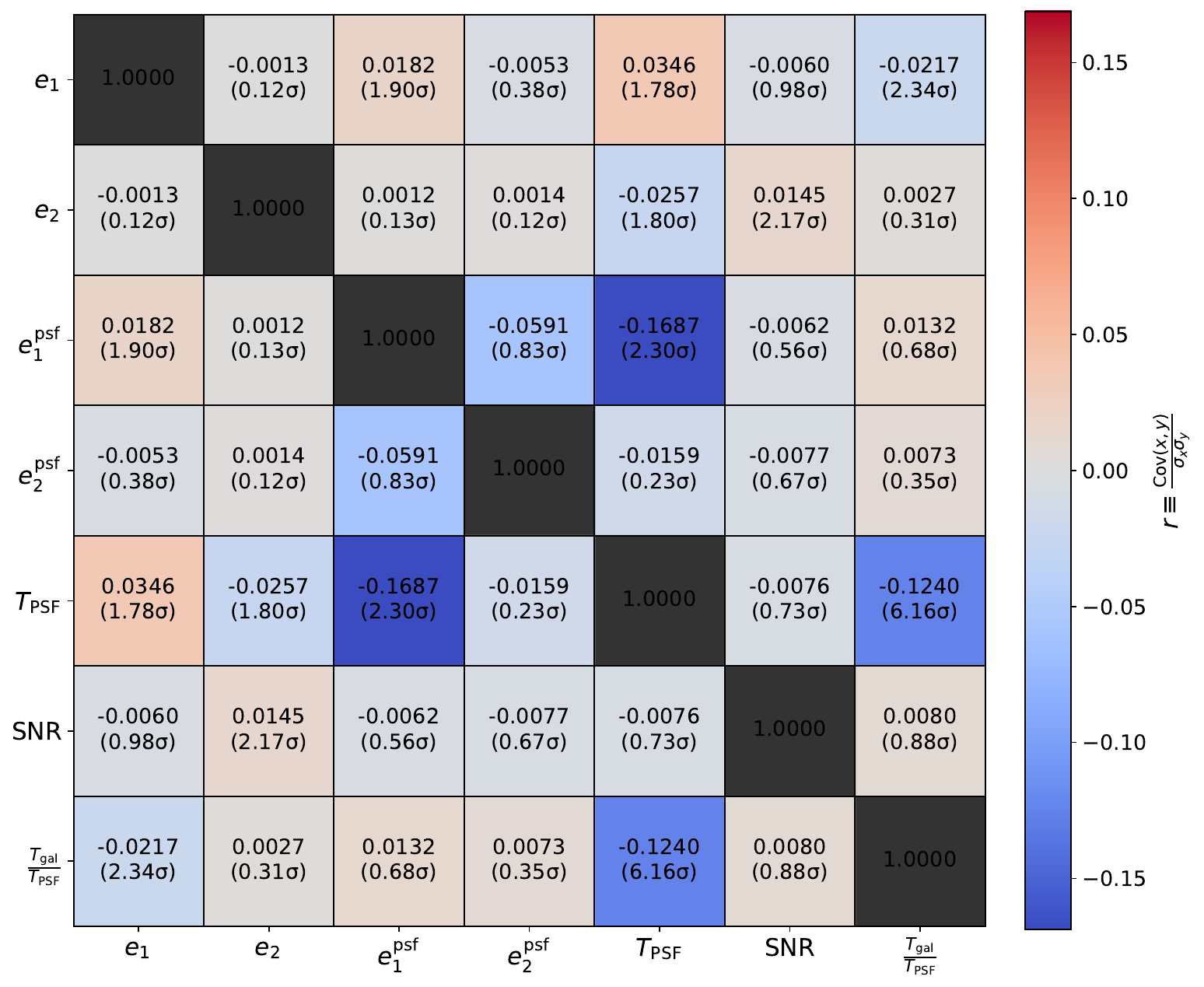}
\caption{The Pearson correlation matrix quantifying the correlations between galaxy shear components and PSF properties, including the PSF ellipticity components ($e_i^{\rm PSF}$) and PSF size ($T_{\rm PSF}$), as well as key \textsc{metacalibration} quantities such as the SNR and the galaxy-PSF size ratio.}
\label{fig:pearson-matrix}
\end{figure*}

\begin{figure*}
\centering
\includegraphics[width=\textwidth]{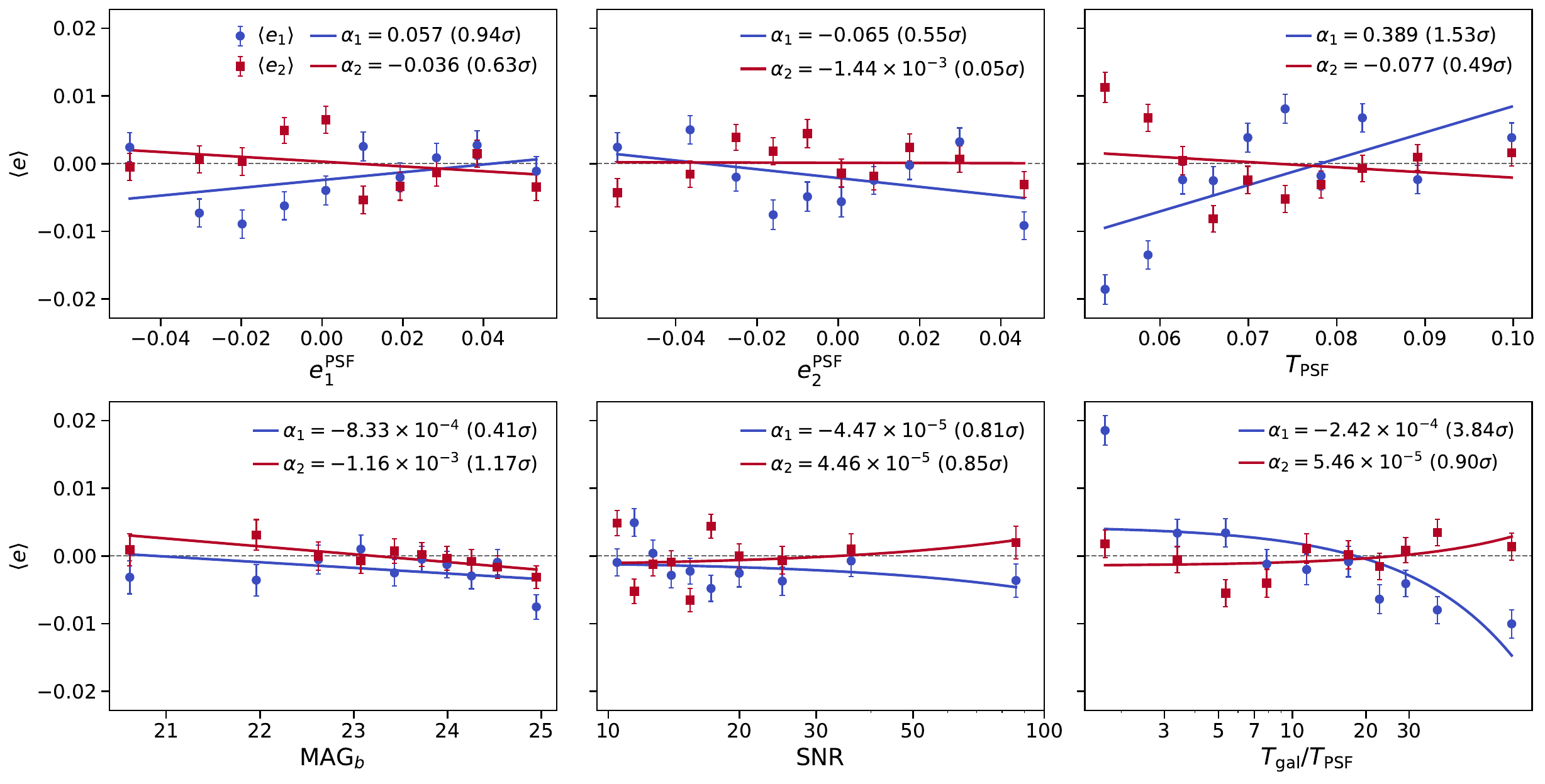}
\caption{Mean Shear dependence as a function of PSF and Galaxy quantities, in percentile bins. The Blue circles and red squares denotes the mean shear $\langle e_1 \rangle$, $\langle e_2 \rangle$ values respectively in each bin. The corresponding solid lines depicts the best fit linear model, showing the slope $\alpha$'s  in the legend}
\label{fig:psf-leakage}
\end{figure*}

An essential requirement for galaxy shape measurements within the \textsc{ngmix} framework is that the best-fit Gaussian mixture model accurately represents the PSF interpolated by \textsc{PSFEx} at the position of each galaxy. We investigate this requirement in detail in Appendix~\ref{appendix:ngmix-workflow}. In this section, we instead examine the shape catalog for potential residual systematics arising from imperfect PSF modeling. Specifically, we test whether the measured galaxy shapes show correlations with PSF properties. We also examine correlations with galaxy properties used to define selection cuts in the \textsc{metacalibration} process. Any significant dependence of the measured shapes on quantities such as the SNR or $T/T_{\text{PSF}}$ would indicate residual systematics in the shape-measurement pipeline.

To assess this, we compute the Pearson correlation matrix between the measured galaxy shear components, PSF ellipticities, and relevant galaxy properties. The Pearson correlation coefficient, $r_{xy}$, between two quantities $x$ and $y$ is defined as
\begin{equation}
    r_{xy} = \frac{\mathrm{Cov}(x,y)}{\sigma_x \sigma_y},
\end{equation}
where $\mathrm{Cov}(x,y)$ denotes the covariance between the two quantities, and $\sigma_x$ and $\sigma_y$ are their respective standard deviations used to normalize the covariance and obtain the correlation coefficient. Prior to computing the correlations, we subtract the global mean shear from the catalog to remove any residual offsets.

The resulting Pearson correlation matrix is presented in Figure~\ref{fig:pearson-matrix}. Statistical uncertainties on each correlation coefficient are estimated using a jackknife resampling of the full catalog into 30 subsets. Each matrix element reports the correlation coefficient along with its significance, quantified as $r/\sigma_r$.

Upon closer inspection of the Pearson correlation matrix, we find a strong correlation between the SNR (Eq.~\ref{eq:snr-ngmix}) of the galaxy parameter fit and $T/T_{\text{PSF}}$. This behavior is expected, as the size ratio quantifies how well resolved a galaxy is relative to the PSF: better-resolved galaxies yield more reliable shape measurements and consequently higher SNR values, and vice versa. We also observe a significant anti-correlation between the PSF ellipticity component $e_1^{\rm PSF}$ and the PSF size $T_{\rm PSF}$. This reflects an intrinsic property of the PSF and is consistent with the spatial PSF residual patterns shown in the upper panel of Figure~\ref{fig:psf-residual-spatial}. In particular, regions where $e_1^{\rm PSF}$ is more positive correspond to smaller PSF sizes near the center of the CCD, whereas regions with more negative $e_1^{\rm PSF}$ exhibit larger PSF sizes toward the corners.

To quantify potential leakage of PSF or galaxy-property correlations into the measured shear, we examine the dependence of the mean galaxy shear on several quantities of interest using a linear model parameterized by a leakage coefficient $\alpha$.
We bin galaxies into percentile bins of quantities including $e_1^{\rm PSF}$, $e_2^{\rm PSF}$, $T_{\rm PSF}$, ${\rm MAG}_b$, SNR, and the size ratio $T_{\rm gal}/T_{\rm PSF}$. Within each bin, we compute the weighted mean galaxy \sayan{shape} components $\langle e_1 \rangle$ and $\langle e_2 \rangle$. We then fit a linear model to the binned measurements,
\begin{equation}
    \langle e_{1,2} \rangle = \alpha_{1,2} X + \beta_{1,2},
\end{equation}
where $X$ denotes the quantity under consideration. A statistically significant non-zero value of $\alpha$ indicates a correlation between the mean shear and the corresponding quantity.

The results of these diagnostics are presented in Figure~\ref{fig:psf-leakage}. Blue and red points show the binned mean shear measurements, while the solid lines of matching colors indicate the best-fitting linear relations. The fitted $\alpha$ values are reported in the legends of each panel. Uncertainties on $\alpha$ are estimated using jackknife resampling of the full catalog, following the same procedure used for the Pearson correlation coefficient uncertainties.
\sayan{We also investigate PSF leakage further} using our fiducial simulations in Section~\ref{sec:psf-leak-sim}.
\section{Pipeline Validation}\label{sec:pipe-val}
\subsection{Image simulation framework}\label{sec:sims}
Our simulation framework for validating the SuperBIT shape-measurement pipeline closely follows the methodology described in \citet{mccleary2023}. In brief, we generate synthetic CCD images by injecting realizations of galaxies and stars into blank exposures. Galaxy injection parameters are drawn from the COSMOS2015 catalog, with object fluxes converted to SuperBIT bands using their spectral energy distributions (SEDs) and the SuperBIT transmission curves. The background galaxy population is lensed using an NFW cluster model with mass $M_{200}=4.1\times10^{14}\,M_{\odot}$ at redshift $z=0.24$. For each injected galaxy, the applied shear and magnification are computed using its corresponding source redshift, drawn from the COSMOS catalog. Instrumental and observational effects are included by adding dark current, detector noise, and an approximate sky background appropriate for stratospheric observing conditions \citep{gill_2020}. We also apply small dithers between exposures to mimic the pointing variations present in real observations. The final images are then processed through the full lensing analysis pipeline described above.

%The subsequent reduction and shape-measurement steps are identical to those applied to the real data. Specifically, we coadd 36 single exposures of 300~s each to construct a deep detection image, perform source detection on the coadd using \textsc{SExtractor}, and generate MEDS cutouts. Galaxy shapes are then measured with \textsc{ngmix} for all \textsc{metacalibration} shear terms, allowing us to estimate both the shear and selection responses. These responses are subsequently applied to calibrate the measured shapes, yielding response-calibrated galaxy shapes for all sources detected in the coadded image.

\subsection{Shear recovery metric}\label{sec:alpha_estimator}
\begin{deluxetable*}{c l cc}
\tabletypesize{\normalsize}
\tablecaption{Estimated $\hat{\zeta}$ and its
$1\sigma$ uncertainty for the sequence of unit-test simulations described in
Section~\ref{sec:unit_tests}.\label{tab:alpha_unit_tests}}
\tablehead{
\colhead{Seq.} &
\colhead{Simulation setup} &
\colhead{$\hat{\zeta}$} &
\colhead{$\sigma_{\hat{\zeta}}$}
}
\startdata
1 & Constant galaxy properties + Gaussian PSF
  & 0.9885 & 0.0612 \\
2 & Constant galaxy properties + SuperBIT PSF
  & 1.0143 & 0.0682 \\
3 & COSMOS size distribution + SuperBIT PSF
  & 1.0078 & 0.0564 \\
4 & COSMOS size + magnitude distributions + SuperBIT PSF
  & 0.9775 & 0.0987 \\
\enddata
\end{deluxetable*}
To quantify the end-to-end performance of the pipeline, we test how accurately it recovers the tangential shear profile of the input NFW lens. For each galaxy at position $(x,y)$ relative to a chosen reference centre $(x_c,y_c)$, the tangential and cross shear components are defined as
\begin{eqnarray}
\gamma_{+} &=& -\left[\gamma_1 \cos(2\phi) + \gamma_2 \sin(2\phi)\right], \\
\gamma_{\times} &=& \gamma_1 \sin(2\phi) - \gamma_2 \cos(2\phi),
\end{eqnarray}
where $\phi$ is the azimuthal angle with respect to the reference centre,
\begin{equation}
\phi = \tan^{-1}\left(\frac{y-y_c}{x-x_c}\right).
\end{equation}
We bin galaxies into annuli and compute the weighted mean tangential shear in each radial bin, denoted $\hat{\gamma}_{+,k}$, where $k$ indexes the radial bins. Because the input lens model is known, we compare the recovered binned profile $\hat{\boldsymbol{\gamma}}_{+}$ to the truth prediction $\boldsymbol{\gamma}^{\rm true}_{+}$ and parametrize the overall multiplicative recovery factor as
\begin{equation}
\zeta \equiv \frac{\langle \hat{\gamma}_{+} \rangle}{\gamma^{\rm true}_{+}}.
\end{equation}
An unbiased recovery corresponds to $\zeta=1$, while deviations from unity indicate residual systematic shear bias.

Following \citet{mccleary2023}, we estimate $\zeta$ using a maximum-likelihood estimator
\begin{equation}
\hat{\zeta} =
\frac{
\boldsymbol{\gamma}_{+}^{\rm true\,T}\,
\mathbf{C}^{-1}\,
\hat{\boldsymbol{\gamma}}_{+}
}{
\boldsymbol{\gamma}_{+}^{\rm true\,T}\,
\mathbf{C}^{-1}\,
\boldsymbol{\gamma}_{+}^{\rm true}
},
\label{eq:alpha_hat}
\end{equation}
with variance
\begin{equation}
\sigma_{\hat{\zeta}}^{2} =
\left(
\boldsymbol{\gamma}_{+}^{\rm true\,T}\,
\mathbf{C}^{-1}\,
\boldsymbol{\gamma}_{+}^{\rm true}
\right)^{-1}.
\label{eq:alpha_var}
\end{equation}
Here $\mathbf{C}$ is the covariance matrix of the binned recovered tangential shear profile $\hat{\boldsymbol{\gamma}}_{+}$. In this work we approximate $\mathbf{C}$ as diagonal, with diagonal elements given by the variance of $\hat{\gamma}_{+,k}$ in each annulus.

To characterize the statistical uncertainty, we generate an ensemble of
$N_{\rm sim}=30$ independent realizations of the same lensed field,
varying (i) the COSMOS galaxies drawn for injection,
(ii) their random orientations, and
(iii) the detector-noise realization.
For each simulation realization $j$, galaxies are binned radially and the
weighted mean tangential shear $\hat{\gamma}^{(j)}_{+,k}$ is computed in
each annular bin $k$.

The covariance matrix $\mathbf{C}$ is constructed to include both the
scatter of galaxy shears within each annulus and the
realization-to-realization variance across simulations.
In practice, the diagonal element for bin $k$ is estimated as the
variance of the mean tangential shear across realizations,
\begin{equation}
C_{kk}
=
\frac{1}{N_{\rm sim}-1}
\sum_{j=1}^{N_{\rm sim}}
\left(
\hat{\gamma}^{(j)}_{+,k}
-
\langle \hat{\gamma}_{+,k} \rangle
\right)^2 ,
\end{equation}
where
\begin{equation}
\langle \hat{\gamma}_{+,k} \ \rangle
=
\frac{1}{N_{\rm sim}}
\sum_{j=1}^{N_{\rm sim}}
\hat{\gamma}^{(j)}_{+,k}
\end{equation}
denotes the mean over realizations.
Since each $\hat{\gamma}^{(j)}_{+,k}$ is itself a weighted mean over
galaxies in that radial bin $k$, this variance naturally captures both the
intrinsic shape noise within each realization and the
scatter between different realizations.

\subsection{Incremental Unit Tests}\label{sec:unit_tests}
We perform a sequence of controlled ``unit-test'' simulations, broadly following the philosophy of the DES Y3 simulation validation \citep{DES:2020lsz}. Starting from highly idealized images, we incrementally increase the realism to isolate the dominant contributors to shear bias.

For these unit tests, galaxies are injected on a square grid lattice to eliminate blending, which is often a leading source of shear bias. Stars are placed on an interleaved grid to avoid overlap with galaxies and to ensure well-conditioned PSF estimation. We consider the following sequence:
\begin{enumerate}
\item \textbf{Constant galaxy properties + Gaussian PSF:} all galaxies have fixed magnitude $m=20$ and fixed half-light radius $r_{1/2}=0.5''$, with a Gaussian PSF of FWHM $0.5''$.
\item \textbf{Constant galaxy properties + SuperBIT PSF:} identical to (1), but replacing the Gaussian PSF with an empirical SuperBIT PSF model derived from real data using \textsc{PSFEx}.
\item \textbf{COSMOS size distribution:} identical to (2), but sampling galaxy sizes from COSMOS (restricted to $r_{1/2}\le 1''$).
\item \textbf{COSMOS magnitude distribution:} identical to (3), but additionally sampling magnitudes from COSMOS, with fluxes converted to SuperBIT bands as described in Section~\ref{sec:sims}.
\end{enumerate}
We report the recovered $\hat{\zeta}$ for each configuration in Table~\ref{tab:alpha_unit_tests}.

\subsection{Fiducial simulation}\label{subsec:fid-sim}

\begin{figure}[t]
\centering
\hspace*{-2em}
\includegraphics[width=0.5\textwidth]{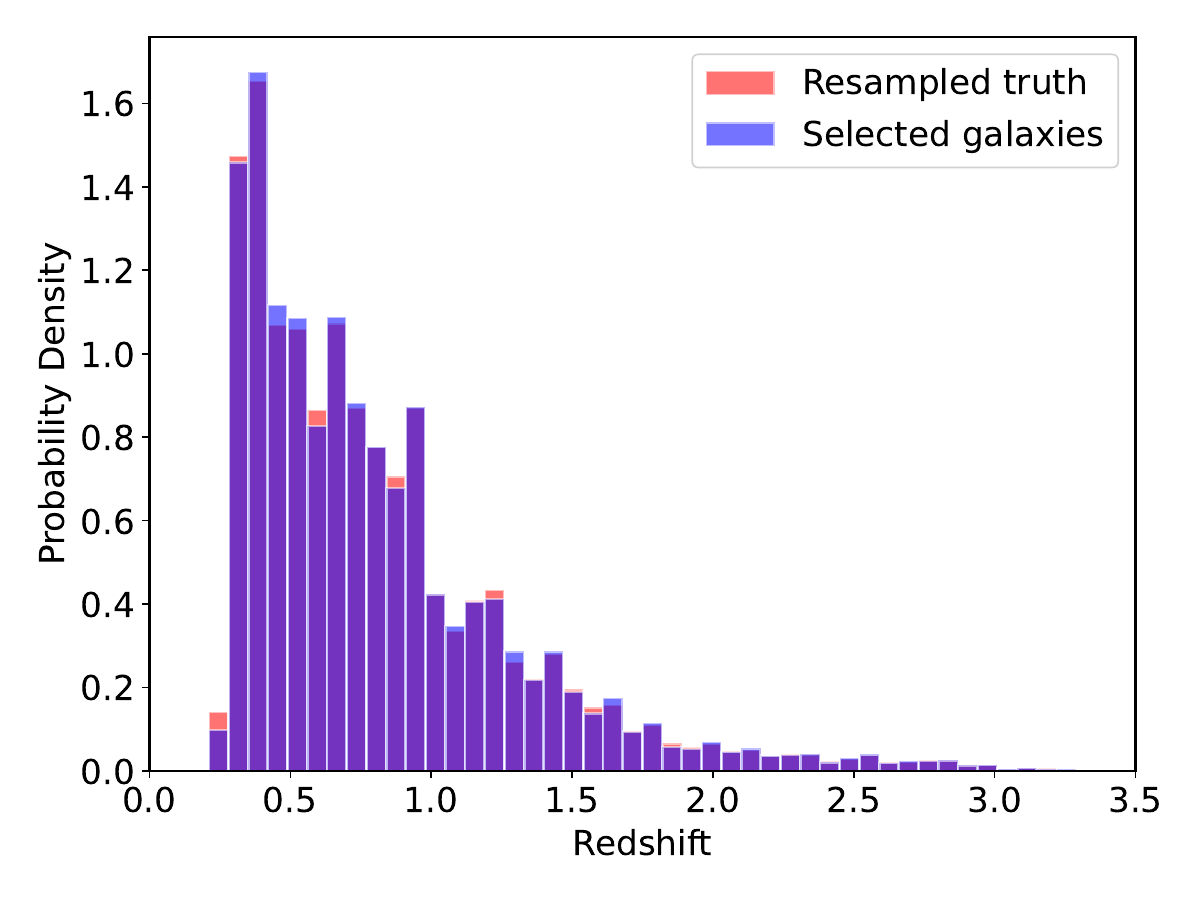}
\caption{Redshift distribution of background galaxies in the fiducial simulations for a cluster at redshift $z=0.245$. The blue histogram shows the redshift distribution of detected galaxies that pass the \textsc{metacalibration} selection criteria, while the red histogram shows the resampled redshift distribution of the injected galaxies used to estimate the true tangential shear, $\gamma_{+}^{\rm true}$.}
\label{fig:redshift-dist}
\end{figure}

As the final step in our progression toward increasing realism, we move from gridded galaxy injections to a random spatial injection of galaxies. This transition naturally introduces source blending. All other aspects of the simulation pipeline remain identical to those used in the final unit tests; in particular, galaxy properties are sampled from the COSMOS catalog.

For the first three unit-test configurations, all injected galaxies are detected in the coadded image. This is a direct consequence of the simplified simulation setup, in which galaxies are assigned constant, relatively bright magnitudes, resulting in a detection efficiency close to unity. In contrast, for unit test~4, galaxy magnitudes are sampled from the COSMOS catalog. As expected, this leads to a significantly reduced detection fraction of $\sim20\%$. A comparable detection fraction is obtained in the fiducial simulations, which adopt the same realistic galaxy population but with random spatial injections.

In these latter two cases, the true tangential shear profile $\gamma_{+}^{\rm true}$ must be evaluated using a source redshift distribution that matches the population of galaxies selected for shape measurement. Accordingly, we resample the injected galaxy redshift distribution to match the redshift distribution of the detected galaxies that pass the \textsc{metacalibration} selection criteria. The redshift distribution of the selected galaxies for our fiducial simulations, together with the corresponding resampled injected redshift distribution, is shown in Figure~\ref{fig:redshift-dist}.
\subsubsection{Representativeness of Fiducial Simulations}

\begin{figure}[t]
\centering
\hspace*{-2em}
\includegraphics[width=0.5\textwidth]{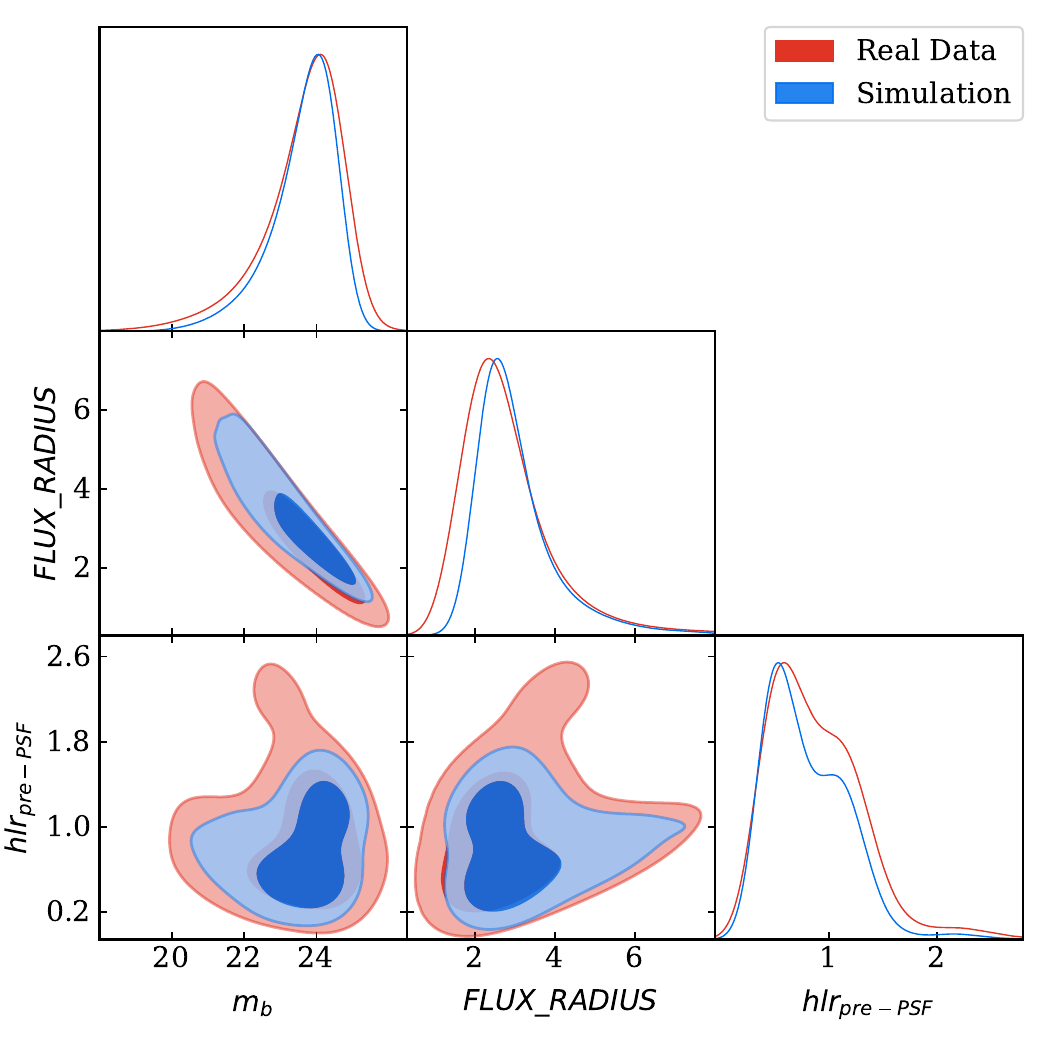}
\caption{Comparison of one-dimensional and joint two-dimensional distributions of galaxy properties in the $b$ band: magnitude ($m_b$), \texttt{FLUX\_RADIUS} (post-PSF half-light radius), and pre-PSF half-light radius (hlr). Red contours correspond to real data, while blue contours correspond to the fiducial simulations.}
\label{fig:representation-plot}
\end{figure}

% \begin{figure}[t]
% \centering
% \hspace*{-2em}
% \includegraphics[width=0.5\textwidth]{response_stability.pdf}
% \caption{Dependence of the shear response $R^{\gamma}$ on the galaxy-fitting SNR as estimated by \textsc{ngmix}. The red curves correspond to $R^{\gamma}_{11}$ and the blue curves to $R^{\gamma}_{22}$. Dashed lines with shaded bands show the results from fiducial simulations, while points with error bars show the measurements from real data.}
% \label{fig:response-stab}
% \end{figure}

To validate the shear reconstruction achieved by our pipeline, it is essential
to establish that the fiducial simulations are statistically representative
of the real data. We perform this validation by comparing one-dimensional and
joint two-dimensional distributions of several key galaxy properties measured
in the main science band F480W ($b$). These include the galaxy magnitude $m_b$, the
\textsc{SExtractor} quantity \texttt{FLUX\_RADIUS} (the post-PSF half-light
radius in units of image pixels), and the pre-PSF half-light radius (hlr),
which is derived from the best-fit galaxy size parameter output by
\textsc{ngmix} as \texttt{T\_noshear}.
\begin{figure*}[t]
\centering
\includegraphics[width=0.8\textwidth]{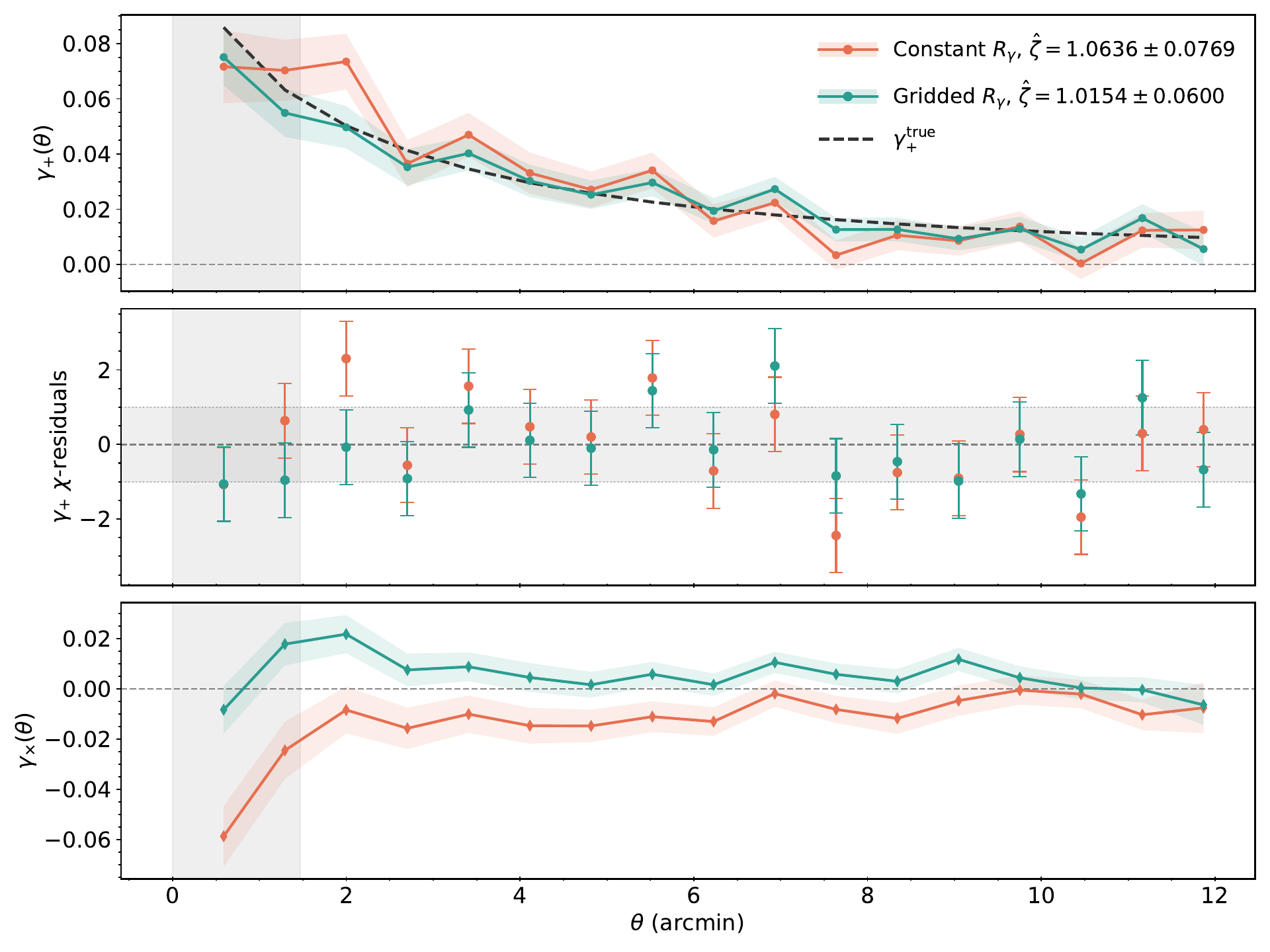}
\caption{Tangential shear bias estimation in the fiducial simulations. The coral color corresponds to the constant shear calibration approach, while the teal color corresponds to the gridded shear calibration approach. The vertical grey shaded region indicates the excluded range of the shear-bias estimation, where the applied shear is sufficiently large that the linearized implementation of \textsc{metacalibration} is no longer valid. 
\emph{Upper panel:} Reconstructed tangential shear $\gamma_+$ for both calibration approaches. The injected NFW shear profile, $\gamma_{+}^{\rm true}$, is shown as a black dashed curve. The corresponding estimates of $\hat{\zeta}$ for both approaches, computed using Eq.~\ref{eq:alpha_hat}, are indicated in the legend. 
\emph{Middle panel:} $\chi$-residuals for all data points for both calibration
approaches. The horizontal shaded band denotes a reference region around zero,
$-1 \le \chi \le +1$. 
\emph{Lower panel:} Reconstructed cross-shear profiles for both approaches.}
\label{fig:shear-bias}
\end{figure*}
The size parameter $T$ is defined as the trace of the second-order moment matrix
of the pre-PSF galaxy light profile (Eq.~\ref{eq:T_admom}). For a Gaussian
surface-brightness profile, the corresponding half-light radius is
\begin{equation}
\mathrm{hlr} = \sqrt{\frac{T}{2}}\,\sqrt{2\ln 2},
\end{equation}
and is expressed in units of arcseconds. The joint and marginal distributions
of these quantities are shown in the triangular (corner) plot presented in
Figure~\ref{fig:representation-plot}\footnote{The figure was generated using
\textsc{GetDist} (\url{https://github.com/cmbant/getdist})~\citep{Lewis:2019xzd}.}. 
\begin{table}[]
\centering
\caption{Comparison of mean $n_{\text{sel}}$ and $n_{\text{eff}}$ between simulations and the real data, using the same selection cuts described in Section~\ref{sec:object-selection}.}
\label{tab:sim_vs_real_density}
\begin{tabular}{lcc}
\hline\hline
Quantity & Simulations & Real data \\
\hline
$n_{\rm sel}$ (arcmin$^{-2}$) & $14.00$ & $13.38$ \\
$n_{\rm eff}$ (arcmin$^{-2}$) & $13.34$ & $10.66$ \\
\hline
\end{tabular}
\end{table}
\sayan{With the same selection cuts applied to the real data (Section~\ref{sec:object-selection}), we compare the mean selected object density and $n_{\text{eff}}$ from the 30 simulations to the mean $n_{\text{sel}}$ and $n_{\text{eff}}$ for real data, obtained from values in Table~\ref{tab:densities}. The comparison is summarized in Table~\ref{tab:sim_vs_real_density}.}

% As an additional consistency test, we compare the dependence of the shear
% response $R^{\gamma}$ on the galaxy-fitting SNR, as estimated by \textsc{ngmix}. The data are binned in percentile bins of SNR, and
% the mean shear response is computed in each bin. The resulting dependence for
% $R^{\gamma}_{11}$ and $R^{\gamma}_{22}$ are shown in
% Figure~\ref{fig:response-stab}, where the dashed curves with shaded bands
% represent the fiducial simulation results and the points with error bars
% represent the measurements from real data. We find good agreement between the
% simulations and observations across the full SNR range, indicating consistent
% shear calibration behavior between the fiducial simulations and the real data.
\subsubsection{Shear-bias estimation}\label{sec:shear_bias}

To validate the shear reconstruction in our fiducial simulations, we apply the
tangential shear bias formalism described in
Section~\ref{sec:alpha_estimator}, and present the results in
Figure~\ref{fig:shear-bias}. We compare two distinct shear-calibration
strategies.

The first approach employs the conventional calibration scheme based on a
global mean shear response derived from the selected galaxy sample, combined
with a global selection response to calibrate individual galaxy shapes. This
conventional calibration method is shown in coral in Figure~\ref{fig:shear-bias}
and yields \sayan{$\hat{\zeta} = 1.0636 \pm 0.0769$}. Since the only structural
difference between Unit Test~4 and the fiducial simulation is the transition
from gridded galaxy injections to randomly distributed galaxies, we attribute
this bias primarily to source blending.

Motivated by this result, we apply the same gridded shear-response calibration
scheme used for the real data (Section~\ref{sec:object-selection}) to the fiducial
simulations. Each galaxy is assigned a shear response based on its location in
the $(\mathrm{SNR},\,T/T_{\rm PSF})$ grid, and a global selection response is
combined with the gridded shear response to obtain the final calibration. The
diagonal elements of the global selection response estimated from the fiducial
simulations are $R_{11}^{S} = -0.022 \pm 0.052$ and
$R_{22}^{S} = 0.058 \pm 0.055$.

This approach leads to a substantial improvement in shear recovery, yielding
\sayan{$\hat{\zeta} = 1.0154 \pm 0.0600$}, consistent with unbiased shear reconstruction.
For both calibration methods, we exclude the regime
$\gamma_{+}^{\rm true} \ge 0.06$, where the applied shear is sufficiently large
that the linearized implementation of \textsc{metacalibration} is no longer valid
\citep{Sheldon:2017szh}. This excluded region is indicated by the vertical grey
shaded band in Figure~\ref{fig:shear-bias}. For the shear-bias estimation in the fiducial simulations, we also exclude
objects for which the PSF ellipticity is large, i.e.,
\sayan{$|e_{i}^{\rm PSF}| > 0.07$}. These objects are predominantly located near the edges of the field of view, where significant PSF ellipticity leakage is expected, as discussed in the following subsection. \sayan{\st{Within our gridded response calibration scheme, applying this PSF ellipticity cut significantly reduces the cross-shear signal in the outer regions of the profile, while leaving the estimated value of $\hat{\zeta}$ nearly unchanged.}} \sayan{We note that the object density reported in Table~\ref{tab:densities} does not include any PSF ellipticity cut, since exposure quality, and hence PSF ellipticity, varies significantly and inhomogeneously across our data, making it difficult to define a universal PSF ellipticity cut for the real data. We plan to explore this in more detail in a forthcoming paper, \textit{Lensing in the Blue IV}, which will examine $\kappa_B$ modes and PSF-leakage-induced $\kappa_E$ on a cluster-specific basis. We also note that we have adopted a single, centered NFW profile for the shear-bias estimation, even though most SuperBIT targets are merging clusters with multiple mass peaks. This choice does not affect our validation because we are not assessing the recovery of cluster masses through NFW profile fitting. Rather, we directly compare the reconstructed shear with the input shear, with the aim of demonstrating that \textsc{metacalibration} provides unbiased shear estimates across a broad range of input shear values.}

\subsubsection{PSF Ellipticity Leakage in Fiducial Simulations}\label{sec:psf-leak-sim}

As noted in the shape-catalog tests for real data
(Section~\ref{sec:shape-cat-test}), we detect a significant correlation between
PSF ellipticity and galaxy shapes in the $e_1$ component, suggestive of a
possible PSF ellipticity leakage. We investigate this effect further using our
fiducial simulations, which employ \textsc{PSFEx} PSF models derived from real
data.

\begin{figure}
\centering
\hspace*{-2em}
\includegraphics[width=0.5\textwidth]{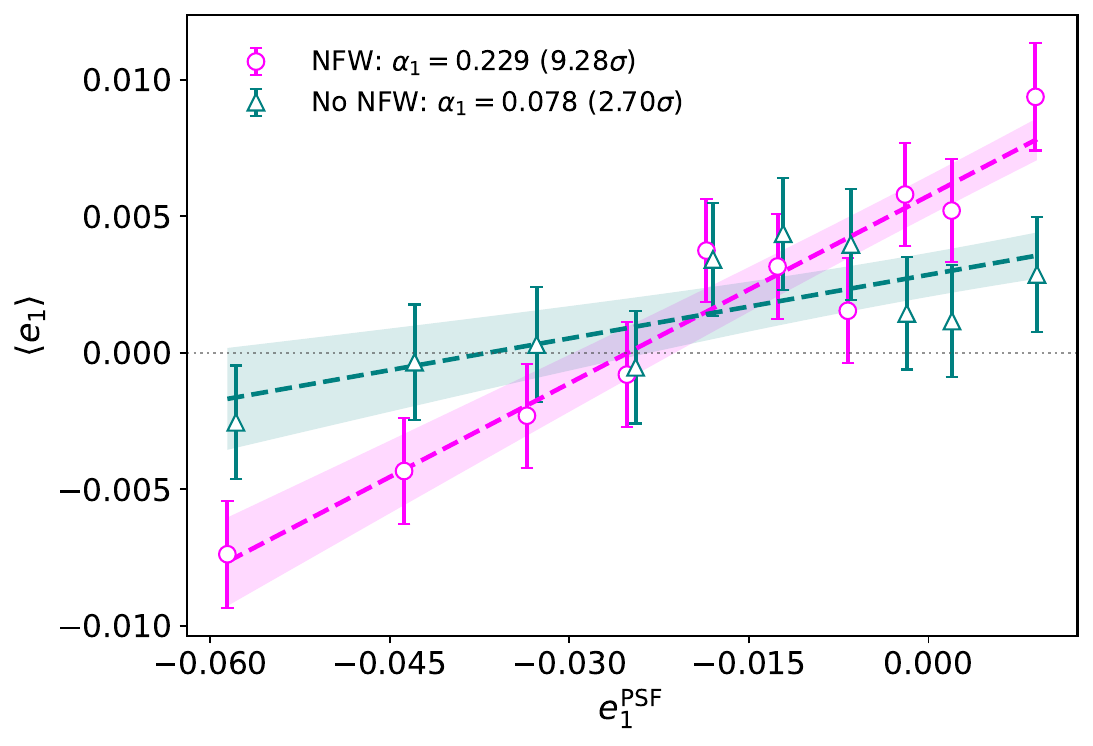}
\caption{Test of PSF ellipticity leakage in simulations. Magenta points show
results from fiducial simulations including NFW shear, while teal points show
results from simulations without NFW shear. A significant correlation is
observed only in the lensed case, indicating that the apparent signal arises
from the lensing-induced shear geometry rather than from true PSF ellipticity
leakage into galaxy shapes.}
\label{fig:psf-leakage-sim}
\end{figure}

The fiducial simulations include lensing by an NFW cluster profile located at
the centre of the field of view, closely mimicking the observational
configuration of the real SuperBIT targets. To disentangle genuine PSF
leakage from correlations induced by lensing geometry, we perform the same
analysis on a control set of simulations that do not include any NFW shear,
corresponding to an unlensed field with randomly oriented galaxies.

The results are shown in Figure~\ref{fig:psf-leakage-sim}. The fiducial
simulations including NFW shear (magenta points) exhibit a significant
correlation, with a measured slope \sayan{$\alpha_1 = 0.229$ at a $9.28\sigma$}
significance. In contrast, simulations without NFW shear (teal points)
show no statistically significant correlation, with \sayan{$\alpha_1 = 0.078$ at a
$2.7\sigma$} significance. For simulations without NFW shear, we also estimate
the correlation between galaxy $e_2$ and $e_2^{\rm PSF}$, obtaining
\sayan{$\alpha_2 = 0.078$ at $2.03\sigma$} significance. These results are obtained
using objects for which the PSF ellipticity satisfies \sayan{$|e| \leq 0.07$},
corresponding to the dominant central region of the field of view.

However, the PSF becomes highly elliptical near the very edges of the field,
where \sayan{$|e| > 0.07$}. When the leakage is evaluated including these regions in
simulations without NFW shear, we detect a significant correlation,
\sayan{$\alpha_1 = 0.149$ at $12.39\sigma$} significance, indicating that PSF
ellipticity leakage becomes important near the edges of the field. We note
this as an avenue for future work, where we plan to investigate this effect
using $B$-mode convergence maps, $\kappa_B$, and to mitigate it by excluding edge regions from science analyses.

These results indicate that the observed correlation in the fiducial
simulations is not due to direct PSF ellipticity leakage into the galaxy
shape measurements within the central field region. Instead, we interpret
this effect as a geometric coincidence: the tangential shear pattern induced
by the NFW lensing profile produces a spatial structure in
$\langle e_1 \rangle$ that aligns with the CCD-coordinate pattern of
$e_1^{\rm PSF}$, thereby generating an apparent correlation without true PSF
contamination in the central portion of the field of view.

\section{Conclusion}\label{sec:conclusion}

We process imaging data obtained during SuperBIT's 2023 flight and present shape catalogs for 30 science targets: primarily merging clusters and filamentary structures, together with 6 COSMOS fields observed mainly for calibration purposes. We obtain a mean density of selected objects of \sayan{$n_{\rm sel}=13.38~\mathrm{arcmin}^{-2}$ and an average effective number density of $n_{\rm eff}^{\rm H12}=10.66~\mathrm{arcmin}^{-2}$} across all targets. The full set of statistics for all fields is provided in Table~\ref{tab:densities}.

The image processing and shape-measurement pipeline consists of multiple steps, including stacking of 36 single exposures, source extraction, masking of hot pixels, and removal of spurious detections associated with diffraction spikes around bright stars. Weak-lensing shapes are measured using the \textsc{Metacalibration} algorithm. We introduce a novel calibration technique based on a gridded shear-response defined in the two-dimensional space of \textsc{metacalibration} quantities, together with galaxy weights constructed in this parameter space.

Two key requirements for precise weak-lensing shape measurements are: (1) accurate modeling of the PSF in the focal plane, and (2) forward modeling of that PSF for unbiased inference of galaxy shapes. To assess the robustness of both steps, we employ a comprehensive set of diagnostics. The PSF interpolation is validated using reserved test stars not used in PSF modeling, and the corresponding PSF residuals and $\rho$-statistics are computed to quantify the impact of PSF mis-modeling on the science analysis. To validate the shape measurements produced by the forward modeling with \textsc{ngmix}, we investigate residual correlations between galaxy shapes, PSF properties, and \textsc{metacalibration} quantities.

In parallel, we develop an image simulation pipeline to test the robustness of the full shape-measurement framework. The fiducial simulations use empirical SuperBIT PSFs obtained from \textsc{PSFEx} runs on real data in order to faithfully capture observational non-idealities. Validation is performed through a hierarchy of unit tests followed by fiducial simulations. Using a conventional global mean shear-response calibration, we find a tangential shear bias of $\sim6.3\%$, which is reduced to $1.1\%$ when employing the gridded shear-response calibration scheme. We further investigate PSF ellipticity leakage using the fiducial simulations and find no significant detection of true PSF-induced leakage.

This work forms part of the \emph{Lensing in the Blue} series. Paper~I \& Paper~II laid the foundations of the lensing analysis framework developed here, and the final convergence maps for the targets presented in this work (Paper~III) will be reported in Paper~IV. The important step of background-foreground separation using color information is not addressed in this paper and will be discussed in detail in Paper~IV, together with an accompanying white paper describing the method.

At present, shapes are measured using only the primary science band F480W ($b$); a multi-band shape-measurement analysis is planned, and corresponding multi-band shape catalogs will be released in future data products. We also note that the current implementation of \textsc{metacalibration} does not account for shear-dependent detection bias. We therefore plan to implement the \textsc{metadetection} framework~\citep{Sheldon:2019uxq} in future versions of the pipeline to mitigate this effect.

%Finally, the same shape-measurement pipeline will be employed for the SuperBIT successor mission GigaBIT~\citep{Voyer2024}. in the GigaBIT analysis.

In the near future, SuperBIT's merging cluster observations will be analyzed in multiple complementary frameworks, including analyses of offsets between X-ray emission, stellar mass, and dark matter distributions for SIDM constraints and an assessment of the dynamical states of SuperBIT clusters using the coherence of different cluster components in Fourier space \citep{Cerini:2022akj}. 

SuperBIT's three square degrees of high-resolution deep NUV and blue imaging adds legacy value to existing NASA data, and the object catalogs produced in this work can support a broad range of investigations beyond cluster cosmology. The three-band photometric data (F400W, F480W and F600W) are already being employed to study circumgalactic dust in the intracluster medium~\citep{2020A&A...633L...7L, McCleary:2025xqb}. Photometric redshift estimation is another natural use case; the inclusion of near-ultraviolet photometry spanning the 3700\,\AA~Balmer and 4000\,\AA~Lyman breaks used in galaxy template fitting can halve uncertainties on photometric redshift estimates \citep{sawicki2019cfht}. Other potential use cases include searches for OB stars and white dwarfs, as well as studies of galaxy evolution.

%% Please use the acknowledgment and contribution environments. This will 
%% be anonomyized when the "anonymous" style option is used. 
\begin{acknowledgments}
This work was completed primarily using the Explorer Cluster, supported by Northeastern University's Research Computing team. Additional computations were performed by utilizing the CANDIDE cluster at the Institut d'Astrophysique de Paris, which was funded through grants from the PNCG, CNES, DIM-ACAV, and the Cosmic Dawn Center and maintained by Stephane Rouberol. SS thanks Dhayaa Anbajagane and Chihway Chang for useful discussions and feedback. \sayan{SS is also grateful to Erin Sheldon for assistance with implementing \textsc{ngmix} and the PSF response correction in the superbit-lensing pipeline.} SS also thanks Leila Ohashi and Marcus Michaud for creating star masks as part of the Young Scholars' Program organized by Northeastern University. The development of analysis pipeline was partially supported by the Jet Propulsion Laboratory, California Institute of Technology, under a contract with the National Aeronautics and Space Administration (80NM0018D0004). The US team acknowledges support
from the NASA APRA grant 80NSSC22K0365. Canadian team
members acknowledge support from the Canadian Institute for
Advanced Research (CIFAR), the Natural Science and Engineering Research Council (NSERC), and the Canadian Space
Agency (CSA). UK coauthors acknowledge funding from Durham University's Astronomy Projects Award, STFC (grants ST/P000541/1, ST/V005766/1 and ST/X001075/1), and UKRI (grant MR/X006069/1).

Based on data products from observations made with ESO Telescopes at the La Silla Paranal Observatory under ESO programme ID 179.A-2005 and on data products produced by TERAPIX and the Cambridge Astronomy Survey Unit on behalf of the UltraVISTA consortium.

This work has made use of data from the European Space Agency (ESA) mission {\it Gaia} (\url{https://www.cosmos.esa.int/gaia}), processed by the {\it Gaia} Data Processing and Analysis Consortium (DPAC, \url{https://www.cosmos.esa.int/web/gaia/dpac/consortium}). Funding for the DPAC has been provided by national institutions, in particular the institutions participating in the {\it Gaia} Multilateral Agreement. \sayan{This work has also made use of the Python package GaiaXPy, developed and maintained by members of DPAC, and in particular, Coordination Unit 5 (CU5), and the Data Processing Centre located at the Institute of Astronomy, Cambridge, UK (DPCI).}

The Pan-STARRS1 Surveys (PS1) and the PS1 public science archive have been made possible through contributions by the Institute for Astronomy, the University of Hawaii, the Pan-STARRS Project Office, the Max-Planck Society and its participating institutes, the Max Planck Institute for Astronomy, Heidelberg and the Max Planck Institute for Extraterrestrial Physics, Garching, The Johns Hopkins University, Durham University, the University of Edinburgh, the Queen's University Belfast, the Harvard-Smithsonian Center for Astrophysics, the Las Cumbres Observatory Global Telescope Network Incorporated, the National Central University of Taiwan, the Space Telescope Science Institute, the National Aeronautics and Space Administration under Grant No. NNX08AR22G issued through the Planetary Science Division of the NASA Science Mission Directorate, the National Science Foundation Grant No. AST-1238877, the University of Maryland, Eotvos Lorand University (ELTE), the Los Alamos National Laboratory, and the Gordon and Betty Moore Foundation. All the {\it Pan-STARRS1} data used in this paper can be found in MAST: \dataset[10.17909/s0zg-jx37]{http://dx.doi.org/10.17909/s0zg-jx37}.

This paper is based on data collected at the Subaru Telescope and retrieved from the HSC data archive system, which is operated by the Subaru Telescope and Astronomy Data Center (ADC) at NAOJ. Data analysis was in part carried out with the cooperation of Center for Computational Astrophysics (CfCA), NAOJ. We are honored and grateful for the opportunity of observing the Universe from Maunakea, which has the cultural, historical and natural significance in Hawaii.

\end{acknowledgments}

\subsection*{Data Availability}

The imaging data, including the single exposures and coadded images, together with supplementary calibration products such as darks, flats, hot-pixel masks, and the photometric data catalogs, will soon be publicly available through the MAST archive (\dataset[10.17909/0hsg-dd76]{http://dx.doi.org/10.17909/0hsg-dd76}). 
The shape catalogs will be made publicly available following the publication of this paper and the convergence-map paper. %The image-processing and shape-measurement code is already publicly available at \url{https://github.com/superbit-collaboration/superbit-lensing}.


\software{Astropy \citep{2013A&A...558A..33A,2018AJ....156..123A,2022ApJ...935..167A}, 
Astroquery~\citep{astroquery},
Swarp~\citep{swarp} 
          \textsc{SExtractor} \citep{sextractor},
          \textsc{PSFEx}~\citep{psfex}
          GalSim~\citep{Rowe:2014cza},
          \textsc{ngmix}~\citep{ngmix},
          NumPy~\citep{harris2020array},
          SciPy~\citep{2020SciPy-NMeth},
          HealPy~\citep{healpy1,healpy2},
          Matplotlib~\citep{Hunter:2007}
          }


\appendix
\section{Per-Cluster PSF Modelling Assessment}\label{app:rho-stats}
\setcounter{figure}{0}
\renewcommand{\thefigure}{\thesection\arabic{figure}}
While in Section~\ref{subsec:psfex} we presented an aggregate assessment of our PSF modelling quality, it is also important to assess PSF modelling on a per-cluster basis before it can be trusted for science use, since, as we discuss in the next subsection, SuperBIT exposures are highly inhomogeneous in quality. In Table~\ref{tab:psf_diagnostics_percluster}, we present the mean ellipticity residuals of the PSF modelling, along with the average values of 
$\rho_1$ and 
$\rho_2$, weighted by the inverse variance at each data point. We report only these two 
$\rho$-statistics, as they contribute most significantly to the two-point correlation function. We also plot 
$\rho_1$ in sky coordinates in Figure~\ref{fig:rho-dist}. We find no correlation with the number of available stars, indicating that even for high-latitude targets, the number of stars remains sufficient and does not significantly contribute to the residual. 

\begin{deluxetable*}{lcccc}
\tablecaption{
Per-cluster PSF diagnostics. Columns show the mean ellipticity residual
components ($\delta e_1$, $\delta e_2$) between observed and PSF-model
shapes, and the $\rho_1$ and $\rho_2$ statistic amplitudes (Eqs.~\ref{eq:rho1}--\ref{eq:rho2}),
each averaged over the fitting range and weighted by inverse variance.
\label{tab:psf_diagnostics_percluster}
}
\tabletypesize{\footnotesize}
\tablewidth{0pt}
\tablehead{
\colhead{Target} &
\colhead{$\delta e_1$} &
\colhead{$\delta e_2$} &
\colhead{$\bar\rho_1$} &
\colhead{$\bar\rho_2$} \\
\colhead{} &
\colhead{($\times 10^{-2}$)} &
\colhead{($\times 10^{-2}$)} &
\colhead{($\times 10^{-5}$)} &
\colhead{($\times 10^{-5}$)}
}
\startdata
1E0657 Bullet & $1.03 \pm 0.08$ & $0.90 \pm 0.08$ & $0.11 \pm 0.07$ & $0.94 \pm 0.08$ \\
ACT-CL J0012.8-0855 & $1.15 \pm 0.21$ & $0.99 \pm 0.21$ & $0.07 \pm 0.57$ & $2.41 \pm 0.46$ \\
Abell 141 & $0.54 \pm 0.25$ & $0.54 \pm 0.23$ & $0.90 \pm 0.78$ & $7.08 \pm 0.64$ \\
Abell 1689 & $1.26 \pm 0.16$ & $0.90 \pm 0.15$ & $2.61 \pm 0.29$ & $21.23 \pm 0.26$ \\
Abell 2163 & $1.27 \pm 0.07$ & $0.95 \pm 0.06$ & $0.68 \pm 0.06$ & $12.34 \pm 0.05$ \\
Abell 2345 & $1.20 \pm 0.09$ & $0.91 \pm 0.09$ & $0.73 \pm 0.10$ & $3.94 \pm 0.09$ \\
Abell 2384a & $1.29 \pm 0.11$ & $1.06 \pm 0.11$ & $0.25 \pm 0.14$ & $4.67 \pm 0.14$ \\
Abell 2384b & $1.13 \pm 0.12$ & $0.68 \pm 0.12$ & $1.79 \pm 0.19$ & $9.80 \pm 0.16$ \\
Abell 3192 & $0.92 \pm 0.20$ & $1.01 \pm 0.22$ & $0.62 \pm 0.61$ & $3.62 \pm 0.53$ \\
Abell 3365 & $1.76 \pm 0.10$ & $1.77 \pm 0.11$ & $0.92 \pm 0.11$ & $20.53 \pm 0.10$ \\
Abell 3411 & $1.15 \pm 0.06$ & $0.86 \pm 0.05$ & $0.16 \pm 0.04$ & $6.08 \pm 0.05$ \\
Abell 3526 & $1.19 \pm 0.05$ & $0.85 \pm 0.05$ & $0.57 \pm 0.03$ & $7.63 \pm 0.03$ \\
Abell 3571 & $0.54 \pm 0.06$ & $0.28 \pm 0.06$ & $0.53 \pm 0.05$ & $5.76 \pm 0.05$ \\
Abell 3667 & $2.02 \pm 0.09$ & $1.38 \pm 0.09$ & $2.05 \pm 0.08$ & $8.03 \pm 0.09$ \\
Abell 3716S & $1.27 \pm 0.10$ & $0.84 \pm 0.09$ & $0.74 \pm 0.11$ & $1.30 \pm 0.10$ \\
Abell 3827 & $1.14 \pm 0.11$ & $1.00 \pm 0.12$ & $1.02 \pm 0.17$ & $17.67 \pm 0.15$ \\
Abell S0592 & $1.51 \pm 0.08$ & $1.10 \pm 0.07$ & $0.93 \pm 0.05$ & $13.37 \pm 0.06$ \\
Abell S780 & $1.04 \pm 0.07$ & $0.59 \pm 0.07$ & $0.60 \pm 0.06$ & $7.82 \pm 0.07$ \\
COSMOS113 & $1.17 \pm 0.17$ & $1.08 \pm 0.17$ & $0.47 \pm 0.21$ & $14.98 \pm 0.29$ \\
COSMOSa & $1.08 \pm 0.15$ & $0.59 \pm 0.15$ & $3.40 \pm 0.29$ & $39.69 \pm 0.31$ \\
COSMOSb & $1.04 \pm 0.17$ & $0.60 \pm 0.15$ & $0.72 \pm 0.23$ & $14.54 \pm 0.30$ \\
COSMOSg & $1.13 \pm 0.15$ & $0.74 \pm 0.14$ & $3.10 \pm 0.29$ & $29.15 \pm 0.26$ \\
COSMOSk & $1.38 \pm 0.20$ & $1.30 \pm 0.20$ & $0.23 \pm 0.34$ & $2.94 \pm 0.42$ \\
COSMOSo & $1.74 \pm 0.20$ & $1.13 \pm 0.19$ & $1.47 \pm 0.40$ & $12.91 \pm 0.36$ \\
MACS J0416.1-2403 & $1.95 \pm 0.18$ & $1.41 \pm 0.19$ & $6.24 \pm 0.49$ & $23.59 \pm 0.40$ \\
SMACSJ0723.3-7327 & $1.47 \pm 0.08$ & $1.08 \pm 0.08$ & $0.52 \pm 0.09$ & $4.23 \pm 0.08$ \\
MACS J1105.7-1014 & $0.73 \pm 0.17$ & $1.02 \pm 0.17$ & $0.79 \pm 0.39$ & $7.98 \pm 0.32$ \\
MACS J1931.8-2635 & $1.25 \pm 0.03$ & $0.94 \pm 0.03$ & $0.29 \pm 0.01$ & $3.52 \pm 0.01$ \\
MS1008.1-1224 & $1.43 \pm 0.12$ & $1.23 \pm 0.12$ & $0.38 \pm 0.17$ & $5.31 \pm 0.15$ \\
MS2137-2353 & $1.42 \pm 0.14$ & $1.02 \pm 0.13$ & $0.54 \pm 0.20$ & $13.65 \pm 0.19$ \\
PLCK G287.0+32.9 & $0.93 \pm 0.11$ & $0.65 \pm 0.11$ & $0.29 \pm 0.16$ & $0.54 \pm 0.16$ \\
RXC J1314.4-2515 & $1.47 \pm 0.12$ & $1.11 \pm 0.12$ & $0.28 \pm 0.16$ & $1.48 \pm 0.17$ \\
RXC J1514.9-1523 & $0.82 \pm 0.07$ & $0.49 \pm 0.07$ & $1.04 \pm 0.07$ & $11.93 \pm 0.07$ \\
RXC J2003.5-2323 & $1.89 \pm 0.07$ & $1.50 \pm 0.07$ & $0.95 \pm 0.05$ & $2.07 \pm 0.04$ \\
RX J1347.5-1145 & $1.11 \pm 0.13$ & $0.61 \pm 0.13$ & $1.36 \pm 0.21$ & $4.55 \pm 0.17$ \\
SMACSJ2031.8-4036 & $1.26 \pm 0.07$ & $0.90 \pm 0.07$ & $0.63 \pm 0.05$ & $2.28 \pm 0.05$ \\
ACT-CL J0411.2-4819 & $1.30 \pm 0.14$ & $0.92 \pm 0.15$ & $0.58 \pm 0.29$ & $2.23 \pm 0.24$ \\
Z20-SPT-CLJ0135-5904 & $0.85 \pm 0.16$ & $0.59 \pm 0.15$ & $0.10 \pm 0.31$ & $1.10 \pm 0.27$ \\
\enddata
\end{deluxetable*}
\section{Exposure Quality Assessment and Selection}\label{appendix:exp-qual}
\begin{figure}[]
\centering
\hspace*{-2em}
\includegraphics[width=0.5\textwidth]{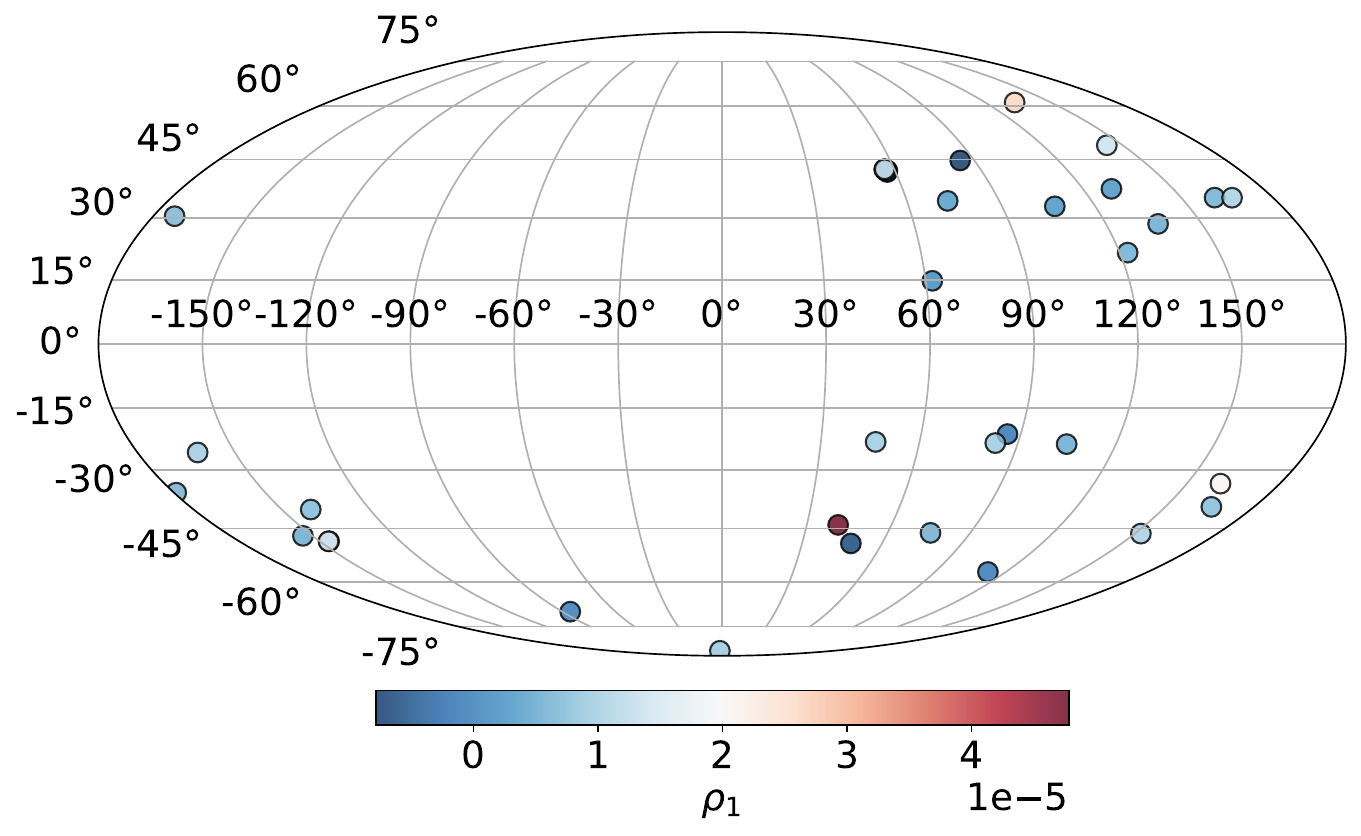}
\caption{$\rho_1$ footprint of SuperBIT targets}
\label{fig:rho-dist}
\end{figure}
\begin{figure*}[t]
\centering
\includegraphics[width=\textwidth]{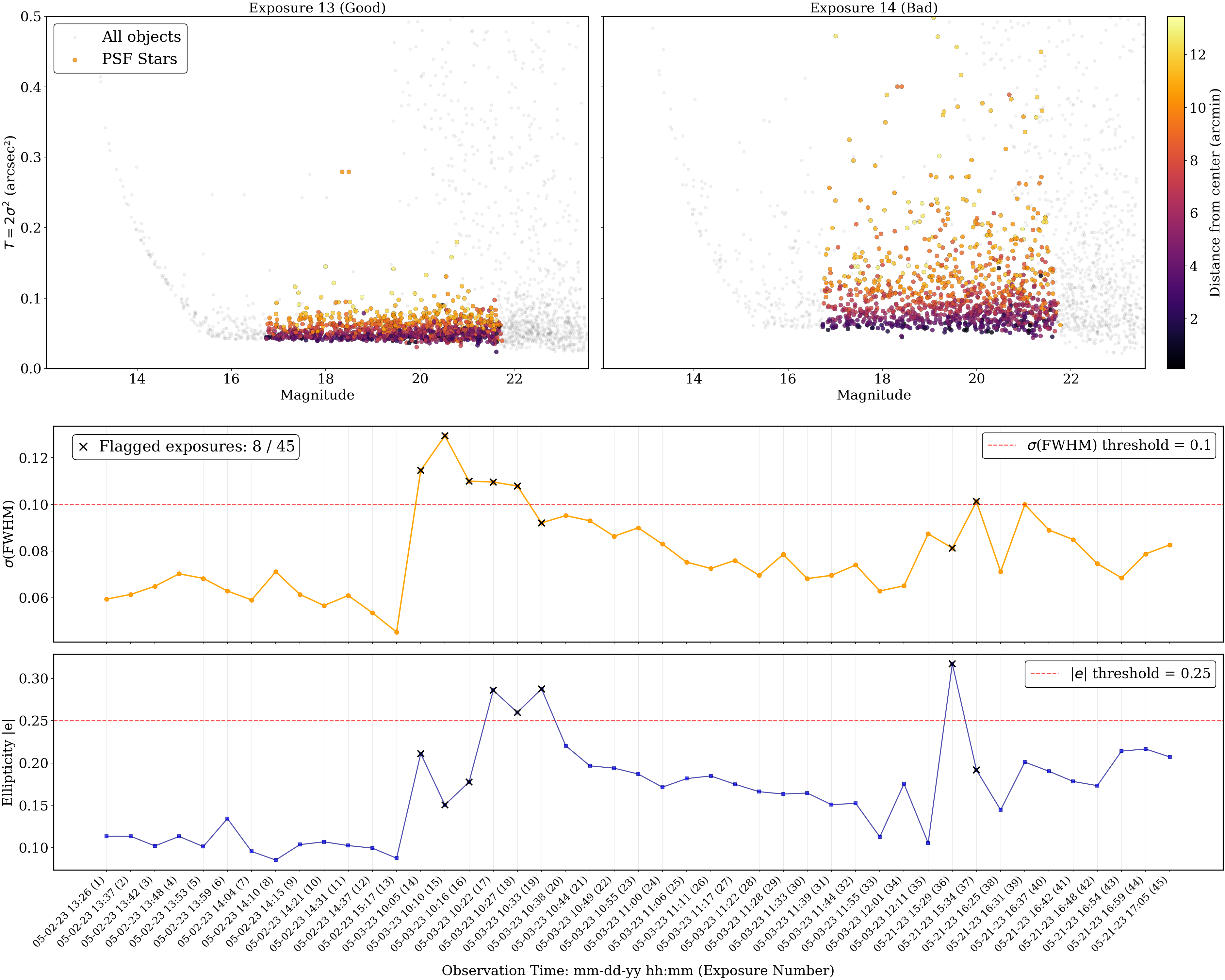}
\caption{Illustration of the exposure-selection scheme used in our pipeline for the target Abell~3411. \emph{Upper panels:} Size-magnitude diagrams for two representative exposures. The left panel shows a ``good'' exposure with a compact, well-defined stellar locus, while the right panel shows a ``bad'' exposure with a more scattered stellar locus. All detected objects are shown in gray, while the stars used for PSF modeling are highlighted with a colorbar, where the colorbar indicates the distance from the exposure center in arcminutes. The increased scatter in the stellar locus is primarily driven by stars located near the outskirts of the exposure. \emph{Lower panels:} Temporal evolution of two exposure-quality diagnostics: the scatter of the PSF-star FWHM (middle panel) and the mean PSF ellipticity measured in the outer 50\% of each exposure (bottom panel). The x-axis shows the UTC observation time together with the exposure number. Red dashed lines indicate the thresholds applied to each diagnostic. Exposures flagged by either criterion are marked with black crosses and are excluded from the shape-measurement analysis.}
\label{fig:exp-qual}
\end{figure*}
While all 300~s single exposures are used to construct deep coadded images for each SuperBIT target, additional care is required when selecting exposures for shape measurement. Accurate PSF deconvolution is essential for unbiased shape estimation with \textsc{ngmix}, and we find that the quality of individual SuperBIT exposures can be highly inhomogeneous due to the unique observing conditions of the balloon-borne platform. This section is therefore dedicated to quantifying and selecting exposures suitable for shape measurement.

To assess exposure quality, we focus on two diagnostic criteria that we find to be most effective. The first is the scatter of the stellar locus in the
size-magnitude diagram. The second is the mean ellipticity of point sources measured in the outer 50\% area of the exposure. We find that the overall PSF ellipticity and stability of PSF quality within an exposure is particularly sensitive to these two quantities.

Figure~\ref{fig:exp-qual} illustrates the application of this exposure-selection scheme for the target Abell~3411. The upper panels show size-magnitude diagrams for two representative exposures: a ``good'' exposure with a compact, well-defined stellar locus (upper left) and a ``bad'' exposure with a more scattered stellar locus (upper right). In both panels, stars are color-coded by their distance from the center of the exposure. As shown, the increased size scatter is primarily associated with stars located near the outskirts of the exposure. The lower panels show the time evolution of the scatter in the PSF-star FWHM (middle panel) and the mean PSF ellipticity $\langle |e| \rangle = \sqrt{e_1^2 + e_2^2}$ measured in the outer regions of the CCD (lower panel) for all exposures of this target. The x-axis denotes the UTC time at which each exposure was taken.

Exposures are flagged and excluded from the shape-measurement pipeline when either of the two selection criteria fails, namely $\sigma(\mathrm{FWHM}) \ge 0.1$ or $\langle |e| \rangle \ge 0.25$. Flagged exposures are indicated by black crosses in Figure~\ref{fig:exp-qual}.

\section{\textsc{ngmix} Shape-Measurement Workflow}\label{appendix:ngmix-workflow}
\begin{figure*}[t]
\centering
% \hspace*{-2em}
\includegraphics[width=0.85\textwidth]{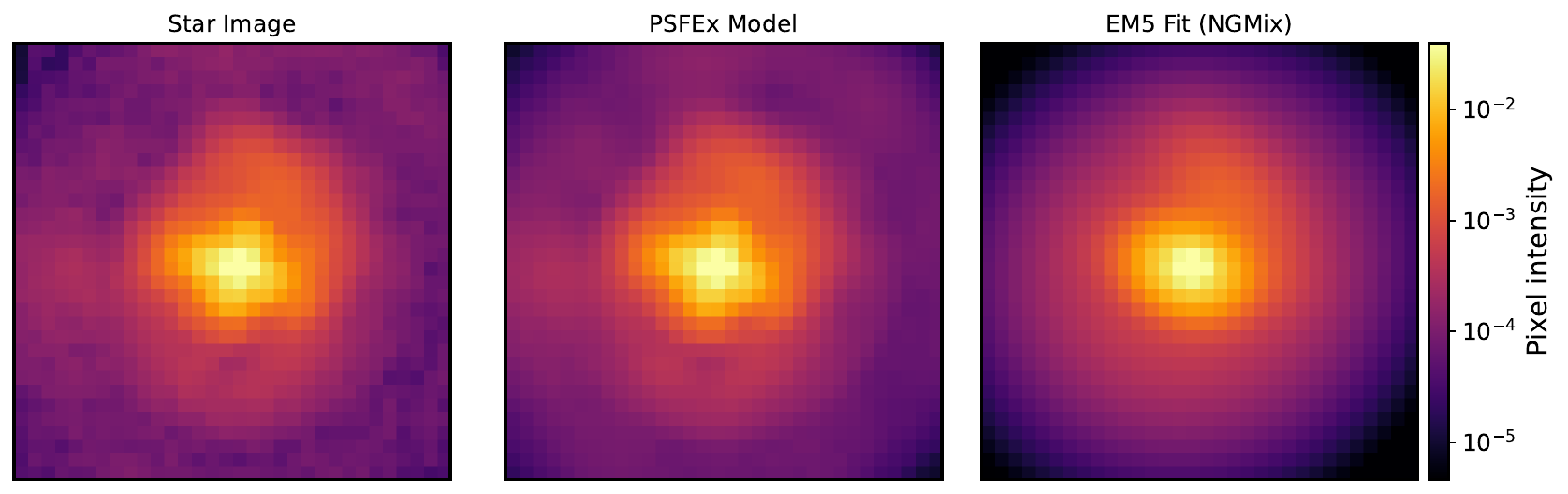}
\caption{Downstream propagation of the PSF model through the pipeline. 
\emph{Left:} normalized image of a test star as detected by the telescope. 
\emph{Middle:} PSF model interpolated at the star position using \textsc{PSFEx}. 
\emph{Right:} best-fit PSF model from \textsc{ngmix}, represented by the
\texttt{em5} Gaussian mixture model (a mixture of five Gaussians), which is used for Bayesian forward modeling in shape measurement. All images are displayed
using a logarithmic color scale.}
\label{fig:psf-downstream}
\end{figure*}
\begin{figure*}[!htbp]
\centering
\hspace*{-2em}
\includegraphics[width=\textwidth]{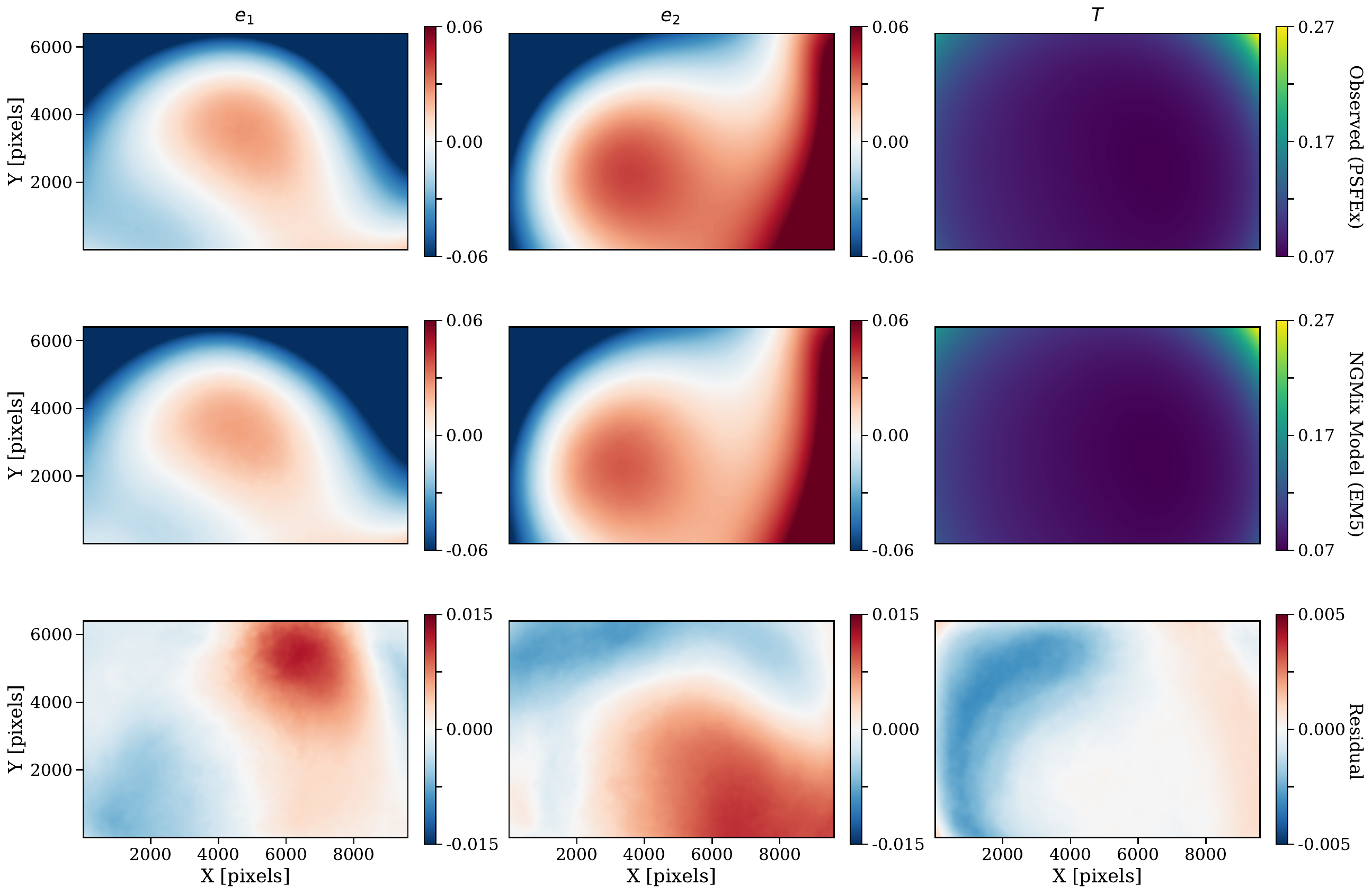}
\caption{Evaluation of the ability of the \textsc{ngmix} \texttt{em5} model to represent the SuperBIT PSF (here given by the PSFEx model). \emph{Top panel:} PSF ellipticity components and size measured from the PSFEx model, rendered across the CCD coordinates of all exposures. \emph{Middle panel:} Corresponding measurements from the best-fit \textsc{ngmix} Gaussian-mixture model (\texttt{em5}). \emph{Bottom panel:} Residuals between the PSFEx measurements and the best-fit \texttt{em5} model across the CCD.}
\label{fig:em5-mean}
\end{figure*}
% \begin{figure*}[t]
% \centering
% \hspace*{-2em}
% \includegraphics[width=\textwidth]{mean_psf_properties_em5_one_panel.pdf}
% \caption{Spatial distribution of PSF modeling residuals across the CCD. The panels show the residual ellipticity components ($\delta e_1$, $\delta e_2$) and size ($\delta T$) between the PSFEx model and the best-fit \textsc{ngmix} \texttt{em5} model.}
% \label{fig:em5-mean}
% \end{figure*}
\begin{figure*}[]
\centering
\hspace*{-2em}
%\vspace*{2em}
\includegraphics[width=0.9\textwidth]{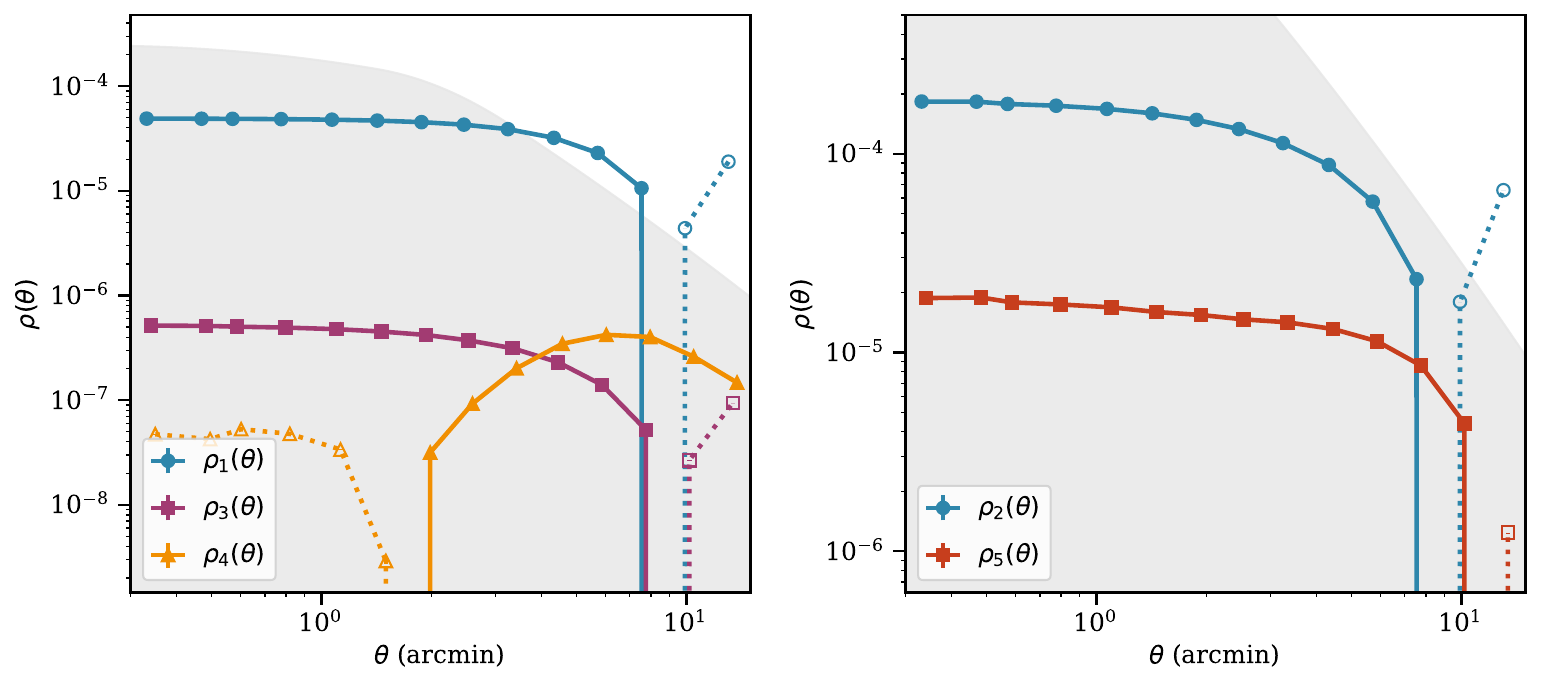}
\caption{The $\rho$-statistics estimated from the ellipticity and size residuals of the best-fit \texttt{em5} model. The left panel shows $\rho_1$, $\rho_3$, and $\rho_4$, while the right panel shows $\rho_2$ and $\rho_5$. The gray shaded region in both panels corresponds to 10\% of the expected two-point shear correlation signal for a representative SuperBIT galaxy cluster, computed using the same conservative assumptions described in Figure~\ref{fig:rho-stats}.}
\label{fig:rhostats_em5}
\end{figure*}

In this appendix, we provide a brief overview of the \textsc{ngmix} shape-measurement workflow used in our analysis. This description is intended to help readers who may be using \textsc{ngmix} for the first time understand how the algorithm operates within the SuperBIT pipeline. In addition, we evaluate the performance of the \textsc{ngmix} PSF modeling in the context of SuperBIT observations, verifying that the Gaussian-mixture representation provides an adequate description of the PSF for our weak-lensing measurements.

\textsc{ngmix} measures pre-PSF galaxy parameters for each object detected in the coadded image. Starting from the detection catalog in the coadd, we construct a Multi-Epoch Data Structure (MEDS)\footnote{ \url{https://github.com/esheldon/meds/wiki}}, which contains the image of each detected object in every exposure, the corresponding PSF image, and the Jacobian of the WCS transformation for each exposure. The Jacobian encodes the
orientation of the CCD plane at the time of each exposure, and due to telescope
dithering, it differs between exposures; this is automatically accounted for in
the MEDS-based shape measurement.

For each exposure, the observed galaxy image is formed by convolution of the
intrinsic (pre-PSF) galaxy light profile with the SuperBIT PSF for that
exposure, followed by pixelization and the addition of detector noise. Thus,
for each galaxy we have a set of observed images
$\{I^{\rm exp}\}$ and corresponding PSF images
$\{\rm PSF^{\rm exp}\}$ across the full exposure stack:
\begin{equation}
I^{\text{exp}}_{i,j} =
\left( G \otimes \rm{PSF}^{\rm exp} \right)_{i,j} + \eta^{\rm exp}_{i,j},
\label{eq:convolution_simple}
\end{equation}
where $(i,j)$ denote pixel coordinates in the two-dimensional cutouts,
$\eta^{\rm exp}_{i,j}$ is the detector noise for that exposure, and $G$ is the
pre-PSF galaxy model, assumed to be constant across all exposures.

The intrinsic galaxy model $G$ is parametrized by the six parameters
$\{c_x, c_y, e_1, e_2, f, T\}$, corresponding to the centroid coordinates
$(c_x, c_y)$, ellipticity components $(e_1, e_2)$, flux $f$, and size $T$.
While the galaxy parameters are constant across exposures, the PSF models and
noise realizations vary between exposures.

The first step in \textsc{ngmix} is a least-squares fit of a Gaussian mixture
model to the PSF images. In our case, the \texttt{em5} model---a mixture of five
Gaussians---is fitted to the PSF images, which are themselves interpolated models
from \textsc{PSFEx}. This procedure yields best-fit PSF parameters for all
$\{\rm PSF^{\rm exp}\}$ across the exposure stack. Figure~\ref{fig:psf-downstream} illustrates the downstream propagation of the PSF model through the pipeline, from the observed stellar image to the \textsc{PSFEx} interpolation and finally to the best-fit \textsc{ngmix} Gaussian-mixture model used for forward modeling. A key concern when using the \textsc{ngmix} Gaussian-mixture representation is its ability to accurately capture the SuperBIT PSF. For ground-based telescopes, atmospheric seeing produces a relatively broad and smooth PSF that is well approximated by Gaussian mixtures. In contrast, for a balloon-borne mission such as SuperBIT, the absence of atmospheric seeing leads to a PSF with significant high-frequency structure (middle panel of Figure~\ref{fig:psf-downstream}). These small-scale features are not fully captured by the best-fit \texttt{em5} Gaussian-mixture model (right panel), which smooths over much of the fine structure present in the PSFEx model. The key question is therefore whether this loss of high-frequency information introduces residual ellipticity in the PSF model that could bias weak-lensing measurements.

To assess this, we evaluate the ellipticity and size residuals between the PSFEx model rendered at the position of each galaxy and the best-fit \texttt{em5} Gaussian-mixture model of that rendering across the CCD. In Figure~\ref{fig:em5-mean}, the top row shows the spatial variation of the PSF ellipticity components ($e_1$, $e_2$) and size $T$ measured from the PSFEx model, while the middle row shows the corresponding measurements from the best-fit \texttt{em5} model. The bottom row displays the residuals between the two as a function of CCD position. In Figure~\ref{fig:rhostats_em5}, we compute the $\rho$ statistics of these residuals to quantify their potential impact on weak-lensing two-point correlation measurements, following the procedure described in Section~\ref{sec:psf-modeling}. The shaded regions indicate the ``safe zone,'' defined as 10\% of the expected shear two-point correlation signal for a SuperBIT-like cluster given the anticipated background galaxy redshift distribution. All $\rho$ statistics fall within this region, indicating that residual modeling errors from the \texttt{em5} representation are sufficiently small for the planned science analysis.

After fitting the PSF models, \textsc{ngmix} performs a least-squares
optimization of the posterior probability distribution for the galaxy model:
\begin{equation}
P(\theta_{\rm gal} \,|\, \{I^{\rm exp}\}) =
\frac{\mathcal{L}(\{I^{\rm exp}\} \,|\, G)\, P(G)}{P(\{I^{\rm exp}\})},
\label{eq:log-posterior}
\end{equation}
where $\mathcal{L}$ is the likelihood and $P(G)$ is the prior on galaxy
properties. The corresponding log-likelihood is given by
\begin{equation}
-2 \log \mathcal{L}
=
\sum_{\rm exp} \sum_{i,j}
\frac{\left[I^{\rm exp}_{i,j} -
(G \otimes \rm{PSF}^{\rm exp})_{i,j}\right]^2}{\sigma_{i,j}^2}.
\end{equation}

\textsc{ngmix} then optimizes the posterior in
Eq.~\ref{eq:log-posterior} to obtain the best-fit galaxy parameters
$G \equiv \{c_x, c_y, e_1, e_2, f, T\}$. This full shape-measurement procedure is
repeated for all \textsc{metacalibration} shear states
$\{\texttt{noshear}, 1+, 1-, 2+, 2-\}$, as well as for the PSF-sheared states
$\{\texttt{PSF1+}, \texttt{PSF1-}, \texttt{PSF2+}, \texttt{PSF2-}\}$. These
measurements are subsequently used to compute the shear and PSF response
matrices, as discussed in detail in Section~\ref{sec:metacal}.

\section{PSF Response Correction}\label{app:psf-correction}
\begin{figure*}[]
\centering
\hspace*{-2em}
\includegraphics[width=\textwidth]{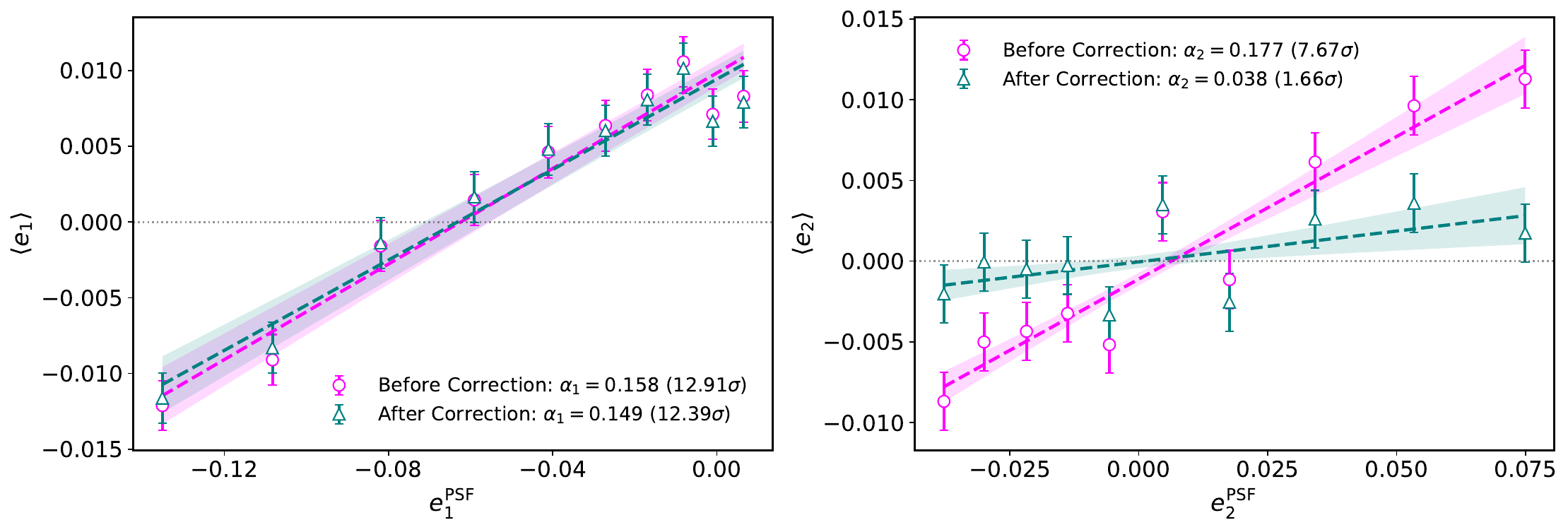}
\caption{PSF leakage results after the PSF response correction mentioned in Eq.~\ref{eq:psf_response_corr}. The magenta line shows the leakage when the response correction has not been applied, and the teal line shows the results after the correction has been applied. The left panel shows the results for the $e_1$ component, and the right panel shows the results for the $e_2$ component. The legends also mention the fitted leakage strength $\alpha$ for each case.}
\label{fig:psf-correction}
\end{figure*}
As mentioned in Section~\ref{sec:metacal}, the balloon-borne platform produces a non-gaussian, unstable PSF, which prevented us from using an analytical reconvolution kernel for metacalibration. Instead, we used a dilated version of the original PSF to ensure that no noisy modes are amplified during metacalibration. Even though this approach is noise-safe, it can introduce PSF ellipticity leakage, which must be corrected using the PSF response, as given in Eq.~\ref{eq:psf_response_corr}. In this appendix, we show how this correction reduces the leakage. Figure~\ref{fig:psf-correction} shows the ellipticity leakage before and after the PSF response correction, with data binned into percentile bins of PSF ellipticity. The PSF response is calculated separately within each bin and used to apply the correction. We find a significant improvement for the $e_2$ component, where $\alpha_2 = 0.177$ ($7.67\sigma$) decreases to $\alpha_2 = 0.038$ ($1.66\sigma$) after correction. In contrast, we do not see a significant improvement for the $e_1$ component: even after PSF response correction, the leakage remains at $\alpha_1 = 0.149$ ($12.39\sigma$). It is important to note that for this result we have not imposed any PSF ellipticity cut, whereas our main calibration results were obtained with a PSF ellipticity cut of $e_{\text{PSF}} < 0.07$ (Section~\ref{sec:shear_bias}); with this cut applied, the leakage remains within our budget, as shown in Figure~\ref{fig:psf-leakage-sim}.

\section{Stability of Gridded Calibration}\label{app:r_stability}
\begin{figure*}[]
\centering
% \hspace*{-2em}
\includegraphics[width=\textwidth]{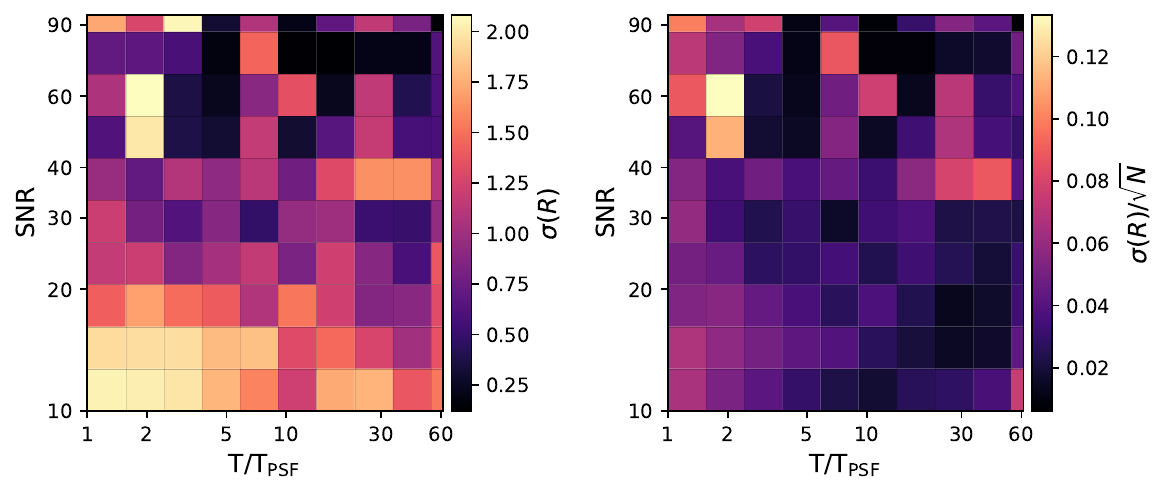}
\caption{Stability of the shear response used to calibrate the shape catalog. The right panel shows the standard deviation ($\sigma$) per bin, and the left panel shows the precision of $R_\gamma$, the quantity shown in Figure~\ref{fig:weights}.}
\label{fig:r_stability}
\end{figure*}
The gridded shear calibration mentioned in Section~\ref{sec:metacal} and Section~\ref{sec:shear_bias} is one of the central results in this work. The gridded response shown in Figure~\ref{fig:weights} is smoothed with a Gaussian kernel of $\sigma=2$ to suppress some of the shot noise. The rationale behind doing this is that we expect the response to be a slowly varying quantity over SNR and $T/T_{\text{PSF}}$. While this can introduce a smoothing bias, it substantially reduces the noise in nearby pixels, and we have validated that this trade-off works using our fiducial simulations. It is still important to track the stability of the response in each bin, as well as the precision of the values plotted in Figure~\ref{fig:weights}. We show this stability in Figure~\ref{fig:r_stability}: the right panel shows the standard deviation of the mean response in each bin, and the left panel shows the precision of the values used to calibrate the shape catalog.

% \section{Shape Comparison with DES Y3}

% To further validate our pipeline, we cross-match the SuperBIT shape catalogs for all targets with the DES Y3 shape catalogs~\citep{DES:2020ekd}. For objects with high signal-to-noise ratio (SNR $\gtrsim 40$), we find excellent consistency between the two datasets. The comparison of ellipticity components is shown in Figure~\ref{fig:des-compare}. The data points are binned using hexagonal binning, and linear fits are shown as red solid lines, with the fitted slopes and intercepts indicated in the legend. The black dashed line denotes the one-to-one relation, $y=x$.

% Quantitatively, we obtain best-fit relations
% $m = 1.006 \pm 0.015$ and $c = 0.004 \pm 0.001$ for the $e_1$ component, and
% $m = 1.004 \pm 0.009$ and $c = 0.002 \pm 0.001$ for the $e_2$ component.

% \begin{figure*}[t]
% \centering
% \hspace*{-2em}
% \includegraphics[width=0.85\textwidth]{figures/shear_comparison_des_sb.pdf}
% \caption{Comparison of galaxy shapes measured with SuperBIT and DES Y3. The data
% are shown in hexagonal bins, with best-fit linear relations overplotted as red
% solid lines and the fitted slopes and intercepts indicated in the legend. The
% black dashed line denotes the one-to-one relation, $y=x$.}
% \label{fig:des-compare}
% \end{figure*}

\bibliography{references}{}

@ARTICLE{2022ApJ...935..167A,
       author = {{Astropy Collaboration} and {Price-Whelan}, Adrian M. and {Lim}, Pey Lian and {Earl}, Nicholas and {Starkman}, Nathaniel and {Bradley}, Larry and {Shupe}, David L. and {Patil}, Aarya A. and {Corrales}, Lia and {Brasseur}, C.~E. and {N{\"o}the}, Maximilian and {Donath}, Axel and {Tollerud}, Erik and {Morris}, Brett M. and {Ginsburg}, Adam and {Vaher}, Eero and {Weaver}, Benjamin A. and {Tocknell}, James and {Jamieson}, William and {van Kerkwijk}, Marten H. and {Robitaille}, Thomas P. and {Merry}, Bruce and {Bachetti}, Matteo and {G{\"u}nther}, H. Moritz and {Aldcroft}, Thomas L. and {Alvarado-Montes}, Jaime A. and {Archibald}, Anne M. and {B{\'o}di}, Attila and {Bapat}, Shreyas and {Barentsen}, Geert and {Baz{\'a}n}, Juanjo and {Biswas}, Manish and {Boquien}, M{\'e}d{\'e}ric and {Burke}, D.~J. and {Cara}, Daria and {Cara}, Mihai and {Conroy}, Kyle E. and {Conseil}, Simon and {Craig}, Matthew W. and {Cross}, Robert M. and {Cruz}, Kelle L. and {D'Eugenio}, Francesco and {Dencheva}, Nadia and {Devillepoix}, Hadrien A.~R. and {Dietrich}, J{\"o}rg P. and {Eigenbrot}, Arthur Davis and {Erben}, Thomas and {Ferreira}, Leonardo and {Foreman-Mackey}, Daniel and {Fox}, Ryan and {Freij}, Nabil and {Garg}, Suyog and {Geda}, Robel and {Glattly}, Lauren and {Gondhalekar}, Yash and {Gordon}, Karl D. and {Grant}, David and {Greenfield}, Perry and {Groener}, Austen M. and {Guest}, Steve and {Gurovich}, Sebastian and {Handberg}, Rasmus and {Hart}, Akeem and {Hatfield-Dodds}, Zac and {Homeier}, Derek and {Hosseinzadeh}, Griffin and {Jenness}, Tim and {Jones}, Craig K. and {Joseph}, Prajwel and {Kalmbach}, J. Bryce and {Karamehmetoglu}, Emir and {Ka{\l}uszy{\'n}ski}, Miko{\l}aj and {Kelley}, Michael S.~P. and {Kern}, Nicholas and {Kerzendorf}, Wolfgang E. and {Koch}, Eric W. and {Kulumani}, Shankar and {Lee}, Antony and {Ly}, Chun and {Ma}, Zhiyuan and {MacBride}, Conor and {Maljaars}, Jakob M. and {Muna}, Demitri and {Murphy}, N.~A. and {Norman}, Henrik and {O'Steen}, Richard and {Oman}, Kyle A. and {Pacifici}, Camilla and {Pascual}, Sergio and {Pascual-Granado}, J. and {Patil}, Rohit R. and {Perren}, Gabriel I. and {Pickering}, Timothy E. and {Rastogi}, Tanuj and {Roulston}, Benjamin R. and {Ryan}, Daniel F. and {Rykoff}, Eli S. and {Sabater}, Jose and {Sakurikar}, Parikshit and {Salgado}, Jes{\'u}s and {Sanghi}, Aniket and {Saunders}, Nicholas and {Savchenko}, Volodymyr and {Schwardt}, Ludwig and {Seifert-Eckert}, Michael and {Shih}, Albert Y. and {Jain}, Anany Shrey and {Shukla}, Gyanendra and {Sick}, Jonathan and {Simpson}, Chris and {Singanamalla}, Sudheesh and {Singer}, Leo P. and {Singhal}, Jaladh and {Sinha}, Manodeep and {Sip{\H{o}}cz}, Brigitta M. and {Spitler}, Lee R. and {Stansby}, David and {Streicher}, Ole and {{\v{S}}umak}, Jani and {Swinbank}, John D. and {Taranu}, Dan S. and {Tewary}, Nikita and {Tremblay}, Grant R. and {de Val-Borro}, Miguel and {Van Kooten}, Samuel J. and {Vasovi{\'c}}, Zlatan and {Verma}, Shresth and {de Miranda Cardoso}, Jos{\'e} Vin{\'\i}cius and {Williams}, Peter K.~G. and {Wilson}, Tom J. and {Winkel}, Benjamin and {Wood-Vasey}, W.~M. and {Xue}, Rui and {Yoachim}, Peter and {Zhang}, Chen and {Zonca}, Andrea and {Astropy Project Contributors}},
        title = "{The Astropy Project: Sustaining and Growing a Community-oriented Open-source Project and the Latest Major Release (v5.0) of the Core Package}",
      journal = {\apj},
         year = 2022,
        month = aug,
       volume = {935},
       number = {2},
          eid = {167},
        pages = {167},
          doi = {10.3847/1538-4357/ac7c74},
archivePrefix = {arXiv},
       eprint = {2206.14220},
 primaryClass = {astro-ph.IM},
       adsurl = {https://ui.adsabs.harvard.edu/abs/2022ApJ...935..167A}
}

@ARTICLE{2018AJ....156..123A,
       author = {{Astropy Collaboration} and {Price-Whelan}, A.~M. and {Sip{\H{o}}cz}, B.~M. and {G{\"u}nther}, H.~M. and {Lim}, P.~L. and {Crawford}, S.~M. and {Conseil}, S. and {Shupe}, D.~L. and {Craig}, M.~W. and {Dencheva}, N. and {Ginsburg}, A. and {VanderPlas}, J.~T. and {Bradley}, L.~D. and {P{\'e}rez-Su{\'a}rez}, D. and {de Val-Borro}, M. and {Aldcroft}, T.~L. and {Cruz}, K.~L. and {Robitaille}, T.~P. and {Tollerud}, E.~J. and {Ardelean}, C. and {Babej}, T. and {Bach}, Y.~P. and {Bachetti}, M. and {Bakanov}, A.~V. and {Bamford}, S.~P. and {Barentsen}, G. and {Barmby}, P. and {Baumbach}, A. and {Berry}, K.~L. and {Biscani}, F. and {Boquien}, M. and {Bostroem}, K.~A. and {Bouma}, L.~G. and {Brammer}, G.~B. and {Bray}, E.~M. and {Breytenbach}, H. and {Buddelmeijer}, H. and {Burke}, D.~J. and {Calderone}, G. and {Cano Rodr{\'\i}guez}, J.~L. and {Cara}, M. and {Cardoso}, J.~V.~M. and {Cheedella}, S. and {Copin}, Y. and {Corrales}, L. and {Crichton}, D. and {D'Avella}, D. and {Deil}, C. and {Depagne}, {\'E}. and {Dietrich}, J.~P. and {Donath}, A. and {Droettboom}, M. and {Earl}, N. and {Erben}, T. and {Fabbro}, S. and {Ferreira}, L.~A. and {Finethy}, T. and {Fox}, R.~T. and {Garrison}, L.~H. and {Gibbons}, S.~L.~J. and {Goldstein}, D.~A. and {Gommers}, R. and {Greco}, J.~P. and {Greenfield}, P. and {Groener}, A.~M. and {Grollier}, F. and {Hagen}, A. and {Hirst}, P. and {Homeier}, D. and {Horton}, A.~J. and {Hosseinzadeh}, G. and {Hu}, L. and {Hunkeler}, J.~S. and {Ivezi{\'c}}, {\v{Z}}. and {Jain}, A. and {Jenness}, T. and {Kanarek}, G. and {Kendrew}, S. and {Kern}, N.~S. and {Kerzendorf}, W.~E. and {Khvalko}, A. and {King}, J. and {Kirkby}, D. and {Kulkarni}, A.~M. and {Kumar}, A. and {Lee}, A. and {Lenz}, D. and {Littlefair}, S.~P. and {Ma}, Z. and {Macleod}, D.~M. and {Mastropietro}, M. and {McCully}, C. and {Montagnac}, S. and {Morris}, B.~M. and {Mueller}, M. and {Mumford}, S.~J. and {Muna}, D. and {Murphy}, N.~A. and {Nelson}, S. and {Nguyen}, G.~H. and {Ninan}, J.~P. and {N{\"o}the}, M. and {Ogaz}, S. and {Oh}, S. and {Parejko}, J.~K. and {Parley}, N. and {Pascual}, S. and {Patil}, R. and {Patil}, A.~A. and {Plunkett}, A.~L. and {Prochaska}, J.~X. and {Rastogi}, T. and {Reddy Janga}, V. and {Sabater}, J. and {Sakurikar}, P. and {Seifert}, M. and {Sherbert}, L.~E. and {Sherwood-Taylor}, H. and {Shih}, A.~Y. and {Sick}, J. and {Silbiger}, M.~T. and {Singanamalla}, S. and {Singer}, L.~P. and {Sladen}, P.~H. and {Sooley}, K.~A. and {Sornarajah}, S. and {Streicher}, O. and {Teuben}, P. and {Thomas}, S.~W. and {Tremblay}, G.~R. and {Turner}, J.~E.~H. and {Terr{\'o}n}, V. and {van Kerkwijk}, M.~H. and {de la Vega}, A. and {Watkins}, L.~L. and {Weaver}, B.~A. and {Whitmore}, J.~B. and {Woillez}, J. and {Zabalza}, V. and {Astropy Contributors}},
        title = "{The Astropy Project: Building an Open-science Project and Status of the v2.0 Core Package}",
      journal = {\aj},
         year = 2018,
        month = sep,
       volume = {156},
       number = {3},
          eid = {123},
        pages = {123},
          doi = {10.3847/1538-3881/aabc4f},
archivePrefix = {arXiv},
       eprint = {1801.02634},
 primaryClass = {astro-ph.IM},
       adsurl = {https://ui.adsabs.harvard.edu/abs/2018AJ....156..123A}
}

@ARTICLE{2013A&A...558A..33A,
       author = {{Astropy Collaboration} and {Robitaille}, Thomas P. and
         {Tollerud}, Erik J. and {Greenfield}, Perry and {Droettboom}, Michael and
         {Bray}, Erik and {Aldcroft}, Tom and {Davis}, Matt and
         {Ginsburg}, Adam and {Price-Whelan}, Adrian M. and
         {Kerzendorf}, Wolfgang E. and {Conley}, Alexander and {Crighton}, Neil and
         {Barbary}, Kyle and {Muna}, Demitri and {Ferguson}, Henry and
         {Grollier}, Fr{\'e}d{\'e}ric and {Parikh}, Madhura M. and
         {Nair}, Prasanth H. and {Unther}, Hans M. and {Deil}, Christoph and
         {Woillez}, Julien and {Conseil}, Simon and {Kramer}, Roban and
         {Turner}, James E.~H. and {Singer}, Leo and {Fox}, Ryan and
         {Weaver}, Benjamin A. and {Zabalza}, Victor and {Edwards}, Zachary I. and
         {Azalee Bostroem}, K. and {Burke}, D.~J. and {Casey}, Andrew R. and
         {Crawford}, Steven M. and {Dencheva}, Nadia and {Ely}, Justin and
         {Jenness}, Tim and {Labrie}, Kathleen and {Lim}, Pey Lian and
         {Pierfederici}, Francesco and {Pontzen}, Andrew and {Ptak}, Andy and
         {Refsdal}, Brian and {Servillat}, Mathieu and {Streicher}, Ole},
        title = "{Astropy: A community Python package for astronomy}",
      journal = {\aap},
         year = "2013",
        month = "Oct",
       volume = {558},
          eid = {A33},
        pages = {A33},
          doi = {10.1051/0004-6361/201322068},
archivePrefix = {arXiv},
       eprint = {1307.6212},
 primaryClass = {astro-ph.IM},
       adsurl = {https://ui.adsabs.harvard.edu/abs/2013A&A...558A..33A}
}

@ARTICLE{sextractor,
       author = {{Bertin}, E. and {Arnouts}, S.},
        title = "{SExtractor: Software for source extraction.}",
      journal = {\aaps},
         year = 1996,
        month = jun,
       volume = {117},
        pages = {393-404},
          doi = {10.1051/aas:1996164},
       adsurl = {https://ui.adsabs.harvard.edu/abs/1996A&AS..117..393B}
}

@ARTICLE{Mandelbaum:2005wv,
       author = {{Mandelbaum}, Rachel and {Hirata}, Christopher M. and {Seljak}, Uro{\v{s}} and {Guzik}, Jacek and {Padmanabhan}, Nikhil and {Blake}, Cullen and {Blanton}, Michael R. and {Lupton}, Robert and {Brinkmann}, Jonathan},
        title = "{Systematic errors in weak lensing: application to SDSS galaxy-galaxy weak lensing}",
      journal = {\mnras},
         year = 2005,
        month = aug,
       volume = {361},
       number = {4},
        pages = {1287-1322},
          doi = {10.1111/j.1365-2966.2005.09282.x},
archivePrefix = {arXiv},
       eprint = {astro-ph/0501201},
 primaryClass = {astro-ph},
       adsurl = {https://ui.adsabs.harvard.edu/abs/2005MNRAS.361.1287M}
}

@ARTICLE{Schneider:1994eq,
       author = {{Schneider}, Peter and {Seitz}, Carolin},
        title = "{Steps towards nonlinear cluster inversion through gravitational distortions. I. Basic considerations and circular clusters.}",
      journal = {\aap},
         year = 1995,
        month = feb,
       volume = {294},
        pages = {411-431},
          doi = {10.48550/arXiv.astro-ph/9407032},
archivePrefix = {arXiv},
       eprint = {astro-ph/9407032},
 primaryClass = {astro-ph},
       adsurl = {https://ui.adsabs.harvard.edu/abs/1995A&A...294..411S}
}

@ARTICLE{Mandelbaum:2017jpr,
       author = {{Mandelbaum}, Rachel},
        title = "{Weak Lensing for Precision Cosmology}",
      journal = {\araa},
         year = 2018,
        month = sep,
       volume = {56},
        pages = {393-433},
          doi = {10.1146/annurev-astro-081817-051928},
archivePrefix = {arXiv},
       eprint = {1710.03235},
 primaryClass = {astro-ph.CO},
       adsurl = {https://ui.adsabs.harvard.edu/abs/2018ARA&A..56..393M}
}

@ARTICLE{Huff:2017qxu,
       author = {{Huff}, Eric and {Mandelbaum}, Rachel},
        title = "{Metacalibration: Direct Self-Calibration of Biases in Shear Measurement}",
      journal = {arXiv e-prints},
         year = 2017,
        month = feb,
          eid = {arXiv:1702.02600},
        pages = {arXiv:1702.02600},
          doi = {10.48550/arXiv.1702.02600},
archivePrefix = {arXiv},
       eprint = {1702.02600},
 primaryClass = {astro-ph.CO},
       adsurl = {https://ui.adsabs.harvard.edu/abs/2017arXiv170202600H}
}

@ARTICLE{Sheldon:2017szh,
       author = {{Sheldon}, Erin S. and {Huff}, Eric M.},
        title = "{Practical Weak-lensing Shear Measurement with Metacalibration}",
      journal = {\apj},
         year = 2017,
        month = may,
       volume = {841},
       number = {1},
          eid = {24},
        pages = {24},
          doi = {10.3847/1538-4357/aa704b},
archivePrefix = {arXiv},
       eprint = {1702.02601},
 primaryClass = {astro-ph.CO},
       adsurl = {https://ui.adsabs.harvard.edu/abs/2017ApJ...841...24S}
}

@ARTICLE{Sheldon:2014vda,
       author = {{Sheldon}, E.~S.},
        title = "{An implementation of Bayesian lensing shear measurement.}",
      journal = {\mnras},
         year = 2014,
        month = oct,
       volume = {444},
        pages = {L25-L29},
          doi = {10.1093/mnrasl/slu104},
archivePrefix = {arXiv},
       eprint = {1403.7669},
 primaryClass = {astro-ph.CO},
       adsurl = {https://ui.adsabs.harvard.edu/abs/2014MNRAS.444L..25S}
}

@ARTICLE{DES:2020ekd,
       author = {{Gatti}, M. and {Sheldon}, E. and {Amon}, A. and {Becker}, M. and {Troxel}, M. and {Choi}, A. and {Doux}, C. and {MacCrann}, N. and {Navarro-Alsina}, A. and {Harrison}, I. and {Gruen}, D. and {Bernstein}, G. and {Jarvis}, M. and {Secco}, L.~F. and {Fert{\'e}}, A. and {Shin}, T. and {McCullough}, J. and {Rollins}, R.~P. and {Chen}, R. and {Chang}, C. and {Pandey}, S. and {Tutusaus}, I. and {Prat}, J. and {Elvin-Poole}, J. and {Sanchez}, C. and {Plazas}, A.~A. and {Roodman}, A. and {Zuntz}, J. and {Abbott}, T.~M.~C. and {Aguena}, M. and {Allam}, S. and {Annis}, J. and {Avila}, S. and {Bacon}, D. and {Bertin}, E. and {Bhargava}, S. and {Brooks}, D. and {Burke}, D.~L. and {Carnero Rosell}, A. and {Carrasco Kind}, M. and {Carretero}, J. and {Castander}, F.~J. and {Conselice}, C. and {Costanzi}, M. and {Crocce}, M. and {da Costa}, L.~N. and {Davis}, T.~M. and {De Vicente}, J. and {Desai}, S. and {Diehl}, H.~T. and {Dietrich}, J.~P. and {Doel}, P. and {Drlica-Wagner}, A. and {Eckert}, K. and {Everett}, S. and {Ferrero}, I. and {Frieman}, J. and {Garc{\'\i}a-Bellido}, J. and {Gerdes}, D.~W. and {Giannantonio}, T. and {Gruendl}, R.~A. and {Gschwend}, J. and {Gutierrez}, G. and {Hartley}, W.~G. and {Hinton}, S.~R. and {Hollowood}, D.~L. and {Honscheid}, K. and {Hoyle}, B. and {Huff}, E.~M. and {Huterer}, D. and {Jain}, B. and {James}, D.~J. and {Jeltema}, T. and {Krause}, E. and {Kron}, R. and {Kuropatkin}, N. and {Lima}, M. and {Maia}, M.~A.~G. and {Marshall}, J.~L. and {Miquel}, R. and {Morgan}, R. and {Myles}, J. and {Palmese}, A. and {Paz-Chinch{\'o}n}, F. and {Rykoff}, E.~S. and {Samuroff}, S. and {Sanchez}, E. and {Scarpine}, V. and {Schubnell}, M. and {Serrano}, S. and {Sevilla-Noarbe}, I. and {Smith}, M. and {Soares-Santos}, M. and {Suchyta}, E. and {Swanson}, M.~E.~C. and {Tarle}, G. and {Thomas}, D. and {To}, C. and {Tucker}, D.~L. and {Varga}, T.~N. and {Wechsler}, R.~H. and {Weller}, J. and {Wester}, W. and {Wilkinson}, R.~D.},
        title = "{Dark energy survey year 3 results: weak lensing shape catalogue}",
      journal = {\mnras},
         year = 2021,
        month = jul,
       volume = {504},
       number = {3},
        pages = {4312-4336},
          doi = {10.1093/mnras/stab918},
archivePrefix = {arXiv},
       eprint = {2011.03408},
 primaryClass = {astro-ph.CO},
       adsurl = {https://ui.adsabs.harvard.edu/abs/2021MNRAS.504.4312G}
}

@ARTICLE{mccleary2023,
       author = {{McCleary}, Jacqueline E. and {Everett}, Spencer W. and {Shaaban}, Mohamed M. and {Gill}, Ajay S. and {Vassilakis}, Georgios N. and {Huff}, Eric M. and {Massey}, Richard J. and {Benton}, Steven J. and {Brown}, Anthony M. and {Clark}, Paul and {Holder}, Bradley and {Fraisse}, Aurelien A. and {Jauzac}, Mathilde and {Jones}, William C. and {Lagattuta}, David and {Leung}, Jason S. -Y. and {Li}, Lun and {T. Luu}, Thuy Vy and {Nagy}, Johanna M. and {Netterfield}, C. Barth and {Paracha}, Emaad and {Redmond}, Susan F. and {Rhodes}, Jason D. and {Schmoll}, J{\"u}rgen and {Sirks}, Ellen and {Tam}, Sut Ieng},
        title = "{Lensing in the Blue. II. Estimating the Sensitivity of Stratospheric Balloons to Weak Gravitational Lensing}",
      journal = "Astron. J.",
         year = 2023,
        month = sep,
       volume = {166},
       number = {3},
          eid = {134},
        pages = {134},
          doi = {10.3847/1538-3881/ace7ca},
archivePrefix = {arXiv},
       eprint = {2307.03295},
 primaryClass = {astro-ph.IM},
       adsurl = {https://ui.adsabs.harvard.edu/abs/2023AJ....166..134M}
}

@ARTICLE{Anbajagane:2025mso,
author = {{Anbajagane}, D. and {Zhang}, Z. and {Chang}, C. and {Tan}, C.~Y. and {Adamow}, M. and {Secco}, L.~F. and {Becker}, M.~R. and {Ferguson}, P.~S. and {Drlica-Wagner}, A. and {Gruendl}, R.~A. and {Herron}, K. and {Tong}, A. and {Troxel}, M.~A. and {Sanchez-Cid}, D. and {Sevilla-Noarbe}, I. and {Chicoine}, N. and {Teixeira}, R. and {Alarcon}, A. and {Suson}, D. and {Alsina}, A.~N. and {Amon}, A. and {Bom}, C.~R. and {Carballo-Bello}, J.~A. and {Cerny}, W. and {Choi}, A. and {Choi}, Y. and {Doux}, C. and {Eckert}, K. and {Gatti}, M. and {Gruen}, D. and {Jarvis}, M. and {James}, D.~J. and {Kuropatkin}, N. and {Mart{\'\i}nez-V{\'a}zquez}, C.~E. and {Massana}, P. and {Mau}, S. and {McCullough}, J. and {Medina}, G.~E. and {Mutlu-Pakdil}, B. and {Navabi}, M. and {No{\"e}l}, N.~E.~D. and {Pace}, A.~B. and {Prat}, J. and {Raveri}, M. and {Riley}, A.~H. and {Rykoff}, E.~S. and {Sakowska}, J.~D. and {Sand}, D.~J. and {Santana-Silva}, L. and {Shin}, T. and {Soares-Santos}, M. and {Stringfellow}, G.~S. and {Vivas}, A.~K. and {Yamamoto}, M.},
        title = "{The DECADE cosmic shear project I: A new weak lensing shape catalog of 107 million galaxies}",
      journal = {The Open Journal of Astrophysics},
         year = 2025,
        month = oct,
       volume = {8},
        pages = {46158},
          doi = {10.33232/001c.146158},
archivePrefix = {arXiv},
       eprint = {2502.17674},
 primaryClass = {astro-ph.CO},
       adsurl = {https://ui.adsabs.harvard.edu/abs/2025OJAp....846158A}
}

@ARTICLE{astrometry.net,
       author = {{Lang}, Dustin and {Hogg}, David W. and {Mierle}, Keir and {Blanton}, Michael and {Roweis}, Sam},
        title = "{Astrometry.net: Blind Astrometric Calibration of Arbitrary Astronomical Images}",
      journal = {\aj},
         year = 2010,
        month = may,
       volume = {139},
       number = {5},
        pages = {1782-1800},
          doi = {10.1088/0004-6256/139/5/1782},
archivePrefix = {arXiv},
       eprint = {0910.2233},
 primaryClass = {astro-ph.IM},
       adsurl = {https://ui.adsabs.harvard.edu/abs/2010AJ....139.1782L}
}

@ARTICLE{Jarvis:2002vs,
       author = {{Jarvis}, M. and {Bernstein}, G.~M. and {Fischer}, P. and {Smith}, D. and {Jain}, B. and {Tyson}, J.~A. and {Wittman}, D.},
        title = "{Weak-Lensing Results from the 75 Square Degree Cerro Tololo Inter-American Observatory Survey}",
      journal = {\aj},
         year = 2003,
        month = mar,
       volume = {125},
       number = {3},
        pages = {1014-1032},
          doi = {10.1086/367799},
archivePrefix = {arXiv},
       eprint = {astro-ph/0210604},
 primaryClass = {astro-ph},
       adsurl = {https://ui.adsabs.harvard.edu/abs/2003AJ....125.1014J}
}

@ARTICLE{DES:2015yfn,
       author = {{Jarvis}, M. and {Sheldon}, E. and {Zuntz}, J. and {Kacprzak}, T. and {Bridle}, S.~L. and {Amara}, A. and {Armstrong}, R. and {Becker}, M.~R. and {Bernstein}, G.~M. and {Bonnett}, C. and {Chang}, C. and {Das}, R. and {Dietrich}, J.~P. and {Drlica-Wagner}, A. and {Eifler}, T.~F. and {Gangkofner}, C. and {Gruen}, D. and {Hirsch}, M. and {Huff}, E.~M. and {Jain}, B. and {Kent}, S. and {Kirk}, D. and {MacCrann}, N. and {Melchior}, P. and {Plazas}, A.~A. and {Refregier}, A. and {Rowe}, B. and {Rykoff}, E.~S. and {Samuroff}, S. and {S{\'a}nchez}, C. and {Suchyta}, E. and {Troxel}, M.~A. and {Vikram}, V. and {Abbott}, T. and {Abdalla}, F.~B. and {Allam}, S. and {Annis}, J. and {Benoit-L{\'e}vy}, A. and {Bertin}, E. and {Brooks}, D. and {Buckley-Geer}, E. and {Burke}, D.~L. and {Capozzi}, D. and {Carnero Rosell}, A. and {Carrasco Kind}, M. and {Carretero}, J. and {Castander}, F.~J. and {Clampitt}, J. and {Crocce}, M. and {Cunha}, C.~E. and {D'Andrea}, C.~B. and {da Costa}, L.~N. and {DePoy}, D.~L. and {Desai}, S. and {Diehl}, H.~T. and {Doel}, P. and {Fausti Neto}, A. and {Flaugher}, B. and {Fosalba}, P. and {Frieman}, J. and {Gaztanaga}, E. and {Gerdes}, D.~W. and {Gruendl}, R.~A. and {Gutierrez}, G. and {Honscheid}, K. and {James}, D.~J. and {Kuehn}, K. and {Kuropatkin}, N. and {Lahav}, O. and {Li}, T.~S. and {Lima}, M. and {March}, M. and {Martini}, P. and {Miquel}, R. and {Mohr}, J.~J. and {Neilsen}, E. and {Nord}, B. and {Ogando}, R. and {Reil}, K. and {Romer}, A.~K. and {Roodman}, A. and {Sako}, M. and {Sanchez}, E. and {Scarpine}, V. and {Schubnell}, M. and {Sevilla-Noarbe}, I. and {Smith}, R.~C. and {Soares-Santos}, M. and {Sobreira}, F. and {Swanson}, M.~E.~C. and {Tarle}, G. and {Thaler}, J. and {Thomas}, D. and {Walker}, A.~R. and {Wechsler}, R.~H.},
        title = "{The DES Science Verification weak lensing shear catalogues}",
      journal = {\mnras},
         year = 2016,
        month = aug,
       volume = {460},
       number = {2},
        pages = {2245-2281},
          doi = {10.1093/mnras/stw990},
archivePrefix = {arXiv},
       eprint = {1507.05603},
 primaryClass = {astro-ph.IM},
       adsurl = {https://ui.adsabs.harvard.edu/abs/2016MNRAS.460.2245J}
}

@ARTICLE{Heymans:2005rv,
       author = {{Heymans}, Catherine and {Van Waerbeke}, Ludovic and {Bacon}, David and {Berge}, Joel and {Bernstein}, Gary and {Bertin}, Emmanuel and {Bridle}, Sarah and {Brown}, Michael L. and {Clowe}, Douglas and {Dahle}, H{\r{a}}kon and {Erben}, Thomas and {Gray}, Meghan and {Hetterscheidt}, Marco and {Hoekstra}, Henk and {Hudelot}, Patrick and {Jarvis}, Mike and {Kuijken}, Konrad and {Margoniner}, Vera and {Massey}, Richard and {Mellier}, Yannick and {Nakajima}, Reiko and {Refregier}, Alexandre and {Rhodes}, Jason and {Schrabback}, Tim and {Wittman}, David},
        title = "{The Shear Testing Programme - I. Weak lensing analysis of simulated ground-based observations}",
      journal = {\mnras},
         year = 2006,
        month = may,
       volume = {368},
       number = {3},
        pages = {1323-1339},
          doi = {10.1111/j.1365-2966.2006.10198.x},
archivePrefix = {arXiv},
       eprint = {astro-ph/0506112},
 primaryClass = {astro-ph},
       adsurl = {https://ui.adsabs.harvard.edu/abs/2006MNRAS.368.1323H}
}

@ARTICLE{DES:2017ibv,
       author = {{Zuntz}, J. and {Sheldon}, E. and {Samuroff}, S. and {Troxel}, M.~A. and {Jarvis}, M. and {MacCrann}, N. and {Gruen}, D. and {Prat}, J. and {S{\'a}nchez}, C. and {Choi}, A. and {Bridle}, S.~L. and {Bernstein}, G.~M. and {Dodelson}, S. and {Drlica-Wagner}, A. and {Fang}, Y. and {Gruendl}, R.~A. and {Hoyle}, B. and {Huff}, E.~M. and {Jain}, B. and {Kirk}, D. and {Kacprzak}, T. and {Krawiec}, C. and {Plazas}, A.~A. and {Rollins}, R.~P. and {Rykoff}, E.~S. and {Sevilla-Noarbe}, I. and {Soergel}, B. and {Varga}, T.~N. and {Abbott}, T.~M.~C. and {Abdalla}, F.~B. and {Allam}, S. and {Annis}, J. and {Bechtol}, K. and {Benoit-L{\'e}vy}, A. and {Bertin}, E. and {Buckley-Geer}, E. and {Burke}, D.~L. and {Carnero Rosell}, A. and {Carrasco Kind}, M. and {Carretero}, J. and {Castander}, F.~J. and {Crocce}, M. and {Cunha}, C.~E. and {D'Andrea}, C.~B. and {da Costa}, L.~N. and {Davis}, C. and {Desai}, S. and {Diehl}, H.~T. and {Dietrich}, J.~P. and {Doel}, P. and {Eifler}, T.~F. and {Estrada}, J. and {Evrard}, A.~E. and {Fausti Neto}, A. and {Fernandez}, E. and {Flaugher}, B. and {Fosalba}, P. and {Frieman}, J. and {Garc{\'\i}a-Bellido}, J. and {Gaztanaga}, E. and {Gerdes}, D.~W. and {Giannantonio}, T. and {Gschwend}, J. and {Gutierrez}, G. and {Hartley}, W.~G. and {Honscheid}, K. and {James}, D.~J. and {Jeltema}, T. and {Johnson}, M.~W.~G. and {Johnson}, M.~D. and {Kuehn}, K. and {Kuhlmann}, S. and {Kuropatkin}, N. and {Lahav}, O. and {Li}, T.~S. and {Lima}, M. and {Maia}, M.~A.~G. and {March}, M. and {Martini}, P. and {Melchior}, P. and {Menanteau}, F. and {Miller}, C.~J. and {Miquel}, R. and {Mohr}, J.~J. and {Neilsen}, E. and {Nichol}, R.~C. and {Ogando}, R.~L.~C. and {Roe}, N. and {Romer}, A.~K. and {Roodman}, A. and {Sanchez}, E. and {Scarpine}, V. and {Schindler}, R. and {Schubnell}, M. and {Smith}, M. and {Smith}, R.~C. and {Soares-Santos}, M. and {Sobreira}, F. and {Suchyta}, E. and {Swanson}, M.~E.~C. and {Tarle}, G. and {Thomas}, D. and {Tucker}, D.~L. and {Vikram}, V. and {Walker}, A.~R. and {Wechsler}, R.~H. and {Zhang}, Y. and {DES Collaboration}},
        title = "{Dark Energy Survey Year 1 results: weak lensing shape catalogues}",
      journal = {\mnras},
         year = 2018,
        month = nov,
       volume = {481},
       number = {1},
        pages = {1149-1182},
          doi = {10.1093/mnras/sty2219},
archivePrefix = {arXiv},
       eprint = {1708.01533},
 primaryClass = {astro-ph.CO},
       adsurl = {https://ui.adsabs.harvard.edu/abs/2018MNRAS.481.1149Z}
}

@ARTICLE{rowe2010,
       author = {{Rowe}, Barnaby},
        title = "{Improving PSF modelling for weak gravitational lensing using new methods in model selection}",
      journal = {\mnras},
         year = 2010,
        month = may,
       volume = {404},
       number = {1},
        pages = {350-366},
          doi = {10.1111/j.1365-2966.2010.16277.x},
archivePrefix = {arXiv},
       eprint = {0904.3056},
 primaryClass = {astro-ph.CO},
       adsurl = {https://ui.adsabs.harvard.edu/abs/2010MNRAS.404..350R}
}

@ARTICLE{Jarvis:2003wq,
       author = {{Jarvis}, M. and {Bernstein}, G. and {Jain}, B.},
        title = "{The skewness of the aperture mass statistic}",
      journal = {\mnras},
         year = 2004,
        month = jul,
       volume = {352},
       number = {1},
        pages = {338-352},
          doi = {10.1111/j.1365-2966.2004.07926.x},
archivePrefix = {arXiv},
       eprint = {astro-ph/0307393},
 primaryClass = {astro-ph},
       adsurl = {https://ui.adsabs.harvard.edu/abs/2004MNRAS.352..338J}
}

@INPROCEEDINGS{psfex,
       author = {{Bertin}, E.},
        title = "{Automated Morphometry with SExtractor and PSFEx}",
    booktitle = {Astronomical Data Analysis Software and Systems XX},
         year = 2011,
       editor = {{Evans}, I.~N. and {Accomazzi}, A. and {Mink}, D.~J. and {Rots}, A.~H.},
       series = {Astronomical Society of the Pacific Conference Series},
       volume = {442},
        month = jul,
        pages = {435},
       adsurl = {https://ui.adsabs.harvard.edu/abs/2011ASPC..442..435B}
}

@ARTICLE{Laigle:2016jxn,
       author = {{Laigle}, C. and {McCracken}, H.~J. and {Ilbert}, O. and {Hsieh}, B.~C. and {Davidzon}, I. and {Capak}, P. and {Hasinger}, G. and {Silverman}, J.~D. and {Pichon}, C. and {Coupon}, J. and {Aussel}, H. and {Le Borgne}, D. and {Caputi}, K. and {Cassata}, P. and {Chang}, Y.-Y. and {Civano}, F. and {Dunlop}, J. and {Fynbo}, J. and {Kartaltepe}, J.~S. and {Koekemoer}, A. and {Le F{\`e}vre}, O. and {Le Floc'h}, E. and {Leauthaud}, A. and {Lilly}, S. and {Lin}, L. and {Marchesi}, S. and {Milvang-Jensen}, B. and {Salvato}, M. and {Sanders}, D.~B. and {Scoville}, N. and {Smolcic}, V. and {Stockmann}, M. and {Taniguchi}, Y. and {Tasca}, L. and {Toft}, S. and {Vaccari}, Mattia and {Zabl}, J.},
        title = "{The COSMOS2015 Catalog: Exploring the 1 < z < 6 Universe with Half a Million Galaxies}",
      journal = {\apjs},
         year = 2016,
        month = jun,
       volume = {224},
       number = {2},
          eid = {24},
        pages = {24},
          doi = {10.3847/0067-0049/224/2/24},
archivePrefix = {arXiv},
       eprint = {1604.02350},
 primaryClass = {astro-ph.GA},
       adsurl = {https://ui.adsabs.harvard.edu/abs/2016ApJS..224...24L}
}

@ARTICLE{sawicki2019cfht,
       author = {{Sawicki}, Marcin and {Arnouts}, Stephane and {Huang}, Jiasheng and {Coupon}, Jean and {Golob}, Anneya and {Gwyn}, Stephen and {Foucaud}, Sebastien and {Moutard}, Thibaud and {Iwata}, Ikuru and {Liu}, Chengze and {Chen}, Lingjian and {Desprez}, Guillaume and {Harikane}, Yuichi and {Ono}, Yoshiaki and {Strauss}, Michael A. and {Tanaka}, Masayuki and {Thibert}, Nathalie and {Balogh}, Michael and {Bundy}, Kevin and {Chapman}, Scott and {Gunn}, James E. and {Hsieh}, Bau-Ching and {Ilbert}, Olivier and {Jing}, Yipeng and {LeF{\`e}vre}, Olivier and {Li}, Cheng and {Matsuda}, Yuichi and {Miyazaki}, Satoshi and {Nagao}, Tohru and {Nishizawa}, Atsushi J. and {Ouchi}, Masami and {Shimasaku}, Kazuhiro and {Silverman}, John and {de la Torre}, Sylvain and {Tresse}, Laurence and {Wang}, Wei-Hao and {Willott}, Chris J. and {Yamada}, Toru and {Yang}, Xiaohu and {Yee}, Howard K.~C.},
        title = "{The CFHT large area U-band deep survey (CLAUDS)}",
      journal = {\mnras},
         year = 2019,
        month = nov,
       volume = {489},
       number = {4},
        pages = {5202-5217},
          doi = {10.1093/mnras/stz2522},
archivePrefix = {arXiv},
       eprint = {1909.05898},
 primaryClass = {astro-ph.GA},
       adsurl = {https://ui.adsabs.harvard.edu/abs/2019MNRAS.489.5202S}
}

@ARTICLE{2020A&A...633L...7L,
       author = {{Longobardi}, A. and {Boselli}, A. and {Boissier}, S. and {Bianchi}, S. and {Andreani}, P. and {Sarpa}, E. and {Nanni}, A. and {Miville-Desch{\^e}nes}, M.},
        title = "{The GALEX Ultraviolet Virgo Cluster Survey (GUViCS). VIII. Diffuse dust in the Virgo intra-cluster space}",
      journal = {\aap},
         year = 2020,
        month = jan,
       volume = {633},
          eid = {L7},
        pages = {L7},
          doi = {10.1051/0004-6361/201937024},
archivePrefix = {arXiv},
       eprint = {2001.05512},
 primaryClass = {astro-ph.GA},
       adsurl = {https://ui.adsabs.harvard.edu/abs/2020A&A...633L...7L}
}

@ARTICLE{H12,
       author = {{Heymans}, Catherine and {Van Waerbeke}, Ludovic and {Miller}, Lance and {Erben}, Thomas and {Hildebrandt}, Hendrik and {Hoekstra}, Henk and {Kitching}, Thomas D. and {Mellier}, Yannick and {Simon}, Patrick and {Bonnett}, Christopher and {Coupon}, Jean and {Fu}, Liping and {Harnois D{\'e}raps}, Joachim and {Hudson}, Michael J. and {Kilbinger}, Martin and {Kuijken}, Koenraad and {Rowe}, Barnaby and {Schrabback}, Tim and {Semboloni}, Elisabetta and {van Uitert}, Edo and {Vafaei}, Sanaz and {Velander}, Malin},
        title = "{CFHTLenS: the Canada-France-Hawaii Telescope Lensing Survey}",
      journal = {\mnras},
         year = 2012,
        month = nov,
       volume = {427},
       number = {1},
        pages = {146-166},
          doi = {10.1111/j.1365-2966.2012.21952.x},
archivePrefix = {arXiv},
       eprint = {1210.0032},
 primaryClass = {astro-ph.CO},
       adsurl = {https://ui.adsabs.harvard.edu/abs/2012MNRAS.427..146H}
}

@ARTICLE{DES:2025zvc,
       author = {{Yamamoto}, M. and {Becker}, M.~R. and {Sheldon}, E. and {Jarvis}, M. and {Gruendl}, R.~A. and {Menanteau}, F. and {Rykoff}, E.~S. and {Mau}, S. and {Schutt}, T. and {Gatti}, M. and {Troxel}, M.~A. and {Amon}, A. and {Anbajagane}, D. and {Bernstein}, G.~M. and {Gruen}, D. and {Huff}, E.~M. and {Tabbutt}, M. and {Tong}, A. and {Yanny}, B. and {Abbott}, T.~M.~C. and {Aguena}, M. and {Alarcon}, A. and {Andrade-Oliveira}, F. and {Bechtol}, K. and {Blazek}, J. and {Brooks}, D. and {Rosell}, A. Carnero and {Carretero}, J. and {Chang}, C. and {Choi}, A. and {Costanzi}, M. and {Crocce}, M. and {da Costa}, L.~N. and {Davis}, T.~M. and {De Vicente}, J. and {Desai}, S. and {Diehl}, H.~T. and {Dodelson}, S. and {Doel}, P. and {Doux}, C. and {Drlica-Wagner}, A. and {Fert{\'e}}, A. and {Flaugher}, B. and {Fosalba}, P. and {Frieman}, J. and {Garc{\'\i}a-Bellido}, J. and {Gaztanaga}, E. and {Giannini}, G. and {Gutierrez}, G. and {Hartley}, W.~G. and {Herner}, K. and {Hinton}, S.~R. and {Hollowood}, D.~L. and {Honscheid}, K. and {Huterer}, D. and {Krause}, E. and {Kuehn}, K. and {Lahav}, O. and {Lima}, M. and {Marshall}, J.~L. and {Mena-Fern{\'a}ndez}, J. and {Miquel}, R. and {Mohr}, J.~J. and {Muir}, J. and {Myles}, J. and {Ogando}, R.~L.~C. and {Pieres}, A. and {Malag{\'o}n}, A.~A. Plazas and {Porredon}, A. and {Prat}, J. and {Raveri}, M. and {Rodriguez-Monroy}, M. and {Roodman}, A. and {Samuroff}, S. and {Sanchez}, E. and {Sanchez Cid}, D. and {Scarpine}, V. and {Sevilla-Noarbe}, I. and {Smith}, M. and {Soares-Santos}, M. and {Suchyta}, E. and {Tarle}, G. and {Vikram}, V. and {Weaverdyck}, N. and {Wiseman}, P. and {Zhang}, Y.},
        title = "{Dark Energy Survey Year 6 results: cell-based coadds and METADETECTION weak lensing shape catalogue}",
      journal = {\mnras},
         year = 2025,
        month = nov,
       volume = {543},
       number = {4},
        pages = {4156-4186},
          doi = {10.1093/mnras/staf1661},
archivePrefix = {arXiv},
       eprint = {2501.05665},
 primaryClass = {astro-ph.CO},
       adsurl = {https://ui.adsabs.harvard.edu/abs/2025MNRAS.543.4156Y}
}

@INPROCEEDINGS{ds9,
       author = {{Joye}, W.~A. and {Mandel}, E.},
        title = "{New Features of SAOImage DS9}",
    booktitle = {Astronomical Data Analysis Software and Systems XII},
         year = 2003,
       editor = {{Payne}, H.~E. and {Jedrzejewski}, R.~I. and {Hook}, R.~N.},
       series = {Astronomical Society of the Pacific Conference Series},
       volume = {295},
        month = jan,
        pages = {489},
       adsurl = {https://ui.adsabs.harvard.edu/abs/2003ASPC..295..489J}
}

@ARTICLE{Gill2024,
       author = {{Gill}, Ajay S. and {Benton}, Steven J. and {Damaren}, Christopher J. and {Everett}, Spencer W. and {Fraisse}, Aurelien A. and {Hartley}, John W. and {Harvey}, David and {Holder}, Bradley and {Huff}, Eric M. and {Jauzac}, Mathilde and {Jones}, William C. and {Lagattuta}, David and {Leung}, Jason S.-Y. and {Li}, Lun and {Luu}, Thuy Vy T. and {Massey}, Richard and {McCleary}, Jacqueline E. and {Nagy}, Johanna M. and {Barth Netterfield}, C. and {Paracha}, Emaad and {Redmond}, Susan F. and {Rhodes}, Jason D. and {Robertson}, Andrew and {Romualdez}, L. Javier and {Schmoll}, J{\"u}rgen and {Shaaban}, Mohamed M. and {Sirks}, Ellen L. and {Vassilakis}, Georgios N. and {Vitorelliand}, Andr{\'e} Z.},
        title = "{SuperBIT Superpressure Flight Instrument Overview and Performance: Near diffraction-limited Astronomical Imaging from the Stratosphere}",
      journal = {arXiv e-prints},
         year = 2024,
        month = aug,
          eid = {arXiv:2408.01847},
        pages = {arXiv:2408.01847},
          doi = {10.48550/arXiv.2408.01847},
archivePrefix = {arXiv},
       eprint = {2408.01847},
 primaryClass = {astro-ph.IM},
       adsurl = {https://ui.adsabs.harvard.edu/abs/2024arXiv240801847G}
}

@ARTICLE{Sirks2020,
       author = {{Sirks}, E.~L. and {Clark}, P. and {Massey}, R.~J. and {Benton}, S.~J. and {Brown}, A.~M. and {Damaren}, C.~J. and {Eifler}, T. and {Fraisse}, A.~A. and {Frenk}, C. and {Funk}, M. and {Galloway}, M.~N. and {Gill}, A. and {Hartley}, J.~W. and {Holder}, B. and {Huff}, E.~M. and {Jauzac}, M. and {Jones}, W.~C. and {Lagattuta}, D. and {Leung}, J.~S.-Y. and {Li}, L. and {Luu}, T.~V.~T. and {McCleary}, J. and {Nagy}, J.~M. and {Netterfield}, C.~B. and {Redmond}, S. and {Rhodes}, J.~D. and {Romualdez}, L.~J. and {Schmoll}, J. and {Shaaban}, M.~M. and {Tam}, S.-I.},
        title = "{Download by parachute: retrieval of assets from high altitude balloons}",
      journal = {Journal of Instrumentation},
         year = 2020,
        month = may,
       volume = {15},
       number = {5},
        pages = {P05014},
          doi = {10.1088/1748-0221/15/05/P05014},
archivePrefix = {arXiv},
       eprint = {2004.10764},
 primaryClass = {astro-ph.IM},
       adsurl = {https://ui.adsabs.harvard.edu/abs/2020JInst..15P5014S}
}

@ARTICLE{Romualdez2020,
       author = {{Romualdez}, L. Javier and {Benton}, Steven J. and {Brown}, Anthony M. and {Clark}, Paul and {Damaren}, Christopher J. and {Eifler}, Tim and {Fraisse}, Aurelien A. and {Galloway}, Mathew N. and {Gill}, Ajay and {Hartley}, John W. and {Holder}, Bradley and {Huff}, Eric M. and {Jauzac}, Mathilde and {Jones}, William C. and {Lagattuta}, David and {Leung}, Jason S.-Y. and {Li}, Lun and {Luu}, Thuy Vy T. and {Massey}, Richard J. and {McCleary}, Jacqueline and {Mullaney}, James and {Nagy}, Johanna M. and {Netterfield}, C. Barth and {Redmond}, Susan and {Rhodes}, Jason D. and {Schmoll}, J{\"u}rgen and {Shaaban}, Mohamed M. and {Sirks}, Ellen and {Tam}, Sut-Ieng},
        title = "{Robust diffraction-limited near-infrared-to-near-ultraviolet wide-field imaging from stratospheric balloon-borne platforms{\textemdash}Super-pressure Balloon-borne Imaging Telescope performance}",
      journal = {Review of Scientific Instruments},
         year = 2020,
        month = mar,
       volume = {91},
       number = {3},
          eid = {034501},
        pages = {034501},
          doi = {10.1063/1.5139711},
archivePrefix = {arXiv},
       eprint = {1911.11210},
 primaryClass = {astro-ph.IM},
       adsurl = {https://ui.adsabs.harvard.edu/abs/2020RScI...91c4501R}
}

@ARTICLE{2015Sci...347.1462H,
       author = {{Harvey}, David and {Massey}, Richard and {Kitching}, Thomas and {Taylor}, Andy and {Tittley}, Eric},
        title = "{The nongravitational interactions of dark matter in colliding galaxy clusters}",
      journal = {Science},
         year = 2015,
        month = mar,
       volume = {347},
       number = {6229},
        pages = {1462-1465},
          doi = {10.1126/science.1261381},
archivePrefix = {arXiv},
       eprint = {1503.07675},
 primaryClass = {astro-ph.CO},
       adsurl = {https://ui.adsabs.harvard.edu/abs/2015Sci...347.1462H}
}

@ARTICLE{2018ApJ...869..104W,
       author = {{Wittman}, David and {Golovich}, Nathan and {Dawson}, William A.},
        title = "{The Mismeasure of Mergers: Revised Limits on Self-interacting Dark Matter in Merging Galaxy Clusters}",
      journal = {\apj},
         year = 2018,
        month = dec,
       volume = {869},
       number = {2},
          eid = {104},
        pages = {104},
          doi = {10.3847/1538-4357/aaee77},
archivePrefix = {arXiv},
       eprint = {1701.05877},
 primaryClass = {astro-ph.CO},
       adsurl = {https://ui.adsabs.harvard.edu/abs/2018ApJ...869..104W}
}

@ARTICLE{2017MNRAS.464.3991H,
       author = {{Harvey}, David and {Robertson}, Andrew and {Massey}, Richard and {Kneib}, Jean-Paul},
        title = "{Looking for dark matter trails in colliding galaxy clusters}",
      journal = {\mnras},
         year = 2017,
        month = feb,
       volume = {464},
       number = {4},
        pages = {3991-3997},
          doi = {10.1093/mnras/stw2671},
archivePrefix = {arXiv},
       eprint = {1610.05327},
 primaryClass = {astro-ph.CO},
       adsurl = {https://ui.adsabs.harvard.edu/abs/2017MNRAS.464.3991H}
}

@ARTICLE{2024ApJ...974...69F,
       author = {{Fu}, Shenming and {Dell'Antonio}, Ian and {Escalante}, Zacharias and {Nelson}, Jessica and {Englert}, Anthony and {Helhoski}, S{\o}ren and {Shinde}, Rahul and {Brockland}, Julia and {LaDuca}, Philip and {Larkin}, Christelyn and {Paris}, Lucca and {Weiner}, Shane and {Black}, William K. and {Chary}, Ranga-Ram and {Clowe}, Douglas and {Cooper}, M.~C. and {Donahue}, Megan and {Evrard}, August and {Lacy}, Mark and {Lauer}, Tod and {Liu}, Binyang and {McCleary}, Jacqueline and {Meneghetti}, Massimo and {Miyatake}, Hironao and {Montes}, Mireia and {Natarajan}, Priyamvada and {Ntampaka}, Michelle and {Pierpaoli}, Elena and {Postman}, Marc and {Sohn}, Jubee and {Turner}, David and {Umetsu}, Keiichi and {Utsumi}, Yousuke and {Wilson}, Gillian},
        title = "{LoVoCCS. II. Weak Lensing Mass Distributions, Red-sequence Galaxy Distributions, and Their Alignment with the Brightest Cluster Galaxy in 58 Nearby X-Ray-luminous Galaxy Clusters}",
      journal = {\apj},
         year = 2024,
        month = oct,
       volume = {974},
       number = {1},
          eid = {69},
        pages = {69},
          doi = {10.3847/1538-4357/ad67c6},
archivePrefix = {arXiv},
       eprint = {2402.10337},
 primaryClass = {astro-ph.CO},
       adsurl = {https://ui.adsabs.harvard.edu/abs/2024ApJ...974...69F}
}

@ARTICLE{2025arXiv250314745D,
       author = {{DESI Collaboration} and {Abdul-Karim}, M. and {Adame}, A.~G. and {Aguado}, D. and {Aguilar}, J. and {Ahlen}, S. and {Alam}, S. and {Aldering}, G. and {Alexander}, D.~M. and {Alfarsy}, R. and {Allen}, L. and {Allende Prieto}, C. and {Alves}, O. and {Anand}, A. and {Andrade}, U. and {Armengaud}, E. and {Avila}, S. and {Aviles}, A. and {Awan}, H. and {Bailey}, S. and {Baleato Lizancos}, A. and {Ballester}, O. and {Bault}, A. and {Bautista}, J. and {BenZvi}, S. and {Beraldo e Silva}, L. and {Bermejo-Climent}, J.~R. and {Beutler}, F. and {Bianchi}, D. and {Blake}, C. and {Blum}, R. and {Bolton}, A.~S. and {Bonici}, M. and {Brieden}, S. and {Brodzeller}, A. and {Brooks}, D. and {Buckley-Geer}, E. and {Burtin}, E. and {Canning}, R. and {Carnero Rosell}, A. and {Carr}, A. and {Carrilho}, P. and {Casas}, L. and {Castander}, F.~J. and {Cereskaite}, R. and {Cervantes-Cota}, J.~L. and {Chaussidon}, E. and {Chaves-Montero}, J. and {Chen}, S. and {Chen}, X. and {Claybaugh}, T. and {Cole}, S. and {Cooper}, A.~P. and {Cousinou}, M.-C. and {Cuceu}, A. and {Davis}, T.~M. and {Dawson}, K.~S. and {de Belsunce}, R. and {de la Cruz}, R. and {de la Macorra}, A. and {de Mattia}, A. and {Deiosso}, N. and {Della Costa}, J. and {Demina}, R. and {Demirbozan}, U. and {DeRose}, J. and {Dey}, A. and {Dey}, B. and {Ding}, J. and {Ding}, Z. and {Doel}, P. and {Douglass}, K. and {Dowicz}, M. and {Ebina}, H. and {Edelstein}, J. and {Eisenstein}, D.~J. and {Elbers}, W. and {Emas}, N. and {Escoffier}, S. and {Fagrelius}, P. and {Fan}, X. and {Fanning}, K. and {Fawcett}, V.~A. and {Fern\textbackslash'andez-Garc\textbackslash'ia}, E. and {Ferraro}, S. and {Findlay}, N. and {Font-Ribera}, A. and {Forero-Romero}, J.~E. and {Forero-S\textbackslash'anchez}, D. and {Frenk}, C.~S. and {G\textbackslash''ansicke}, B.~T. and {Galbany}, L. and {Garc\textbackslash'ia-Bellido}, J. and {Garcia-Quintero}, C. and {Garrison}, L.~H. and {Gazta\~{n}aga}, E. and {Gil-Mar\textbackslash'in}, H. and {Gnedin}, O.~Y. and {Gontcho}, S. Gontcho A and {Gonzalez-Morales}, A.~X. and {Gonzalez-Perez}, V. and {Gordon}, C. and {Graur}, O. and {Green}, D. and {Gruen}, D. and {Gsponer}, R. and {Guandalin}, C. and {Gutierrez}, G. and {Guy}, J. and {Hahn}, C. and {Han}, J.~J. and {Han}, J. and {He}, S. and {Herrera-Alcantar}, H.~K. and {Honscheid}, K. and {Hou}, J. and {Howlett}, C. and {Huterer}, D. and {Ir\'{s}i\'{c}}, V. and {Ishak}, M. and {Jacques}, A. and {Jimenez}, J. and {Jing}, Y.~P. and {Joachimi}, B. and {Joudaki}, S. and {Joyce}, R. and {Jullo}, E. and {Juneau}, S. and {Kara\'{c}ayl\'{i}}, N.~G. and {Karim}, T. and {Kehoe}, R. and {Kent}, S. and {Khederlarian}, A. and {Kirkby}, D. and {Kisner}, T. and {Kitaura}, F.-S. and {Kizhuprakkat}, N. and {Kong}, H. and {Koposov}, S.~E. and {Kremin}, A. and {Krolewski}, A. and {Lahav}, O. and {Lai}, Y. and {Lamman}, C. and {Lan}, T.-W. and {Landriau}, M. and {Lang}, D. and {Lange}, J.~U. and {Lasker}, J. and {Le Goff}, J.~M. and {Le Guillou}, L. and {Leauthaud}, A. and {Levi}, M.~E. and {Li}, S. and {Li}, T.~S. and {Lodha}, K. and {Lokken}, M. and {Luo}, Y. and {Magneville}, C. and {Manera}, M. and {Manser}, C.~J. and {Margala}, D. and {Martini}, P. and {Maus}, M. and {McCullough}, J. and {McDonald}, P. and {Medina}, G.~E. and {Medina-Varela}, L. and {Meisner}, A. and {Mena-Fern\textbackslash'andez}, J. and {Menegas}, A. and {Mezcua}, M. and {Miquel}, R. and {Montero-Camacho}, P. and {Moon}, J. and {Moustakas}, J. and {Mu\~{n}oz-Guti\textbackslash'errez}, A. and {Mu\~{n}oz-Santos}, D. and {Myers}, A.~D. and {Myles}, J. and {Nadathur}, S. and {Najita}, J. and {Napolitano}, L. and {Newman}, J.~A. and {Nikakhtar}, F. and {Nikutta}, R. and {Niz}, G. and {Noriega}, H.~E. and {Padmanabhan}, N. and {Paillas}, E. and {Palanque-Delabrouille}, N. and {Palmese}, A. and {Pan}, J. and {Pan}, Z. and {Parkinson}, D. and {Peacock}, J. and {Percival}, W.~J. and {P\textbackslash'erez-Fern\textbackslash'andez}, A. and {P\textbackslash'erez-R\textbackslash`afols}, I. and {Peterson}, P.},
        title = "{Data Release 1 of the Dark Energy Spectroscopic Instrument}",
      journal = {arXiv e-prints},
         year = 2025,
        month = mar,
          eid = {arXiv:2503.14745},
        pages = {arXiv:2503.14745},
          doi = {10.48550/arXiv.2503.14745},
archivePrefix = {arXiv},
       eprint = {2503.14745},
 primaryClass = {astro-ph.CO},
       adsurl = {https://ui.adsabs.harvard.edu/abs/2025arXiv250314745D}
}

@ARTICLE{2022PhRvD.105h3528P,
       author = {{Prat}, J. and {Blazek}, J. and {S{\'a}nchez}, C. and {Tutusaus}, I. and {Pandey}, S. and {Elvin-Poole}, J. and {Krause}, E. and {Troxel}, M.~A. and {Secco}, L.~F. and {Amon}, A. and {DeRose}, J. and {Zacharegkas}, G. and {Chang}, C. and {Jain}, B. and {MacCrann}, N. and {Park}, Y. and {Sheldon}, E. and {Giannini}, G. and {Bocquet}, S. and {To}, C. and {Alarcon}, A. and {Alves}, O. and {Andrade-Oliveira}, F. and {Baxter}, E. and {Bechtol}, K. and {Becker}, M.~R. and {Bernstein}, G.~M. and {Camacho}, H. and {Campos}, A. and {Carnero Rosell}, A. and {Carrasco Kind}, M. and {Cawthon}, R. and {Chen}, R. and {Choi}, A. and {Cordero}, J. and {Crocce}, M. and {Davis}, C. and {De Vicente}, J. and {Diehl}, H.~T. and {Dodelson}, S. and {Doux}, C. and {Drlica-Wagner}, A. and {Eckert}, K. and {Eifler}, T.~F. and {Elsner}, F. and {Everett}, S. and {Fang}, X. and {Farahi}, A. and {Fert{\'e}}, A. and {Fosalba}, P. and {Friedrich}, O. and {Gatti}, M. and {Gruen}, D. and {Gruendl}, R.~A. and {Harrison}, I. and {Hartley}, W.~G. and {Herner}, K. and {Huang}, H. and {Huff}, E.~M. and {Huterer}, D. and {Jarvis}, M. and {Kuropatkin}, N. and {Leget}, P.-F. and {Lemos}, P. and {Liddle}, A.~R. and {McCullough}, J. and {Muir}, J. and {Myles}, J. and {Navarro-Alsina}, A. and {Porredon}, A. and {Raveri}, M. and {Rodriguez-Monroy}, M. and {Rollins}, R.~P. and {Roodman}, A. and {Rosenfeld}, R. and {Ross}, A.~J. and {Rykoff}, E.~S. and {Sanchez}, J. and {Sevilla-Noarbe}, I. and {Shin}, T. and {Troja}, A. and {Varga}, T.~N. and {Weaverdyck}, N. and {Wechsler}, R.~H. and {Yanny}, B. and {Yin}, B. and {Zuntz}, J. and {Abbott}, T.~M.~C. and {Aguena}, M. and {Allam}, S. and {Annis}, J. and {Bacon}, D. and {Brooks}, D. and {Burke}, D.~L. and {Carretero}, J. and {Conselice}, C. and {Costanzi}, M. and {da Costa}, L.~N. and {Pereira}, M.~E.~S. and {Desai}, S. and {Dietrich}, J.~P. and {Doel}, P. and {Evrard}, A.~E. and {Ferrero}, I. and {Flaugher}, B. and {Frieman}, J. and {Garc{\'\i}a-Bellido}, J. and {Gaztanaga}, E. and {Gerdes}, D.~W. and {Giannantonio}, T. and {Gschwend}, J. and {Gutierrez}, G. and {Hinton}, S.~R. and {Hollowood}, D.~L. and {Honscheid}, K. and {James}, D.~J. and {Kuehn}, K. and {Lahav}, O. and {Lin}, H. and {Maia}, M.~A.~G. and {Marshall}, J.~L. and {Martini}, P. and {Melchior}, P. and {Menanteau}, F. and {Miller}, C.~J. and {Miquel}, R. and {Mohr}, J.~J. and {Morgan}, R. and {Ogando}, R.~L.~C. and {Palmese}, A. and {Paz-Chinch{\'o}n}, F. and {Petravick}, D. and {Plazas Malag{\'o}n}, A.~A. and {Sanchez}, E. and {Serrano}, S. and {Smith}, M. and {Soares-Santos}, M. and {Suchyta}, E. and {Tarle}, G. and {Thomas}, D. and {Weller}, J. and {DES Collaboration}},
        title = "{Dark energy survey year 3 results: High-precision measurement and modeling of galaxy-galaxy lensing}",
      journal = {\prd},
         year = 2022,
        month = apr,
       volume = {105},
       number = {8},
          eid = {083528},
        pages = {083528},
          doi = {10.1103/PhysRevD.105.083528},
archivePrefix = {arXiv},
       eprint = {2105.13541},
 primaryClass = {astro-ph.CO},
       adsurl = {https://ui.adsabs.harvard.edu/abs/2022PhRvD.105h3528P}
}

@ARTICLE{Voyer2024,
       author = {{Voyer}, Philippe and {Benton}, Steven J. and {Damaren}, Christopher J. and {Everett}, Spencer W. and {Fraisse}, Aurelien A. and {Gill}, Ajay S. and {Hartley}, John W. and {Harvey}, David and {Henderson}, Michael and {Holder}, Bradley and {Huff}, Eric M. and {Jauzac}, Mathilde and {Jones}, William C. and {Lagattuta}, David and {Leung}, Jason S.-Y. and {Li}, Lun and {Luu}, Thuy Vy T. and {Massey}, Richard and {McCleary}, Jacqueline E. and {Nagy}, Johanna M. and {Barth Netterfield}, C. and {Paracha}, Emaad and {Redmond}, Susan F. and {Rhodes}, Jason D. and {Robertson}, Andrew and {Romualdez}, L. Javier and {Schmoll}, J{\"u}rgen and {Shaaban}, Mohamed M. and {Sirks}, Ellen L. and {Vassilakis}, Georgios N. and {Vitorelli}, Andr{\'e} Z.},
        title = "{From SuperBIT to GigaBIT: Informing next-generation balloon-borne telescope design with Fine Guidance System flight data}",
      journal = {Proc. SPIE},
        series = {Ground-Based and Airborne Telescopes X},
       volume = {13094},
          eid = {130944Z},
         year = 2024,
         month = jul,
          doi = {10.1117/12.3017869},
          url = {https://www.spiedigitallibrary.org/conference-proceedings-of-spie/13094/3017869/From-SuperBIT-to-GigaBIT--informing-next-generation-balloon-borne/10.1117/12.3017869.full},
 archivePrefix = {arXiv},
        eprint = {2407.10103},
   primaryClass = {astro-ph.IM},
        adsurl = {https://ui.adsabs.harvard.edu/abs/2024arXiv240710103V}
}

@INPROCEEDINGS{Voyer2026,
       author = {Voyer, Philippe and Amit, Maya and Benton, Steven J. and Boyd, Benjamin E. and Brown, Anthony M. and Cerini, Giulia and Clark, Paul and Craigie, Matthew and Damaren, Christopher J. and Eifler, Tim and Everett, Spencer W. and Fraisse, Aurelien A. and Fung, Leo W. H. and Gill, Ajay S. and Gourapura, Suren and Habjan, Eric and Hartley, John W. and Harvey, David and Holder, Bradley and Huff, Eric M. and Jauzac, Mathilde and Jones, William C. and Lagattuta, David and Leroy, Gavin and Leung, Jason S.-Y. and Li, Lun and Luu, Thuy Vy T. and Massey, Richard and McCleary, Jacqueline E. and Miles, Adyn and Nagy, Johanna M. and Netterfield, C. Barth and Paracha, Emaad and Redmond, Susan F. and Rhodes, Jason D. and Robertson, Andrew and Romualdez, L. Javier and Saha, Sayan and Schmoll, J{\"u}rgen and Shaaban, Mohamed M. and Sirks, Ellen and Tam, Sut Ieng and Tartakovsky, Simon and Vassilakis, Georgios N. and Vitorelli, Andr{\'e} Z. and Warren, Bryce and Zenteno, Alfredo},
        title = {{Multi-sensor fusion for fine-guidance and milliarcsecond-level attitude estimation of balloon-borne telescope}},
    booktitle = {Proc. SPIE},
       volume = {14152},
          eid = {1415229},
         year = {2026},
          doi = {10.1117/12.3103046},
          url = {https://www.spiedigitallibrary.org/conference-proceedings-of-spie/14152/3103046/Multi-sensor-fusion-for-fine-guidance-and-milliarcsecond-level-attitude/10.1117/12.3103046.full},
  archivePrefix = {arXiv},
        eprint = {2607.28816},
   primaryClass = {astro-ph.IM},
         note = {SPIE Astronomical Telescopes + Instrumentation, Copenhagen, Denmark},
        adsurl = {https://ui.adsabs.harvard.edu/abs/2026arXiv260728816V}
}

@inproceedings{Paracha2024,
    title = {Overview of communications and observation strategies in the stratosphere for the {Super}-pressure {Balloon}-borne {Imaging} {Telescope} ({SuperBIT})},
	language = {en},
	booktitle = {{Proc. SPIE}},
	author = {Paracha, Emaad and Benton, Steven J. and Damaren, Christopher J. and Everett, Spencer W. and Fraisse, Aurelien A. and Gill, Ajay and Hartley, John W. and Harvey, David and Holder, Bradley and Huff, Eric M. and Jauzac, Mathilde and Jones, William C. and Lagattuta, David and Leung, Jason S.-Y. and Li, Lun and Luu, Thuy Vy T. and Massey, Richard and McCleary, Jacqueline E. and Nagy, Johanna M. and Netterfield, C. Barth and Redmond, Susan F. and Rhodes, Jason D. and Robertson, Andrew and Romualdez, L. Javier and Schmoll, Jürgen and Shaaban, Mohamed M. and Sirks, Ellen and Tam, Sut Ieng and Vassilakis, Georgios N. and Vitorelli, Andre},
    journal = {Proc. SPIE},
    series = {Ground-based and Airborne Telescopes X},
    volume = {13094},
    eid = {1309414},
    year = 2024,
    doi          = {10.1117/12.3019360},
    event        = {SPIE Astronomical Telescopes + Instrumentation, 2024, Yokohama, Japan}
}

@ARTICLE{gill_2020,
       author = {{Gill}, Ajay and {Benton}, Steven J. and {Brown}, Anthony M. and {Clark}, Paul and {Damaren}, Christopher J. and {Eifler}, Tim and {Fraisse}, Aurelien A. and {Galloway}, Mathew N. and {Hartley}, John W. and {Holder}, Bradley and {Huff}, Eric M. and {Jauzac}, Mathilde and {Jones}, William C. and {Lagattuta}, David and {Leung}, Jason S.-Y. and {Li}, Lun and {Luu}, Thuy Vy T. and {Massey}, Richard J. and {McCleary}, Jacqueline and {Mullaney}, James and {Nagy}, Johanna M. and {Netterfield}, C. Barth and {Redmond}, Susan and {Rhodes}, Jason D. and {Romualdez}, L. Javier and {Schmoll}, J{\"u}rgen and {Shaaban}, Mohamed M. and {Sirks}, Ellen and {Sivanandam}, Suresh and {Tam}, Sut-Ieng},
        title = "{Optical Night Sky Brightness Measurements from the Stratosphere}",
      journal = {\aj},
         year = 2020,
        month = dec,
       volume = {160},
       number = {6},
          eid = {266},
        pages = {266},
          doi = {10.3847/1538-3881/abbffb},
archivePrefix = {arXiv},
       eprint = {2010.05145},
 primaryClass = {astro-ph.IM},
       adsurl = {https://ui.adsabs.harvard.edu/abs/2020AJ....160..266G}
}

@ARTICLE{DES:2020lsz,
       author = {{MacCrann}, N. and {Becker}, M.~R. and {McCullough}, J. and {Amon}, A. and {Gruen}, D. and {Jarvis}, M. and {Choi}, A. and {Troxel}, M.~A. and {Sheldon}, E. and {Yanny}, B. and {Herner}, K. and {Dodelson}, S. and {Zuntz}, J. and {Eckert}, K. and {Rollins}, R.~P. and {Varga}, T.~N. and {Bernstein}, G.~M. and {Gruendl}, R.~A. and {Harrison}, I. and {Hartley}, W.~G. and {Sevilla-Noarbe}, I. and {Pieres}, A. and {Bridle}, S.~L. and {Myles}, J. and {Alarcon}, A. and {Everett}, S. and {S{\'a}nchez}, C. and {Huff}, E.~M. and {Tarsitano}, F. and {Gatti}, M. and {Secco}, L.~F. and {Abbott}, T.~M.~C. and {Aguena}, M. and {Allam}, S. and {Annis}, J. and {Bacon}, D. and {Bertin}, E. and {Brooks}, D. and {Burke}, D.~L. and {Carnero Rosell}, A. and {Carrasco Kind}, M. and {Carretero}, J. and {Costanzi}, M. and {Crocce}, M. and {Pereira}, M.~E.~S. and {De Vicente}, J. and {Desai}, S. and {Diehl}, H.~T. and {Dietrich}, J.~P. and {Doel}, P. and {Eifler}, T.~F. and {Ferrero}, I. and {Fert{\'e}}, A. and {Flaugher}, B. and {Fosalba}, P. and {Frieman}, J. and {Garc{\'\i}a-Bellido}, J. and {Gaztanaga}, E. and {Gerdes}, D.~W. and {Giannantonio}, T. and {Gschwend}, J. and {Gutierrez}, G. and {Hinton}, S.~R. and {Hollowood}, D.~L. and {Honscheid}, K. and {James}, D.~J. and {Lahav}, O. and {Lima}, M. and {Maia}, M.~A.~G. and {March}, M. and {Marshall}, J.~L. and {Martini}, P. and {Melchior}, P. and {Menanteau}, F. and {Miquel}, R. and {Mohr}, J.~J. and {Morgan}, R. and {Muir}, J. and {Ogando}, R.~L.~C. and {Palmese}, A. and {Paz-Chinch{\'o}n}, F. and {Plazas}, A.~A. and {Rodriguez-Monroy}, M. and {Roodman}, A. and {Samuroff}, S. and {Sanchez}, E. and {Scarpine}, V. and {Serrano}, S. and {Smith}, M. and {Soares-Santos}, M. and {Suchyta}, E. and {Swanson}, M.~E.~C. and {Tarle}, G. and {Thomas}, D. and {To}, C. and {Wilkinson}, R.~D. and {Wilkinson}, R.~D. and {DES Collaboration}},
        title = "{Dark Energy Survey Y3 results: blending shear and redshift biases in image simulations}",
      journal = {\mnras},
         year = 2022,
        month = jan,
       volume = {509},
       number = {3},
        pages = {3371-3394},
          doi = {10.1093/mnras/stab2870},
archivePrefix = {arXiv},
       eprint = {2012.08567},
 primaryClass = {astro-ph.CO},
       adsurl = {https://ui.adsabs.harvard.edu/abs/2022MNRAS.509.3371M}
}

@ARTICLE{Lewis:2019xzd,
       author = {{Lewis}, Antony},
        title = "{GetDist: a Python package for analysing Monte Carlo samples}",
      journal = {\jcap},
         year = 2025,
        month = aug,
       volume = {2025},
       number = {8},
          eid = {025},
        pages = {025},
          doi = {10.1088/1475-7516/2025/08/025},
archivePrefix = {arXiv},
       eprint = {1910.13970},
 primaryClass = {astro-ph.IM},
       adsurl = {https://ui.adsabs.harvard.edu/abs/2025JCAP...08..025L}
}

@ARTICLE{Sheldon:2019uxq,
       author = {{Sheldon}, Erin S. and {Becker}, Matthew R. and {MacCrann}, Niall and {Jarvis}, Michael},
        title = "{Mitigating Shear-dependent Object Detection Biases with Metacalibration}",
      journal = {\apj},
         year = 2020,
        month = oct,
       volume = {902},
       number = {2},
          eid = {138},
        pages = {138},
          doi = {10.3847/1538-4357/abb595},
archivePrefix = {arXiv},
       eprint = {1911.02505},
 primaryClass = {astro-ph.CO},
       adsurl = {https://ui.adsabs.harvard.edu/abs/2020ApJ...902..138S}
}

@ARTICLE{McCleary:2025xqb,
       author = {{McCleary}, Jacqueline E. and {Huff}, Eric M. and {Bartlett}, James G. and {Hensley}, Brandon S.},
        title = "{A Detection of Circumgalactic Dust at Megaparsec Scales with Maximum Likelihood Estimation}",
      journal = {\apj},
         year = 2026,
        month = apr,
       volume = {1000},
       number = {2},
          eid = {313},
        pages = {313},
          doi = {10.3847/1538-4357/ae4c3d},
archivePrefix = {arXiv},
       eprint = {2503.04098},
 primaryClass = {astro-ph.GA},
       adsurl = {https://ui.adsabs.harvard.edu/abs/2026ApJ..1000..313M}
}

@ARTICLE{shaaban2022weak,
       author = {{Shaaban}, Mohamed M. and {Gill}, Ajay S. and {McCleary}, Jacqueline and {Massey}, Richard J. and {Benton}, Steven J. and {Brown}, Anthony M. and {Damaren}, Christopher J. and {Eifler}, Tim and {Fraisse}, Aurelien A. and {Everett}, Spencer and {Galloway}, Mathew N. and {Henderson}, Michael and {Holder}, Bradley and {Huff}, Eric M. and {Jauzac}, Mathilde and {Jones}, William C. and {Lagattuta}, David and {Leung}, Jason S. -Y. and {Li}, Lun and {T. Luu}, Thuy Vy and {Nagy}, Johanna M. and {Netterfield}, C. Barth and {Redmond}, Susan F. and {Rhodes}, Jason D. and {Robertson}, Andrew and {Schmoll}, J{\"u}rgen and {Sirks}, Ellen and {Sivanandam}, Suresh},
        title = "{Weak Lensing in the Blue: A Counter-intuitive Strategy for Stratospheric Observations}",
      journal = {\aj},
         year = 2022,
        month = dec,
       volume = {164},
       number = {6},
          eid = {245},
        pages = {245},
          doi = {10.3847/1538-3881/ac9b1c},
archivePrefix = {arXiv},
       eprint = {2210.09182},
 primaryClass = {astro-ph.IM},
       adsurl = {https://ui.adsabs.harvard.edu/abs/2022AJ....164..245S}
}

@ARTICLE{Randall2008,
       author = {{Randall}, Scott W. and {Markevitch}, Maxim and {Clowe}, Douglas and {Gonzalez}, Anthony H. and {Brada{\v{c}}}, Marusa},
        title = "{Constraints on the Self-Interaction Cross Section of Dark Matter from Numerical Simulations of the Merging Galaxy Cluster 1E 0657-56}",
      journal = {\apj},
         year = 2008,
        month = jun,
       volume = {679},
       number = {2},
        pages = {1173-1180},
          doi = {10.1086/587859},
archivePrefix = {arXiv},
       eprint = {0704.0261},
 primaryClass = {astro-ph},
       adsurl = {https://ui.adsabs.harvard.edu/abs/2008ApJ...679.1173R}
}

@ARTICLE{bulleticity,
       author = {{Massey}, Richard and {Kitching}, Thomas and {Nagai}, Daisuke},
        title = "{Cluster bulleticity}",
      journal = {\mnras},
         year = 2011,
        month = may,
       volume = {413},
       number = {3},
        pages = {1709-1716},
          doi = {10.1111/j.1365-2966.2011.18246.x},
archivePrefix = {arXiv},
       eprint = {1007.1924},
 primaryClass = {astro-ph.CO},
       adsurl = {https://ui.adsabs.harvard.edu/abs/2011MNRAS.413.1709M}
}

@ARTICLE{Sirks:2024njj,
       author = {{Sirks}, Ellen L. and {Harvey}, David and {Massey}, Richard and {Oman}, Kyle A. and {Robertson}, Andrew and {Frenk}, Carlos and {Everett}, Spencer and {Gill}, Ajay S. and {Lagattuta}, David and {McCleary}, Jacqueline},
        title = "{Hydrodynamical simulations of merging galaxy clusters: giant dark matter particle colliders, powered by gravity}",
      journal = {\mnras},
         year = 2024,
        month = may,
       volume = {530},
       number = {3},
        pages = {3160-3170},
          doi = {10.1093/mnras/stae1012},
archivePrefix = {arXiv},
       eprint = {2405.00140},
 primaryClass = {astro-ph.CO},
       adsurl = {https://ui.adsabs.harvard.edu/abs/2024MNRAS.530.3160S}
}

@ARTICLE{Cerini:2022akj,
       author = {{Cerini}, Giulia and {Cappelluti}, Nico and {Natarajan}, Priyamvada},
        title = "{New Metrics to Probe the Dynamical State of Galaxy Clusters}",
      journal = {\apj},
         year = 2023,
        month = mar,
       volume = {945},
       number = {2},
          eid = {152},
        pages = {152},
          doi = {10.3847/1538-4357/acbccb},
archivePrefix = {arXiv},
       eprint = {2209.06831},
 primaryClass = {astro-ph.CO},
       adsurl = {https://ui.adsabs.harvard.edu/abs/2023ApJ...945..152C}
}

@ARTICLE{Rowe:2014cza,
       author = {{Rowe}, B.~T.~P. and {Jarvis}, M. and {Mandelbaum}, R. and {Bernstein}, G.~M. and {Bosch}, J. and {Simet}, M. and {Meyers}, J.~E. and {Kacprzak}, T. and {Nakajima}, R. and {Zuntz}, J. and {Miyatake}, H. and {Dietrich}, J.~P. and {Armstrong}, R. and {Melchior}, P. and {Gill}, M.~S.~S.},
        title = "{GALSIM: The modular galaxy image simulation toolkit}",
      journal = {Astronomy and Computing},
         year = 2015,
        month = apr,
       volume = {10},
        pages = {121-150},
          doi = {10.1016/j.ascom.2015.02.002},
archivePrefix = {arXiv},
       eprint = {1407.7676},
 primaryClass = {astro-ph.IM},
       adsurl = {https://ui.adsabs.harvard.edu/abs/2015A&C....10..121R}
}

@software{ngmix,
       author = {{Sheldon}, Erin},
        title = "{NGMIX: Gaussian mixture models for 2D images}",
 howpublished = {Astrophysics Source Code Library, record ascl:1508.008},
         year = 2015,
        month = aug,
          eid = {ascl:1508.008},
archivePrefix = {ascl},
       eprint = {1508.008},
       adsurl = {https://ui.adsabs.harvard.edu/abs/2015ascl.soft08008S}
}

@ARTICLE{Hunter:2007,
       author = {{Hunter}, John D.},
        title = "{Matplotlib: A 2D Graphics Environment}",
      journal = {Computing in Science and Engineering},
         year = 2007,
        month = jan,
       volume = {9},
       number = {3},
        pages = {90-95},
          doi = {10.1109/MCSE.2007.55},
       adsurl = {https://ui.adsabs.harvard.edu/abs/2007CSE.....9...90H}
}

@ARTICLE{2020SciPy-NMeth,
  author  = {Virtanen, Pauli and Gommers, Ralf and Oliphant, Travis E. and
            Haberland, Matt and Reddy, Tyler and Cournapeau, David and
            Burovski, Evgeni and Peterson, Pearu and Weckesser, Warren and
            Bright, Jonathan and {van der Walt}, St{\'e}fan J. and
            Brett, Matthew and Wilson, Joshua and Millman, K. Jarrod and
            Mayorov, Nikolay and Nelson, Andrew R. J. and Jones, Eric and
            Kern, Robert and Larson, Eric and Carey, C J and
            Polat, {\.I}lhan and Feng, Yu and Moore, Eric W. and
            {VanderPlas}, Jake and Laxalde, Denis and Perktold, Josef and
            Cimrman, Robert and Henriksen, Ian and Quintero, E. A. and
            Harris, Charles R. and Archibald, Anne M. and
            Ribeiro, Ant{\^o}nio H. and Pedregosa, Fabian and
            {van Mulbregt}, Paul and {SciPy 1.0 Contributors}},
  title   = {{{SciPy} 1.0: Fundamental Algorithms for Scientific
            Computing in Python}},
  journal = {Nature Methods},
  year    = {2020},
  volume  = {17},
  pages   = {261--272},
  adsurl  = {https://rdcu.be/b08Wh},
  doi     = {10.1038/s41592-019-0686-2},
}

@ARTICLE{harris2020array,
       author = {{Harris}, Charles R. and {Millman}, K. Jarrod and {van der Walt}, St{\'e}fan J. and {Gommers}, Ralf and {Virtanen}, Pauli and {Cournapeau}, David and {Wieser}, Eric and {Taylor}, Julian and {Berg}, Sebastian and {Smith}, Nathaniel J. and {Kern}, Robert and {Picus}, Matti and {Hoyer}, Stephan and {van Kerkwijk}, Marten H. and {Brett}, Matthew and {Haldane}, Allan and {del R{\'\i}o}, Jaime Fern{\'a}ndez and {Wiebe}, Mark and {Peterson}, Pearu and {G{\'e}rard-Marchant}, Pierre and {Sheppard}, Kevin and {Reddy}, Tyler and {Weckesser}, Warren and {Abbasi}, Hameer and {Gohlke}, Christoph and {Oliphant}, Travis E.},
        title = "{Array programming with NumPy}",
      journal = {\nat},
         year = 2020,
        month = sep,
       volume = {585},
       number = {7825},
        pages = {357-362},
          doi = {10.1038/s41586-020-2649-2},
archivePrefix = {arXiv},
       eprint = {2006.10256},
 primaryClass = {cs.MS},
       adsurl = {https://ui.adsabs.harvard.edu/abs/2020Natur.585..357H}
}

@ARTICLE{healpy2,
       author = {{Zonca}, Andrea and {Singer}, Leo and {Lenz}, Daniel and {Reinecke}, Martin and {Rosset}, Cyrille and {Hivon}, Eric and {Gorski}, Krzysztof},
        title = "{healpy: equal area pixelization and spherical harmonics transforms for data on the sphere in Python}",
      journal = {The Journal of Open Source Software},
         year = 2019,
        month = mar,
       volume = {4},
       number = {35},
          eid = {1298},
        pages = {1298},
          doi = {10.21105/joss.01298},
       adsurl = {https://ui.adsabs.harvard.edu/abs/2019JOSS....4.1298Z}
}

@ARTICLE{healpy1,
   author = {{G{\'o}rski}, K.~M. and {Hivon}, E. and {Banday}, A.~J. and 
	{Wandelt}, B.~D. and {Hansen}, F.~K. and {Reinecke}, M. and 
	{Bartelmann}, M.},
    title = "{HEALPix: A Framework for High-Resolution Discretization and Fast Analysis of Data Distributed on the Sphere}",
  journal = {\apj},
   eprint = {arXiv:astro-ph/0409513},
     year = 2005,
    month = apr,
   volume = 622,
    pages = {759-771},
      doi = {10.1086/427976},
   adsurl = {http://adsabs.harvard.edu/abs/2005ApJ...622..759G}
}

@software{swarp,
       author = {{Bertin}, Emmanuel},
        title = "{SWarp: Resampling and Co-adding FITS Images Together}",
 howpublished = {Astrophysics Source Code Library, record ascl:1010.068},
         year = 2010,
        month = oct,
          eid = {ascl:1010.068},
archivePrefix = {ascl},
       eprint = {1010.068},
       adsurl = {https://ui.adsabs.harvard.edu/abs/2010ascl.soft10068B}
}

@ARTICLE{Wittman:2000tc,
       author = {{Wittman}, David M. and {Tyson}, J. Anthony and {Kirkman}, David and {Dell'Antonio}, Ian and {Bernstein}, Gary},
        title = "{Detection of weak gravitational lensing distortions of distant galaxies by cosmic dark matter at large scales}",
      journal = {\nat},
         year = 2000,
        month = may,
       volume = {405},
       number = {6783},
        pages = {143-148},
          doi = {10.1038/35012001},
archivePrefix = {arXiv},
       eprint = {astro-ph/0003014},
 primaryClass = {astro-ph},
       adsurl = {https://ui.adsabs.harvard.edu/abs/2000Natur.405..143W}
}

@ARTICLE{Bacon:2000sy,
       author = {{Bacon}, David J. and {Refregier}, Alexandre R. and {Ellis}, Richard S.},
        title = "{Detection of weak gravitational lensing by large-scale structure}",
      journal = {\mnras},
         year = 2000,
        month = oct,
       volume = {318},
       number = {2},
        pages = {625-640},
          doi = {10.1046/j.1365-8711.2000.03851.x},
archivePrefix = {arXiv},
       eprint = {astro-ph/0003008},
 primaryClass = {astro-ph},
       adsurl = {https://ui.adsabs.harvard.edu/abs/2000MNRAS.318..625B}
}

@ARTICLE{Li:2022,
       author = {{Li}, Xiangchong and {Li}, Yin and {Massey}, Richard},
        title = "{Weak gravitational lensing shear measurement with FPFS: analytical mitigation of noise bias and selection bias}",
      journal = {\mnras},
         year = 2022,
        month = apr,
       volume = {511},
       number = {4},
        pages = {4850-4860},
          doi = {10.1093/mnras/stac342},
archivePrefix = {arXiv},
       eprint = {2110.01214},
 primaryClass = {astro-ph.CO},
       adsurl = {https://ui.adsabs.harvard.edu/abs/2022MNRAS.511.4850L}
}

@ARTICLE{Li:2022qzu,
       author = {{Li}, Xiangchong and {Mandelbaum}, Rachel},
        title = "{Analytical weak-lensing shear responses of galaxy properties and galaxy detection}",
      journal = {\mnras},
         year = 2023,
        month = jun,
       volume = {521},
       number = {4},
        pages = {4904-4926},
          doi = {10.1093/mnras/stad890},
archivePrefix = {arXiv},
       eprint = {2208.10522},
 primaryClass = {astro-ph.CO},
       adsurl = {https://ui.adsabs.harvard.edu/abs/2023MNRAS.521.4904L}
}

@ARTICLE{firstwl3,
       author = {{Kaiser}, Nick and {Wilson}, Gillian and {Luppino}, Gerard A.},
        title = "{Large-Scale Cosmic Shear Measurements}",
      journal = {arXiv e-prints},
         year = 2000,
        month = mar,
          eid = {astro-ph/0003338},
        pages = {astro-ph/0003338},
          doi = {10.48550/arXiv.astro-ph/0003338},
archivePrefix = {arXiv},
       eprint = {astro-ph/0003338},
 primaryClass = {astro-ph},
       adsurl = {https://ui.adsabs.harvard.edu/abs/2000astro.ph..3338K}
}

@ARTICLE{firstwl4,
       author = {{Van Waerbeke}, L. and {Mellier}, Y. and {Erben}, T. and {Cuillandre}, J.~C. and {Bernardeau}, F. and {Maoli}, R. and {Bertin}, E. and {McCracken}, H.~J. and {Le F{\`e}vre}, O. and {Fort}, B. and {Dantel-Fort}, M. and {Jain}, B. and {Schneider}, P.},
        title = "{Detection of correlated galaxy ellipticities from CFHT data: first evidence for gravitational lensing by large-scale structures}",
      journal = {\aap},
         year = 2000,
        month = jun,
       volume = {358},
        pages = {30-44},
          doi = {10.48550/arXiv.astro-ph/0002500},
archivePrefix = {arXiv},
       eprint = {astro-ph/0002500},
 primaryClass = {astro-ph},
       adsurl = {https://ui.adsabs.harvard.edu/abs/2000A&A...358...30V}
}

@ARTICLE{step2,
       author = {{Massey}, Richard and {Heymans}, Catherine and {Berg{\'e}}, Joel and {Bernstein}, Gary and {Bridle}, Sarah and {Clowe}, Douglas and {Dahle}, H{\r{a}}kon and {Ellis}, Richard and {Erben}, Thomas and {Hetterscheidt}, Marco and {High}, F. William and {Hirata}, Christopher and {Hoekstra}, Henk and {Hudelot}, Patrick and {Jarvis}, Mike and {Johnston}, David and {Kuijken}, Konrad and {Margoniner}, Vera and {Mandelbaum}, Rachel and {Mellier}, Yannick and {Nakajima}, Reiko and {Paulin-Henriksson}, Stephane and {Peeples}, Molly and {Roat}, Chris and {Refregier}, Alexandre and {Rhodes}, Jason and {Schrabback}, Tim and {Schirmer}, Mischa and {Seljak}, Uro{\v{s}} and {Semboloni}, Elisabetta and {van Waerbeke}, Ludovic},
        title = "{The Shear Testing Programme 2: Factors affecting high-precision weak-lensing analyses}",
      journal = {\mnras},
         year = 2007,
        month = mar,
       volume = {376},
       number = {1},
        pages = {13-38},
          doi = {10.1111/j.1365-2966.2006.11315.x},
archivePrefix = {arXiv},
       eprint = {astro-ph/0608643},
 primaryClass = {astro-ph},
       adsurl = {https://ui.adsabs.harvard.edu/abs/2007MNRAS.376...13M}
}

@ARTICLE{bridle2010great08,
       author = {{Bridle}, Sarah and {Balan}, Sreekumar T. and {Bethge}, Matthias and {Gentile}, Marc and {Harmeling}, Stefan and {Heymans}, Catherine and {Hirsch}, Michael and {Hosseini}, Reshad and {Jarvis}, Mike and {Kirk}, Donnacha and {Kitching}, Thomas and {Kuijken}, Konrad and {Lewis}, Antony and {Paulin-Henriksson}, Stephane and {Sch{\"o}lkopf}, Bernhard and {Velander}, Malin and {Voigt}, Lisa and {Witherick}, Dugan and {Amara}, Adam and {Bernstein}, Gary and {Courbin}, Fr{\'e}d{\'e}ric and {Gill}, Mandeep and {Heavens}, Alan and {Mandelbaum}, Rachel and {Massey}, Richard and {Moghaddam}, Baback and {Rassat}, Anais and {R{\'e}fr{\'e}gier}, Alexandre and {Rhodes}, Jason and {Schrabback}, Tim and {Shawe-Taylor}, John and {Shmakova}, Marina and {van Waerbeke}, Ludovic and {Wittman}, David},
        title = "{Results of the GREAT08 Challenge: an image analysis competition for cosmological lensing}",
      journal = {\mnras},
         year = 2010,
        month = jul,
       volume = {405},
       number = {3},
        pages = {2044-2061},
          doi = {10.1111/j.1365-2966.2010.16598.x},
archivePrefix = {arXiv},
       eprint = {0908.0945},
 primaryClass = {astro-ph.CO},
       adsurl = {https://ui.adsabs.harvard.edu/abs/2010MNRAS.405.2044B}
}

@ARTICLE{Mandelbaum:2014fta,
       author = {{Mandelbaum}, Rachel and {Rowe}, Barnaby and {Armstrong}, Robert and {Bard}, Deborah and {Bertin}, Emmanuel and {Bosch}, James and {Boutigny}, Dominique and {Courbin}, Frederic and {Dawson}, William A. and {Donnarumma}, Annamaria and {Fenech Conti}, Ian and {Gavazzi}, Rapha{\"e}l and {Gentile}, Marc and {Gill}, Mandeep S.~S. and {Hogg}, David W. and {Huff}, Eric M. and {Jee}, M. James and {Kacprzak}, Tomasz and {Kilbinger}, Martin and {Kuntzer}, Thibault and {Lang}, Dustin and {Luo}, Wentao and {March}, Marisa C. and {Marshall}, Philip J. and {Meyers}, Joshua E. and {Miller}, Lance and {Miyatake}, Hironao and {Nakajima}, Reiko and {Ngol{\'e} Mboula}, Fred Maurice and {Nurbaeva}, Guldariya and {Okura}, Yuki and {Paulin-Henriksson}, St{\'e}phane and {Rhodes}, Jason and {Schneider}, Michael D. and {Shan}, Huanyuan and {Sheldon}, Erin S. and {Simet}, Melanie and {Starck}, Jean-Luc and {Sureau}, Florent and {Tewes}, Malte and {Zarb Adami}, Kristian and {Zhang}, Jun and {Zuntz}, Joe},
        title = "{GREAT3 results - I. Systematic errors in shear estimation and the impact of real galaxy morphology}",
      journal = {\mnras},
         year = 2015,
        month = jul,
       volume = {450},
       number = {3},
        pages = {2963-3007},
          doi = {10.1093/mnras/stv781},
archivePrefix = {arXiv},
       eprint = {1412.1825},
 primaryClass = {astro-ph.CO},
       adsurl = {https://ui.adsabs.harvard.edu/abs/2015MNRAS.450.2963M}
}

@ARTICLE{Kitching2012,
       author = {{Kitching}, T.~D. and {Balan}, S.~T. and {Bridle}, S. and {Cantale}, N. and {Courbin}, F. and {Eifler}, T. and {Gentile}, M. and {Gill}, M.~S.~S. and {Harmeling}, S. and {Heymans}, C. and {Hirsch}, M. and {Honscheid}, K. and {Kacprzak}, T. and {Kirkby}, D. and {Margala}, D. and {Massey}, R.~J. and {Melchior}, P. and {Nurbaeva}, G. and {Patton}, K. and {Rhodes}, J. and {Rowe}, B.~T.~P. and {Taylor}, A.~N. and {Tewes}, M. and {Viola}, M. and {Witherick}, D. and {Voigt}, L. and {Young}, J. and {Zuntz}, J.},
        title = "{Image analysis for cosmology: results from the GREAT10 Galaxy Challenge}",
      journal = {\mnras},
         year = 2012,
        month = jul,
       volume = {423},
       number = {4},
        pages = {3163-3208},
          doi = {10.1111/j.1365-2966.2012.21095.x},
archivePrefix = {arXiv},
       eprint = {1202.5254},
 primaryClass = {astro-ph.CO},
       adsurl = {https://ui.adsabs.harvard.edu/abs/2012MNRAS.423.3163K}
}

@ARTICLE{1993ApJ...404..441K,
       author = {{Kaiser}, Nick and {Squires}, Gordon},
        title = "{Mapping the Dark Matter with Weak Gravitational Lensing}",
      journal = {\apj},
         year = 1993,
        month = feb,
       volume = {404},
        pages = {441},
          doi = {10.1086/172297},
       adsurl = {https://ui.adsabs.harvard.edu/abs/1993ApJ...404..441K}
}

@ARTICLE{Bartelmann:1999yn,
       author = {{Bartelmann}, M. and {Schneider}, P.},
        title = "{Weak gravitational lensing}",
      journal = {\physrep},
         year = 2001,
        month = jan,
       volume = {340},
       number = {4-5},
        pages = {291-472},
          doi = {10.1016/S0370-1573(00)00082-X},
archivePrefix = {arXiv},
       eprint = {astro-ph/9912508},
 primaryClass = {astro-ph},
       adsurl = {https://ui.adsabs.harvard.edu/abs/2001PhR...340..291B}
}

@ARTICLE{Bernstein:2001nz,
       author = {{Bernstein}, G.~M. and {Jarvis}, M.},
        title = "{Shapes and Shears, Stars and Smears: Optimal Measurements for Weak Lensing}",
      journal = {\aj},
         year = 2002,
        month = feb,
       volume = {123},
       number = {2},
        pages = {583-618},
          doi = {10.1086/338085},
archivePrefix = {arXiv},
       eprint = {astro-ph/0107431},
 primaryClass = {astro-ph},
       adsurl = {https://ui.adsabs.harvard.edu/abs/2002AJ....123..583B}
}

@article{Hirata:2003cv,
       author = {{Hirata}, Christopher and {Seljak}, Uro{\v{s}}},
        title = "{Shear calibration biases in weak-lensing surveys}",
      journal = {\mnras},
         year = 2003,
        month = aug,
       volume = {343},
       number = {2},
        pages = {459-480},
          doi = {10.1046/j.1365-8711.2003.06683.x},
archivePrefix = {arXiv},
       eprint = {astro-ph/0301054},
 primaryClass = {astro-ph},
       adsurl = {https://ui.adsabs.harvard.edu/abs/2003MNRAS.343..459H}
}

@ARTICLE{Schneider:2005ka,
       author = {{Schneider}, Peter},
        title = "{Weak Gravitational Lensing}",
      journal = {arXiv e-prints},
         year = 2005,
        month = sep,
          eid = {astro-ph/0509252},
        pages = {astro-ph/0509252},
          doi = {10.48550/arXiv.astro-ph/0509252},
archivePrefix = {arXiv},
       eprint = {astro-ph/0509252},
 primaryClass = {astro-ph},
       adsurl = {https://ui.adsabs.harvard.edu/abs/2005astro.ph..9252S}
}

@ARTICLE{Clowe:2003tk,
       author = {{Clowe}, Douglas and {Gonzalez}, Anthony and {Markevitch}, Maxim},
        title = "{Weak-Lensing Mass Reconstruction of the Interacting Cluster 1E 0657-558: Direct Evidence for the Existence of Dark Matter}",
      journal = {\apj},
         year = 2004,
        month = apr,
       volume = {604},
       number = {2},
        pages = {596-603},
          doi = {10.1086/381970},
archivePrefix = {arXiv},
       eprint = {astro-ph/0312273},
 primaryClass = {astro-ph},
       adsurl = {https://ui.adsabs.harvard.edu/abs/2004ApJ...604..596C}
}

@ARTICLE{Robertson:2016xjh,
       author = {{Robertson}, Andrew and {Massey}, Richard and {Eke}, Vincent},
        title = "{What does the Bullet Cluster tell us about self-interacting dark matter?}",
      journal = {\mnras},
         year = 2017,
        month = feb,
       volume = {465},
       number = {1},
        pages = {569-587},
          doi = {10.1093/mnras/stw2670},
archivePrefix = {arXiv},
       eprint = {1605.04307},
 primaryClass = {astro-ph.CO},
       adsurl = {https://ui.adsabs.harvard.edu/abs/2017MNRAS.465..569R}
}

@ARTICLE{2001PASP..113.1420V,
       author = {{van Dokkum}, Pieter G.},
        title = "{Cosmic-Ray Rejection by Laplacian Edge Detection}",
      journal = {\pasp},
         year = 2001,
        month = nov,
       volume = {113},
       number = {789},
        pages = {1420-1427},
          doi = {10.1086/323894},
archivePrefix = {arXiv},
       eprint = {astro-ph/0108003},
 primaryClass = {astro-ph},
       adsurl = {https://ui.adsabs.harvard.edu/abs/2001PASP..113.1420V}
}

@ARTICLE{Zenteno2020,
       author = {{Zenteno}, A. and {Hern{\'a}ndez-Lang}, D. and {Klein}, M. and {Vergara Cervantes}, C. and {Hollowood}, D.~L. and {Bhargava}, S. and {Palmese}, A. and {Strazzullo}, V. and {Romer}, A.~K. and {Mohr}, J.~J. and {Jeltema}, T. and {Saro}, A. and {Lidman}, C. and {Gruen}, D. and {Ojeda}, V. and {Katzenberger}, A. and {Aguena}, M. and {Allam}, S. and {Avila}, S. and {Bayliss}, M. and {Bertin}, E. and {Brooks}, D. and {Buckley-Geer}, E. and {Burke}, D.~L. and {Capasso}, R. and {Carnero Rosell}, A. and {Carrasco Kind}, M. and {Carretero}, J. and {Castander}, F.~J. and {Costanzi}, M. and {da Costa}, L.~N. and {De Vicente}, J. and {Desai}, S. and {Diehl}, H.~T. and {Doel}, P. and {Eifler}, T.~F. and {Evrard}, A.~E. and {Flaugher}, B. and {Floyd}, B. and {Fosalba}, P. and {Frieman}, J. and {Garc{\'\i}a-Bellido}, J. and {Gerdes}, D.~W. and {Gonzalez}, J.~R. and {Gruendl}, R.~A. and {Gschwend}, J. and {Gutierrez}, G. and {Hartley}, W.~G. and {Hinton}, S.~R. and {Honscheid}, K. and {James}, D.~J. and {Kuehn}, K. and {Lahav}, O. and {Lima}, M. and {McDonald}, M. and {Maia}, M.~A.~G. and {March}, M. and {Melchior}, P. and {Menanteau}, F. and {Miquel}, R. and {Ogando}, R.~L.~C. and {Paz-Chinch{\'o}n}, F. and {Plazas}, A.~A. and {Roodman}, A. and {Rykoff}, E.~S. and {Sanchez}, E. and {Scarpine}, V. and {Schubnell}, M. and {Serrano}, S. and {Sevilla-Noarbe}, I. and {Smith}, M. and {Soares-Santos}, M. and {Suchyta}, E. and {Swanson}, M.~E.~C. and {Tarle}, G. and {Thomas}, D. and {Varga}, T.~N. and {Walker}, A.~R. and {Wilkinson}, R.~D. and {DES Collaboration}},
        title = "{A joint SZ-X-ray-optical analysis of the dynamical state of 288 massive galaxy clusters}",
      journal = {\mnras},
         year = 2020,
        month = jun,
       volume = {495},
       number = {1},
        pages = {705-725},
          doi = {10.1093/mnras/staa1157},
archivePrefix = {arXiv},
       eprint = {2004.01721},
 primaryClass = {astro-ph.GA},
       adsurl = {https://ui.adsabs.harvard.edu/abs/2020MNRAS.495..705Z}
}

@ARTICLE{intro1,
       author = {{Vikhlinin}, A. and {Kravtsov}, A.~V. and {Burenin}, R.~A. and {Ebeling}, H. and {Forman}, W.~R. and {Hornstrup}, A. and {Jones}, C. and {Murray}, S.~S. and {Nagai}, D. and {Quintana}, H. and {Voevodkin}, A.},
        title = "{Chandra Cluster Cosmology Project III: Cosmological Parameter Constraints}",
      journal = {\apj},
         year = 2009,
        month = feb,
       volume = {692},
       number = {2},
        pages = {1060-1074},
          doi = {10.1088/0004-637X/692/2/1060},
archivePrefix = {arXiv},
       eprint = {0812.2720},
 primaryClass = {astro-ph},
       adsurl = {https://ui.adsabs.harvard.edu/abs/2009ApJ...692.1060V}
}

@ARTICLE{intro2,
       author = {{Allen}, Steven W. and {Evrard}, August E. and {Mantz}, Adam B.},
        title = "{Cosmological Parameters from Observations of Galaxy Clusters}",
      journal = {\araa},
         year = 2011,
        month = sep,
       volume = {49},
       number = {1},
        pages = {409-470},
          doi = {10.1146/annurev-astro-081710-102514},
archivePrefix = {arXiv},
       eprint = {1103.4829},
 primaryClass = {astro-ph.CO},
       adsurl = {https://ui.adsabs.harvard.edu/abs/2011ARA&A..49..409A}
}

@ARTICLE{intro3,
       author = {{Mantz}, A.~B. and {Allen}, S.~W. and {Morris}, R.~G. and {Rapetti}, D.~A. and {Applegate}, D.~E. and {Kelly}, P.~L. and {von der Linden}, A. and {Schmidt}, R.~W.},
        title = "{Cosmology and astrophysics from relaxed galaxy clusters - II. Cosmological constraints}",
      journal = {\mnras},
         year = 2014,
        month = may,
       volume = {440},
       number = {3},
        pages = {2077-2098},
          doi = {10.1093/mnras/stu368},
archivePrefix = {arXiv},
       eprint = {1402.6212},
 primaryClass = {astro-ph.CO},
       adsurl = {https://ui.adsabs.harvard.edu/abs/2014MNRAS.440.2077M}
}

@ARTICLE{intro4,
       author = {{Mantz}, Adam B. and {von der Linden}, Anja and {Allen}, Steven W. and {Applegate}, Douglas E. and {Kelly}, Patrick L. and {Morris}, R. Glenn and {Rapetti}, David A. and {Schmidt}, Robert W. and {Adhikari}, Saroj and {Allen}, Mark T. and {Burchat}, Patricia R. and {Burke}, David L. and {Cataneo}, Matteo and {Donovan}, David and {Ebeling}, Harald and {Shandera}, Sarah and {Wright}, Adam},
        title = "{Weighing the giants - IV. Cosmology and neutrino mass}",
      journal = {\mnras},
         year = 2015,
        month = jan,
       volume = {446},
       number = {3},
        pages = {2205-2225},
          doi = {10.1093/mnras/stu2096},
archivePrefix = {arXiv},
       eprint = {1407.4516},
 primaryClass = {astro-ph.CO},
       adsurl = {https://ui.adsabs.harvard.edu/abs/2015MNRAS.446.2205M}
}

@ARTICLE{richness1,
       author = {{Koester}, Benjamin P. and {McKay}, Timothy A. and {Annis}, James and {Wechsler}, Risa H. and {Evrard}, August E. and {Rozo}, Eduardo and {Bleem}, Lindsey and {Sheldon}, Erin S. and {Johnston}, David},
        title = "{MaxBCG: A Red-Sequence Galaxy Cluster Finder}",
      journal = {\apj},
         year = 2007,
        month = may,
       volume = {660},
       number = {1},
        pages = {221-238},
          doi = {10.1086/512092},
archivePrefix = {arXiv},
       eprint = {astro-ph/0701268},
 primaryClass = {astro-ph},
       adsurl = {https://ui.adsabs.harvard.edu/abs/2007ApJ...660..221K}
}

@ARTICLE{richness2,
       author = {{Andreon}, S.},
        title = "{Richness-based masses of rich and famous galaxy clusters}",
      journal = {\aap},
         year = 2016,
        month = mar,
       volume = {587},
          eid = {A158},
        pages = {A158},
          doi = {10.1051/0004-6361/201526852},
archivePrefix = {arXiv},
       eprint = {1601.06912},
 primaryClass = {astro-ph.CO},
       adsurl = {https://ui.adsabs.harvard.edu/abs/2016A&A...587A.158A}
}

@ARTICLE{xray2,
       author = {{Arnaud}, M. and {Pointecouteau}, E. and {Pratt}, G.~W.},
        title = "{The structural and scaling properties of nearby galaxy clusters.  II. The M-T relation}",
      journal = {\aap},
         year = 2005,
        month = oct,
       volume = {441},
       number = {3},
        pages = {893-903},
          doi = {10.1051/0004-6361:20052856},
archivePrefix = {arXiv},
       eprint = {astro-ph/0502210},
 primaryClass = {astro-ph},
       adsurl = {https://ui.adsabs.harvard.edu/abs/2005A&A...441..893A}
}

@ARTICLE{sz1,
       author = {{Vanderlinde}, K. and {Crawford}, T.~M. and {de Haan}, T. and {Dudley}, J.~P. and {Shaw}, L. and {Ade}, P.~A.~R. and {Aird}, K.~A. and {Benson}, B.~A. and {Bleem}, L.~E. and {Brodwin}, M. and {Carlstrom}, J.~E. and {Chang}, C.~L. and {Crites}, A.~T. and {Desai}, S. and {Dobbs}, M.~A. and {Foley}, R.~J. and {George}, E.~M. and {Gladders}, M.~D. and {Hall}, N.~R. and {Halverson}, N.~W. and {High}, F.~W. and {Holder}, G.~P. and {Holzapfel}, W.~L. and {Hrubes}, J.~D. and {Joy}, M. and {Keisler}, R. and {Knox}, L. and {Lee}, A.~T. and {Leitch}, E.~M. and {Loehr}, A. and {Lueker}, M. and {Marrone}, D.~P. and {McMahon}, J.~J. and {Mehl}, J. and {Meyer}, S.~S. and {Mohr}, J.~J. and {Montroy}, T.~E. and {Ngeow}, C.-C. and {Padin}, S. and {Plagge}, T. and {Pryke}, C. and {Reichardt}, C.~L. and {Rest}, A. and {Ruel}, J. and {Ruhl}, J.~E. and {Schaffer}, K.~K. and {Shirokoff}, E. and {Song}, J. and {Spieler}, H.~G. and {Stalder}, B. and {Staniszewski}, Z. and {Stark}, A.~A. and {Stubbs}, C.~W. and {van Engelen}, A. and {Vieira}, J.~D. and {Williamson}, R. and {Yang}, Y. and {Zahn}, O. and {Zenteno}, A.},
        title = "{Galaxy Clusters Selected with the Sunyaev-Zel'dovich Effect from 2008 South Pole Telescope Observations}",
      journal = {\apj},
         year = 2010,
        month = oct,
       volume = {722},
       number = {2},
        pages = {1180-1196},
          doi = {10.1088/0004-637X/722/2/1180},
archivePrefix = {arXiv},
       eprint = {1003.0003},
 primaryClass = {astro-ph.CO},
       adsurl = {https://ui.adsabs.harvard.edu/abs/2010ApJ...722.1180V}
}

@ARTICLE{sz2,
       author = {{Planck Collaboration} and {Ade}, P.~A.~R. and {Aghanim}, N. and {Arnaud}, M. and {Ashdown}, M. and {Aumont}, J. and {Baccigalupi}, C. and {Banday}, A.~J. and {Barreiro}, R.~B. and {Bartlett}, J.~G. and {Bartolo}, N. and {Battaner}, E. and {Battye}, R. and {Benabed}, K. and {Beno{\^\i}t}, A. and {Benoit-L{\'e}vy}, A. and {Bernard}, J.-P. and {Bersanelli}, M. and {Bielewicz}, P. and {Bock}, J.~J. and {Bonaldi}, A. and {Bonavera}, L. and {Bond}, J.~R. and {Borrill}, J. and {Bouchet}, F.~R. and {Bucher}, M. and {Burigana}, C. and {Butler}, R.~C. and {Calabrese}, E. and {Cardoso}, J.-F. and {Catalano}, A. and {Challinor}, A. and {Chamballu}, A. and {Chary}, R.-R. and {Chiang}, H.~C. and {Christensen}, P.~R. and {Church}, S. and {Clements}, D.~L. and {Colombi}, S. and {Colombo}, L.~P.~L. and {Combet}, C. and {Comis}, B. and {Couchot}, F. and {Coulais}, A. and {Crill}, B.~P. and {Curto}, A. and {Cuttaia}, F. and {Danese}, L. and {Davies}, R.~D. and {Davis}, R.~J. and {de Bernardis}, P. and {de Rosa}, A. and {de Zotti}, G. and {Delabrouille}, J. and {D{\'e}sert}, F.-X. and {Diego}, J.~M. and {Dolag}, K. and {Dole}, H. and {Donzelli}, S. and {Dor{\'e}}, O. and {Douspis}, M. and {Ducout}, A. and {Dupac}, X. and {Efstathiou}, G. and {Elsner}, F. and {En{\ss}lin}, T.~A. and {Eriksen}, H.~K. and {Falgarone}, E. and {Fergusson}, J. and {Finelli}, F. and {Forni}, O. and {Frailis}, M. and {Fraisse}, A.~A. and {Franceschi}, E. and {Frejsel}, A. and {Galeotta}, S. and {Galli}, S. and {Ganga}, K. and {Giard}, M. and {Giraud-H{\'e}raud}, Y. and {Gjerl{\o}w}, E. and {Gonz{\'a}lez-Nuevo}, J. and {G{\'o}rski}, K.~M. and {Gratton}, S. and {Gregorio}, A. and {Gruppuso}, A. and {Gudmundsson}, J.~E. and {Hansen}, F.~K. and {Hanson}, D. and {Harrison}, D.~L. and {Henrot-Versill{\'e}}, S. and {Hern{\'a}ndez-Monteagudo}, C. and {Herranz}, D. and {Hildebrandt}, S.~R. and {Hivon}, E. and {Hobson}, M. and {Holmes}, W.~A. and {Hornstrup}, A. and {Hovest}, W. and {Huffenberger}, K.~M. and {Hurier}, G. and {Jaffe}, A.~H. and {Jaffe}, T.~R. and {Jones}, W.~C. and {Juvela}, M. and {Keih{\"a}nen}, E. and {Keskitalo}, R. and {Kisner}, T.~S. and {Kneissl}, R. and {Knoche}, J. and {Kunz}, M. and {Kurki-Suonio}, H. and {Lagache}, G. and {L{\"a}hteenm{\"a}ki}, A. and {Lamarre}, J.-M. and {Lasenby}, A. and {Lattanzi}, M. and {Lawrence}, C.~R. and {Leonardi}, R. and {Lesgourgues}, J. and {Levrier}, F. and {Liguori}, M. and {Lilje}, P.~B. and {Linden-V{\o}rnle}, M. and {L{\'o}pez-Caniego}, M. and {Lubin}, P.~M. and {Mac{\'\i}as-P{\'e}rez}, J.~F. and {Maggio}, G. and {Maino}, D. and {Mandolesi}, N. and {Mangilli}, A. and {Maris}, M. and {Martin}, P.~G. and {Mart{\'\i}nez-Gonz{\'a}lez}, E. and {Masi}, S. and {Matarrese}, S. and {McGehee}, P. and {Meinhold}, P.~R. and {Melchiorri}, A. and {Melin}, J.-B. and {Mendes}, L. and {Mennella}, A. and {Migliaccio}, M. and {Mitra}, S. and {Miville-Desch{\^e}nes}, M.-A. and {Moneti}, A. and {Montier}, L. and {Morgante}, G. and {Mortlock}, D. and {Moss}, A. and {Munshi}, D. and {Murphy}, J.~A. and {Naselsky}, P. and {Nati}, F. and {Natoli}, P. and {Netterfield}, C.~B. and {N{\o}rgaard-Nielsen}, H.~U. and {Noviello}, F. and {Novikov}, D. and {Novikov}, I. and {Oxborrow}, C.~A. and {Paci}, F. and {Pagano}, L. and {Pajot}, F. and {Paoletti}, D. and {Partridge}, B. and {Pasian}, F. and {Patanchon}, G. and {Pearson}, T.~J. and {Perdereau}, O. and {Perotto}, L. and {Perrotta}, F. and {Pettorino}, V. and {Piacentini}, F. and {Piat}, M. and {Pierpaoli}, E. and {Pietrobon}, D. and {Plaszczynski}, S. and {Pointecouteau}, E. and {Polenta}, G. and {Popa}, L. and {Pratt}, G.~W. and {Pr{\'e}zeau}, G. and {Prunet}, S. and {Puget}, J.-L. and {Rachen}, J.~P. and {Rebolo}, R. and {Reinecke}, M. and {Remazeilles}, M. and {Renault}, C. and {Renzi}, A. and {Ristorcelli}, I. and {Rocha}, G. and {Roman}, M. and {Rosset}, C. and {Rossetti}, M. and {Roudier}, G. and {Rubi{\~n}o-Mart{\'\i}n}, J.~A. and {Rusholme}, B. and {Sandri}, M.},
        title = "{Planck 2015 results. XXIV. Cosmology from Sunyaev-Zeldovich cluster counts}",
      journal = {\aap},
         year = 2016,
        month = sep,
       volume = {594},
          eid = {A24},
        pages = {A24},
          doi = {10.1051/0004-6361/201525833},
archivePrefix = {arXiv},
       eprint = {1502.01597},
 primaryClass = {astro-ph.CO},
       adsurl = {https://ui.adsabs.harvard.edu/abs/2016A&A...594A..24P}
}

@INPROCEEDINGS{Gill:2022bej,
       author = {{Gill}, Ajay S. and {Shaaban}, Mohamed M. and {Tohuvavohu}, Aaron and {Sivanandam}, Suresh and {Abraham}, Roberto G. and {Chen}, Seery and {Drout}, Maria R. and {Lokhorst}, Deborah and {Matzner}, Christopher D. and {Mochnacki}, Stefan W. and {Netterfield}, Calvin B.},
        title = "{A low-cost ultraviolet-to-infrared absolute quantum efficiency characterization system of detectors}",
    booktitle = {X-Ray, Optical, and Infrared Detectors for Astronomy X},
         year = 2022,
       editor = {{Holland}, Andrew D. and {Beletic}, James},
       series = {Society of Photo-Optical Instrumentation Engineers (SPIE) Conference Series},
       volume = {12191},
        month = aug,
          eid = {1219114},
        pages = {1219114},
          doi = {10.1117/12.2627564},
archivePrefix = {arXiv},
       eprint = {2207.13052},
 primaryClass = {astro-ph.IM},
       adsurl = {https://ui.adsabs.harvard.edu/abs/2022SPIE12191E..14G}
}

@misc{cosmos2015,
  doi = {10.26131/IRSA527},
  url = {https://catcopy.ipac.caltech.edu/dois/doi.php?id=10.26131/IRSA527},
    author = {{Laigle}, C. and {McCracken}, H. J. and {Ilbert}, O. and {Hsieh}, B. C. and {Davidzon}, I. and {Capak}, P. and {Hasinger}, G. and {Silverman}, J. D. and {Pichon}, C. and {Coupon}, J. and {Aussel}, H. and {Le Borgne}, D. and {Caputi}, K. and {Cassata}, P. and {Chang}, Y.-Y. and {Civano}, F. and {Dunlop}, J. and {Fynbo}, J. and {Kartaltepe}, J. S. and {Koekemoer}, A. and {Le Fèvre}, O. and {Le Floc'h}, E. and {Leauthaud}, A. and {Lilly}, S. and {Lin}, L. and {Marchesi}, S. and {Milvang-Jensen}, B. and {Salvato}, M. and {Sanders}, D. B. and {Scoville}, N. and {Smolcic}, V. and {Stockmann}, M. and {Taniguchi}, Y. and {Tasca}, L. and {Toft}, S. and {Vaccari}, M. and {Zabl}, J.},
  title = {COSMOS2015 Catalog},
  publisher = {IPAC},
  year = {2021}
}

@ARTICLE{gaiadr3,
       author = {{Gaia Collaboration} and {Vallenari}, A. and {Brown}, A.~G.~A. and {Prusti}, T. and {de Bruijne}, J.~H.~J. and {Arenou}, F. and {Babusiaux}, C. and {Biermann}, M. and {Creevey}, O.~L. and {Ducourant}, C. and {Evans}, D.~W. and {Eyer}, L. and {Guerra}, R. and {Hutton}, A. and {Jordi}, C. and {Klioner}, S.~A. and {Lammers}, U.~L. and {Lindegren}, L. and {Luri}, X. and {Mignard}, F. and {Panem}, C. and {Pourbaix}, D. and {Randich}, S. and {Sartoretti}, P. and {Soubiran}, C. and {Tanga}, P. and {Walton}, N.~A. and {Bailer-Jones}, C.~A.~L. and {Bastian}, U. and {Drimmel}, R. and {Jansen}, F. and {Katz}, D. and {Lattanzi}, M.~G. and {van Leeuwen}, F. and {Bakker}, J. and {Cacciari}, C. and {Casta{\~n}eda}, J. and {De Angeli}, F. and {Fabricius}, C. and {Fouesneau}, M. and {Fr{\'e}mat}, Y. and {Galluccio}, L. and {Guerrier}, A. and {Heiter}, U. and {Masana}, E. and {Messineo}, R. and {Mowlavi}, N. and {Nicolas}, C. and {Nienartowicz}, K. and {Pailler}, F. and {Panuzzo}, P. and {Riclet}, F. and {Roux}, W. and {Seabroke}, G.~M. and {Sordo}, R. and {Th{\'e}venin}, F. and {Gracia-Abril}, G. and {Portell}, J. and {Teyssier}, D. and {Altmann}, M. and {Andrae}, R. and {Audard}, M. and {Bellas-Velidis}, I. and {Benson}, K. and {Berthier}, J. and {Blomme}, R. and {Burgess}, P.~W. and {Busonero}, D. and {Busso}, G. and {C{\'a}novas}, H. and {Carry}, B. and {Cellino}, A. and {Cheek}, N. and {Clementini}, G. and {Damerdji}, Y. and {Davidson}, M. and {de Teodoro}, P. and {Nu{\~n}ez Campos}, M. and {Delchambre}, L. and {Dell'Oro}, A. and {Esquej}, P. and {Fern{\'a}ndez-Hern{\'a}ndez}, J. and {Fraile}, E. and {Garabato}, D. and {Garc{\'\i}a-Lario}, P. and {Gosset}, E. and {Haigron}, R. and {Halbwachs}, J.-L. and {Hambly}, N.~C. and {Harrison}, D.~L. and {Hern{\'a}ndez}, J. and {Hestroffer}, D. and {Hodgkin}, S.~T. and {Holl}, B. and {Jan{\ss}en}, K. and {Jevardat de Fombelle}, G. and {Jordan}, S. and {Krone-Martins}, A. and {Lanzafame}, A.~C. and {L{\"o}ffler}, W. and {Marchal}, O. and {Marrese}, P.~M. and {Moitinho}, A. and {Muinonen}, K. and {Osborne}, P. and {Pancino}, E. and {Pauwels}, T. and {Recio-Blanco}, A. and {Reyl{\'e}}, C. and {Riello}, M. and {Rimoldini}, L. and {Roegiers}, T. and {Rybizki}, J. and {Sarro}, L.~M. and {Siopis}, C. and {Smith}, M. and {Sozzetti}, A. and {Utrilla}, E. and {van Leeuwen}, M. and {Abbas}, U. and {{\'A}brah{\'a}m}, P. and {Abreu Aramburu}, A. and {Aerts}, C. and {Aguado}, J.~J. and {Ajaj}, M. and {Aldea-Montero}, F. and {Altavilla}, G. and {{\'A}lvarez}, M.~A. and {Alves}, J. and {Anders}, F. and {Anderson}, R.~I. and {Anglada Varela}, E. and {Antoja}, T. and {Baines}, D. and {Baker}, S.~G. and {Balaguer-N{\'u}{\~n}ez}, L. and {Balbinot}, E. and {Balog}, Z. and {Barache}, C. and {Barbato}, D. and {Barros}, M. and {Barstow}, M.~A. and {Bartolom{\'e}}, S. and {Bassilana}, J.-L. and {Bauchet}, N. and {Becciani}, U. and {Bellazzini}, M. and {Berihuete}, A. and {Bernet}, M. and {Bertone}, S. and {Bianchi}, L. and {Binnenfeld}, A. and {Blanco-Cuaresma}, S. and {Blazere}, A. and {Boch}, T. and {Bombrun}, A. and {Bossini}, D. and {Bouquillon}, S. and {Bragaglia}, A. and {Bramante}, L. and {Breedt}, E. and {Bressan}, A. and {Brouillet}, N. and {Brugaletta}, E. and {Bucciarelli}, B. and {Burlacu}, A. and {Butkevich}, A.~G. and {Buzzi}, R. and {Caffau}, E. and {Cancelliere}, R. and {Cantat-Gaudin}, T. and {Carballo}, R. and {Carlucci}, T. and {Carnerero}, M.~I. and {Carrasco}, J.~M. and {Casamiquela}, L. and {Castellani}, M. and {Castro-Ginard}, A. and {Chaoul}, L. and {Charlot}, P. and {Chemin}, L. and {Chiaramida}, V. and {Chiavassa}, A. and {Chornay}, N. and {Comoretto}, G. and {Contursi}, G. and {Cooper}, W.~J. and {Cornez}, T. and {Cowell}, S. and {Crifo}, F. and {Cropper}, M. and {Crosta}, M. and {Crowley}, C. and {Dafonte}, C. and {Dapergolas}, A. and {David}, M. and {David}, P. and {de Laverny}, P. and {De Luise}, F. and {De March}, R.},
        title = "{Gaia Data Release 3. Summary of the content and survey properties}",
      journal = {\aap},
         year = 2023,
        month = jun,
       volume = {674},
          eid = {A1},
        pages = {A1},
          doi = {10.1051/0004-6361/202243940},
archivePrefix = {arXiv},
       eprint = {2208.00211},
 primaryClass = {astro-ph.GA},
       adsurl = {https://ui.adsabs.harvard.edu/abs/2023A&A...674A...1G}
}

@ARTICLE{gaia2016,
       author = {{Gaia Collaboration} and {Prusti}, T. and {de Bruijne}, J.~H.~J. and {Brown}, A.~G.~A. and {Vallenari}, A. and {Babusiaux}, C. and {Bailer-Jones}, C.~A.~L. and {Bastian}, U. and {Biermann}, M. and {Evans}, D.~W. and {Eyer}, L. and {Jansen}, F. and {Jordi}, C. and {Klioner}, S.~A. and {Lammers}, U. and {Lindegren}, L. and {Luri}, X. and {Mignard}, F. and {Milligan}, D.~J. and {Panem}, C. and {Poinsignon}, V. and {Pourbaix}, D. and {Randich}, S. and {Sarri}, G. and {Sartoretti}, P. and {Siddiqui}, H.~I. and {Soubiran}, C. and {Valette}, V. and {van Leeuwen}, F. and {Walton}, N.~A. and {Aerts}, C. and {Arenou}, F. and {Cropper}, M. and {Drimmel}, R. and {H{\o}g}, E. and {Katz}, D. and {Lattanzi}, M.~G. and {O'Mullane}, W. and {Grebel}, E.~K. and {Holland}, A.~D. and {Huc}, C. and {Passot}, X. and {Bramante}, L. and {Cacciari}, C. and {Casta{\~n}eda}, J. and {Chaoul}, L. and {Cheek}, N. and {De Angeli}, F. and {Fabricius}, C. and {Guerra}, R. and {Hern{\'a}ndez}, J. and {Jean-Antoine-Piccolo}, A. and {Masana}, E. and {Messineo}, R. and {Mowlavi}, N. and {Nienartowicz}, K. and {Ord{\'o}{\~n}ez-Blanco}, D. and {Panuzzo}, P. and {Portell}, J. and {Richards}, P.~J. and {Riello}, M. and {Seabroke}, G.~M. and {Tanga}, P. and {Th{\'e}venin}, F. and {Torra}, J. and {Els}, S.~G. and {Gracia-Abril}, G. and {Comoretto}, G. and {Garcia-Reinaldos}, M. and {Lock}, T. and {Mercier}, E. and {Altmann}, M. and {Andrae}, R. and {Astraatmadja}, T.~L. and {Bellas-Velidis}, I. and {Benson}, K. and {Berthier}, J. and {Blomme}, R. and {Busso}, G. and {Carry}, B. and {Cellino}, A. and {Clementini}, G. and {Cowell}, S. and {Creevey}, O. and {Cuypers}, J. and {Davidson}, M. and {De Ridder}, J. and {de Torres}, A. and {Delchambre}, L. and {Dell'Oro}, A. and {Ducourant}, C. and {Fr{\'e}mat}, Y. and {Garc{\'\i}a-Torres}, M. and {Gosset}, E. and {Halbwachs}, J.-L. and {Hambly}, N.~C. and {Harrison}, D.~L. and {Hauser}, M. and {Hestroffer}, D. and {Hodgkin}, S.~T. and {Huckle}, H.~E. and {Hutton}, A. and {Jasniewicz}, G. and {Jordan}, S. and {Kontizas}, M. and {Korn}, A.~J. and {Lanzafame}, A.~C. and {Manteiga}, M. and {Moitinho}, A. and {Muinonen}, K. and {Osinde}, J. and {Pancino}, E. and {Pauwels}, T. and {Petit}, J.-M. and {Recio-Blanco}, A. and {Robin}, A.~C. and {Sarro}, L.~M. and {Siopis}, C. and {Smith}, M. and {Smith}, K.~W. and {Sozzetti}, A. and {Thuillot}, W. and {van Reeven}, W. and {Viala}, Y. and {Abbas}, U. and {Abreu Aramburu}, A. and {Accart}, S. and {Aguado}, J.~J. and {Allan}, P.~M. and {Allasia}, W. and {Altavilla}, G. and {{\'A}lvarez}, M.~A. and {Alves}, J. and {Anderson}, R.~I. and {Andrei}, A.~H. and {Anglada Varela}, E. and {Antiche}, E. and {Antoja}, T. and {Ant{\'o}n}, S. and {Arcay}, B. and {Atzei}, A. and {Ayache}, L. and {Bach}, N. and {Baker}, S.~G. and {Balaguer-N{\'u}{\~n}ez}, L. and {Barache}, C. and {Barata}, C. and {Barbier}, A. and {Barblan}, F. and {Baroni}, M. and {Barrado y Navascu{\'e}s}, D. and {Barros}, M. and {Barstow}, M.~A. and {Becciani}, U. and {Bellazzini}, M. and {Bellei}, G. and {Bello Garc{\'\i}a}, A. and {Belokurov}, V. and {Bendjoya}, P. and {Berihuete}, A. and {Bianchi}, L. and {Bienaym{\'e}}, O. and {Billebaud}, F. and {Blagorodnova}, N. and {Blanco-Cuaresma}, S. and {Boch}, T. and {Bombrun}, A. and {Borrachero}, R. and {Bouquillon}, S. and {Bourda}, G. and {Bouy}, H. and {Bragaglia}, A. and {Breddels}, M.~A. and {Brouillet}, N. and {Br{\"u}semeister}, T. and {Bucciarelli}, B. and {Budnik}, F. and {Burgess}, P. and {Burgon}, R. and {Burlacu}, A. and {Busonero}, D. and {Buzzi}, R. and {Caffau}, E. and {Cambras}, J. and {Campbell}, H. and {Cancelliere}, R. and {Cantat-Gaudin}, T. and {Carlucci}, T. and {Carrasco}, J.~M. and {Castellani}, M. and {Charlot}, P. and {Charnas}, J. and {Charvet}, P. and {Chassat}, F. and {Chiavassa}, A. and {Clotet}, M. and {Cocozza}, G. and {Collins}, R.~S. and {Collins}, P. and {Costigan}, G.},
        title = "{The Gaia mission}",
      journal = {\aap},
         year = 2016,
        month = nov,
       volume = {595},
          eid = {A1},
        pages = {A1},
          doi = {10.1051/0004-6361/201629272},
archivePrefix = {arXiv},
       eprint = {1609.04153},
 primaryClass = {astro-ph.IM},
       adsurl = {https://ui.adsabs.harvard.edu/abs/2016A&A...595A...1G}
}

@ARTICLE{astroquery,
   author = {{Ginsburg}, A. and {Sip{\H o}cz}, B.~M. and {Brasseur}, C.~E. and
	{Cowperthwaite}, P.~S. and {Craig}, M.~W. and {Deil}, C. and
	{Guillochon}, J. and {Guzman}, G. and {Liedtke}, S. and {Lian Lim}, P. and
	{Lockhart}, K.~E. and {Mommert}, M. and {Morris}, B.~M. and
	{Norman}, H. and {Parikh}, M. and {Persson}, M.~V. and {Robitaille}, T.~P. and
	{Segovia}, J.-C. and {Singer}, L.~P. and {Tollerud}, E.~J. and
	{de Val-Borro}, M. and {Valtchanov}, I. and {Woillez}, J. and
	{The Astroquery collaboration} and {a subset of the astropy collaboration}
	},
    title = "{astroquery: An Astronomical Web-querying Package in Python}",
  journal = {\aj},
archivePrefix = "arXiv",
   eprint = {1901.04520},
 primaryClass = "astro-ph.IM",
     year = 2019,
    month = mar,
   volume = 157,
      eid = {98},
    pages = {98},
      doi = {10.3847/1538-3881/aafc33},
   adsurl = {https://adsabs.harvard.edu/abs/2019AJ....157...98G}
}

@ARTICLE{tyson_wl,
       author = {{Tyson}, J.~A. and {Valdes}, F. and {Wenk}, R.~A.},
        title = "{Detection of Systematic Gravitational Lens Galaxy Image Alignments: Mapping Dark Matter in Galaxy Clusters}",
      journal = {\apjl},
         year = 1990,
        month = jan,
       volume = {349},
        pages = {L1},
          doi = {10.1086/185636},
       adsurl = {https://ui.adsabs.harvard.edu/abs/1990ApJ...349L...1T}
}

@ARTICLE{SDSS:2005jin,
       author = {{Scranton}, Ryan and {M{\'e}nard}, Brice and {Richards}, Gordon T. and {Nichol}, Robert C. and {Myers}, Adam D. and {Jain}, Bhuvnesh and {Gray}, Alex and {Bartelmann}, Matthias and {Brunner}, Robert J. and {Connolly}, Andrew J. and {Gunn}, James E. and {Sheth}, Ravi K. and {Bahcall}, Neta A. and {Brinkman}, John and {Loveday}, Jon and {Schneider}, Donald P. and {Thakar}, Aniruddha and {York}, Donald G.},
        title = "{Detection of Cosmic Magnification with the Sloan Digital Sky Survey}",
      journal = {\apj},
         year = 2005,
        month = nov,
       volume = {633},
       number = {2},
        pages = {589-602},
          doi = {10.1086/431358},
archivePrefix = {arXiv},
       eprint = {astro-ph/0504510},
 primaryClass = {astro-ph},
       adsurl = {https://ui.adsabs.harvard.edu/abs/2005ApJ...633..589S}
}

@ARTICLE{CARS,
       author = {{Hildebrandt}, H. and {van Waerbeke}, L. and {Erben}, T.},
        title = "{CARS: The CFHTLS-Archive-Research Survey. III. First detection of cosmic magnification in samples of normal high-z galaxies}",
      journal = {\aap},
         year = 2009,
        month = nov,
       volume = {507},
       number = {2},
        pages = {683-691},
          doi = {10.1051/0004-6361/200912655},
archivePrefix = {arXiv},
       eprint = {0906.1580},
 primaryClass = {astro-ph.CO},
       adsurl = {https://ui.adsabs.harvard.edu/abs/2009A&A...507..683H}
}

@ARTICLE{Li:2021mvq,
       author = {{Li}, Xiangchong and {Miyatake}, Hironao and {Luo}, Wentao and {More}, Surhud and {Oguri}, Masamune and {Hamana}, Takashi and {Mandelbaum}, Rachel and {Shirasaki}, Masato and {Takada}, Masahiro and {Armstrong}, Robert and {Kannawadi}, Arun and {Takita}, Satoshi and {Miyazaki}, Satoshi and {Nishizawa}, Atsushi J. and {Plazas Malagon}, Andres A. and {Strauss}, Michael A. and {Tanaka}, Masayuki and {Yoshida}, Naoki},
        title = "{The three-year shear catalog of the Subaru Hyper Suprime-Cam SSP Survey}",
      journal = {\pasj},
         year = 2022,
        month = apr,
       volume = {74},
       number = {2},
        pages = {421-459},
          doi = {10.1093/pasj/psac006},
archivePrefix = {arXiv},
       eprint = {2107.00136},
 primaryClass = {astro-ph.CO},
       adsurl = {https://ui.adsabs.harvard.edu/abs/2022PASJ...74..421L}
}

@ARTICLE{Koekemoer:2007sr,
       author = {{Koekemoer}, A.~M. and {Aussel}, H. and {Calzetti}, D. and {Capak}, P. and {Giavalisco}, M. and {Kneib}, J.-P. and {Leauthaud}, A. and {Le F{\`e}vre}, O. and {McCracken}, H.~J. and {Massey}, R. and {Mobasher}, B. and {Rhodes}, J. and {Scoville}, N. and {Shopbell}, P.~L.},
        title = "{The COSMOS Survey: Hubble Space Telescope Advanced Camera for Surveys Observations and Data Processing}",
      journal = {\apjs},
         year = 2007,
        month = sep,
       volume = {172},
       number = {1},
        pages = {196-202},
          doi = {10.1086/520086},
archivePrefix = {arXiv},
       eprint = {astro-ph/0703095},
 primaryClass = {astro-ph},
       adsurl = {https://ui.adsabs.harvard.edu/abs/2007ApJS..172..196K}
}

@ARTICLE{Leauthaud:2007fb,
       author = {{Leauthaud}, Alexie and {Massey}, Richard and {Kneib}, Jean-Paul and {Rhodes}, Jason and {Johnston}, David E. and {Capak}, Peter and {Heymans}, Catherine and {Ellis}, Richard S. and {Koekemoer}, Anton M. and {Le F{\`e}vre}, Oliver and {Mellier}, Yannick and {R{\'e}fr{\'e}gier}, Alexandre and {Robin}, Annie C. and {Scoville}, Nick and {Tasca}, Lidia and {Taylor}, James E. and {Van Waerbeke}, Ludovic},
        title = "{Weak Gravitational Lensing with COSMOS: Galaxy Selection and Shape Measurements}",
      journal = {\apjs},
         year = 2007,
        month = sep,
       volume = {172},
       number = {1},
        pages = {219-238},
          doi = {10.1086/516598},
archivePrefix = {arXiv},
       eprint = {astro-ph/0702359},
 primaryClass = {astro-ph},
       adsurl = {https://ui.adsabs.harvard.edu/abs/2007ApJS..172..219L}
}

@ARTICLE{Casey:2022amu,
       author = {{Casey}, Caitlin M. and {Kartaltepe}, Jeyhan S. and {Drakos}, Nicole E. and {Franco}, Maximilien and {Harish}, Santosh and {Paquereau}, Louise and {Ilbert}, Olivier and {Rose}, Caitlin and {Cox}, Isabella G. and {Nightingale}, James W. and {Robertson}, Brant E. and {Silverman}, John D. and {Koekemoer}, Anton M. and {Massey}, Richard and {McCracken}, Henry Joy and {Rhodes}, Jason and {Akins}, Hollis B. and {Allen}, Natalie and {Amvrosiadis}, Aristeidis and {Arango-Toro}, Rafael C. and {Bagley}, Micaela B. and {Bongiorno}, Angela and {Capak}, Peter L. and {Champagne}, Jaclyn B. and {Chartab}, Nima and {Ch{\'a}vez Ortiz}, {\'O}scar A. and {Chworowsky}, Katherine and {Cooke}, Kevin C. and {Cooper}, Olivia R. and {Darvish}, Behnam and {Ding}, Xuheng and {Faisst}, Andreas L. and {Finkelstein}, Steven L. and {Fujimoto}, Seiji and {Gentile}, Fabrizio and {Gillman}, Steven and {Gould}, Katriona M.~L. and {Gozaliasl}, Ghassem and {Hayward}, Christopher C. and {He}, Qiuhan and {Hemmati}, Shoubaneh and {Hirschmann}, Michaela and {Jahnke}, Knud and {Jin}, Shuowen and {Khostovan}, Ali Ahmad and {Kokorev}, Vasily and {Lambrides}, Erini and {Laigle}, Clotilde and {Larson}, Rebecca L. and {Leung}, Gene C.~K. and {Liu}, Daizhong and {Liaudat}, Tobias and {Long}, Arianna S. and {Magdis}, Georgios and {Mahler}, Guillaume and {Mainieri}, Vincenzo and {Manning}, Sinclaire M. and {Maraston}, Claudia and {Martin}, Crystal L. and {McCleary}, Jacqueline E. and {McKinney}, Jed and {McPartland}, Conor J.~R. and {Mobasher}, Bahram and {Pattnaik}, Rohan and {Renzini}, Alvio and {Rich}, R. Michael and {Sanders}, David B. and {Sattari}, Zahra and {Scognamiglio}, Diana and {Scoville}, Nick and {Sheth}, Kartik and {Shuntov}, Marko and {Sparre}, Martin and {Suzuki}, Tomoko L. and {Talia}, Margherita and {Toft}, Sune and {Trakhtenbrot}, Benny and {Urry}, C. Megan and {Valentino}, Francesco and {Vanderhoof}, Brittany N. and {Vardoulaki}, Eleni and {Weaver}, John R. and {Whitaker}, Katherine E. and {Wilkins}, Stephen M. and {Yang}, Lilan and {Zavala}, Jorge A.},
        title = "{COSMOS-Web: An Overview of the JWST Cosmic Origins Survey}",
      journal = {\apj},
         year = 2023,
        month = sep,
       volume = {954},
       number = {1},
          eid = {31},
        pages = {31},
          doi = {10.3847/1538-4357/acc2bc},
archivePrefix = {arXiv},
       eprint = {2211.07865},
 primaryClass = {astro-ph.GA},
       adsurl = {https://ui.adsabs.harvard.edu/abs/2023ApJ...954...31C}
}

@ARTICLE{Scognamiglio:2026phv,
       author = {{Scognamiglio}, Diana and {Leroy}, Gavin and {Harvey}, David and {Massey}, Richard and {Rhodes}, Jason and {Akins}, Hollis B. and {Brinch}, Malte and {Berman}, Edward and {Casey}, Caitlin M. and {Drakos}, Nicole E. and {Faisst}, Andreas L. and {Franco}, Maximilien and {Fung}, Leo W.~H. and {Gozaliasl}, Ghassem and {He}, Qiuhan and {Hatamnia}, Hossein and {Huff}, Eric and {Hogg}, Natalie B. and {Ilbert}, Olivier and {Kartaltepe}, Jeyhan S. and {Koekemoer}, Anton M. and {Jin}, Shouwen and {Lambrides}, Erini and {Leauthaud}, Alexie and {Lentz}, Zane D. and {Liu}, Daizhong and {Mahler}, Guillaume and {Maraston}, Claudia and {Martin}, Crystal L. and {McCleary}, Jacqueline and {Nightingale}, James and {Mobasher}, Bahram and {Paquereau}, Louise and {Pires}, Sandrine and {Robertson}, Brant E. and {Sanders}, David B. and {Scarlata}, Claudia and {Shuntov}, Marko and {Toni}, Greta and {von Wietersheim-Kramsta}, Maximilian and {Weaver}, John R.},
        title = "{An ultra-high-resolution map of (dark) matter}",
      journal = {Nature Astronomy},
         year = 2026,
        month = apr,
       volume = {10},
        pages = {573-582},
          doi = {10.1038/s41550-025-02763-9},
archivePrefix = {arXiv},
       eprint = {2601.17239},
 primaryClass = {astro-ph.CO},
       adsurl = {https://ui.adsabs.harvard.edu/abs/2026NatAs..10..573S}
}

@ARTICLE{lsst2018,
       author = {{The LSST Dark Energy Science Collaboration} and {Mandelbaum}, Rachel and {Eifler}, Tim and {Hlo{\v{z}}ek}, Ren{\'e}e and {Collett}, Thomas and {Gawiser}, Eric and {Scolnic}, Daniel and {Alonso}, David and {Awan}, Humna and {Biswas}, Rahul and {Blazek}, Jonathan and {Burchat}, Patricia and {Chisari}, Nora Elisa and {Dell'Antonio}, Ian and {Digel}, Seth and {Frieman}, Josh and {Goldstein}, Daniel A. and {Hook}, Isobel and {Ivezi{\'c}}, {\v{Z}}eljko and {Kahn}, Steven M. and {Kamath}, Sowmya and {Kirkby}, David and {Kitching}, Thomas and {Krause}, Elisabeth and {Leget}, Pierre-Fran{\c{c}}ois and {Marshall}, Philip J. and {Meyers}, Joshua and {Miyatake}, Hironao and {Newman}, Jeffrey A. and {Nichol}, Robert and {Rykoff}, Eli and {Sanchez}, F. Javier and {Slosar}, An{\v{z}}e and {Sullivan}, Mark and {Troxel}, M.~A.},
        title = "{The LSST Dark Energy Science Collaboration (DESC) Science Requirements Document}",
      journal = {arXiv e-prints},
         year = 2018,
        month = sep,
          eid = {arXiv:1809.01669},
        pages = {arXiv:1809.01669},
    reportNumber = "FERMILAB-PUB-18-465-A",
    doi = "10.2172/1471560",
archivePrefix = {arXiv},
       eprint = {1809.01669},
 primaryClass = {astro-ph.CO},
       adsurl = {https://ui.adsabs.harvard.edu/abs/2018arXiv180901669T}
}

@ARTICLE{Chang:2013xja,
       author = {{Chang}, C. and {Jarvis}, M. and {Jain}, B. and {Kahn}, S.~M. and {Kirkby}, D. and {Connolly}, A. and {Krughoff}, S. and {Peng}, E.-H. and {Peterson}, J.~R.},
        title = "{The effective number density of galaxies for weak lensing measurements in the LSST project}",
      journal = {\mnras},
         year = 2013,
        month = sep,
       volume = {434},
       number = {3},
        pages = {2121-2135},
          doi = {10.1093/mnras/stt1156},
archivePrefix = {arXiv},
       eprint = {1305.0793},
 primaryClass = {astro-ph.CO},
       adsurl = {https://ui.adsabs.harvard.edu/abs/2013MNRAS.434.2121C}
}

@ARTICLE{euclid2011,
       author = {{Laureijs}, R. and {Amiaux}, J. and {Arduini}, S. and {Augu{\`e}res}, J.  -L. and {Brinchmann}, J. and {Cole}, R. and {Cropper}, M. and {Dabin}, C. and {Duvet}, L. and {Ealet}, A. and {Garilli}, B. and {Gondoin}, P. and {Guzzo}, L. and {Hoar}, J. and {Hoekstra}, H. and {Holmes}, R. and {Kitching}, T. and {Maciaszek}, T. and {Mellier}, Y. and {Pasian}, F. and {Percival}, W. and {Rhodes}, J. and {Saavedra Criado}, G. and {Sauvage}, M. and {Scaramella}, R. and {Valenziano}, L. and {Warren}, S. and {Bender}, R. and {Castander}, F. and {Cimatti}, A. and {Le F{\`e}vre}, O. and {Kurki-Suonio}, H. and {Levi}, M. and {Lilje}, P. and {Meylan}, G. and {Nichol}, R. and {Pedersen}, K. and {Popa}, V. and {Rebolo Lopez}, R. and {Rix}, H.  -W. and {Rottgering}, H. and {Zeilinger}, W. and {Grupp}, F. and {Hudelot}, P. and {Massey}, R. and {Meneghetti}, M. and {Miller}, L. and {Paltani}, S. and {Paulin-Henriksson}, S. and {Pires}, S. and {Saxton}, C. and {Schrabback}, T. and {Seidel}, G. and {Walsh}, J. and {Aghanim}, N. and {Amendola}, L. and {Bartlett}, J. and {Baccigalupi}, C. and {Beaulieu}, J.  -P. and {Benabed}, K. and {Cuby}, J.  -G. and {Elbaz}, D. and {Fosalba}, P. and {Gavazzi}, G. and {Helmi}, A. and {Hook}, I. and {Irwin}, M. and {Kneib}, J.  -P. and {Kunz}, M. and {Mannucci}, F. and {Moscardini}, L. and {Tao}, C. and {Teyssier}, R. and {Weller}, J. and {Zamorani}, G. and {Zapatero Osorio}, M.~R. and {Boulade}, O. and {Foumond}, J.~J. and {Di Giorgio}, A. and {Guttridge}, P. and {James}, A. and {Kemp}, M. and {Martignac}, J. and {Spencer}, A. and {Walton}, D. and {Bl{\"u}mchen}, T. and {Bonoli}, C. and {Bortoletto}, F. and {Cerna}, C. and {Corcione}, L. and {Fabron}, C. and {Jahnke}, K. and {Ligori}, S. and {Madrid}, F. and {Martin}, L. and {Morgante}, G. and {Pamplona}, T. and {Prieto}, E. and {Riva}, M. and {Toledo}, R. and {Trifoglio}, M. and {Zerbi}, F. and {Abdalla}, F. and {Douspis}, M. and {Grenet}, C. and {Borgani}, S. and {Bouwens}, R. and {Courbin}, F. and {Delouis}, J.  -M. and {Dubath}, P. and {Fontana}, A. and {Frailis}, M. and {Grazian}, A. and {Koppenh{\"o}fer}, J. and {Mansutti}, O. and {Melchior}, M. and {Mignoli}, M. and {Mohr}, J. and {Neissner}, C. and {Noddle}, K. and {Poncet}, M. and {Scodeggio}, M. and {Serrano}, S. and {Shane}, N. and {Starck}, J.  -L. and {Surace}, C. and {Taylor}, A. and {Verdoes-Kleijn}, G. and {Vuerli}, C. and {Williams}, O.~R. and {Zacchei}, A. and {Altieri}, B. and {Escudero Sanz}, I. and {Kohley}, R. and {Oosterbroek}, T. and {Astier}, P. and {Bacon}, D. and {Bardelli}, S. and {Baugh}, C. and {Bellagamba}, F. and {Benoist}, C. and {Bianchi}, D. and {Biviano}, A. and {Branchini}, E. and {Carbone}, C. and {Cardone}, V. and {Clements}, D. and {Colombi}, S. and {Conselice}, C. and {Cresci}, G. and {Deacon}, N. and {Dunlop}, J. and {Fedeli}, C. and {Fontanot}, F. and {Franzetti}, P. and {Giocoli}, C. and {Garcia-Bellido}, J. and {Gow}, J. and {Heavens}, A. and {Hewett}, P. and {Heymans}, C. and {Holland}, A. and {Huang}, Z. and {Ilbert}, O. and {Joachimi}, B. and {Jennins}, E. and {Kerins}, E. and {Kiessling}, A. and {Kirk}, D. and {Kotak}, R. and {Krause}, O. and {Lahav}, O. and {van Leeuwen}, F. and {Lesgourgues}, J. and {Lombardi}, M. and {Magliocchetti}, M. and {Maguire}, K. and {Majerotto}, E. and {Maoli}, R. and {Marulli}, F. and {Maurogordato}, S. and {McCracken}, H. and {McLure}, R. and {Melchiorri}, A. and {Merson}, A. and {Moresco}, M. and {Nonino}, M. and {Norberg}, P. and {Peacock}, J. and {Pello}, R. and {Penny}, M. and {Pettorino}, V. and {Di Porto}, C. and {Pozzetti}, L. and {Quercellini}, C. and {Radovich}, M. and {Rassat}, A. and {Roche}, N. and {Ronayette}, S. and {Rossetti}, E.},
        title = "{Euclid Definition Study Report}",
      journal = {arXiv e-prints},
         year = 2011,
        month = oct,
          eid = {arXiv:1110.3193},
        pages = {arXiv:1110.3193},
          doi = {10.48550/arXiv.1110.3193},
archivePrefix = {arXiv},
       eprint = {1110.3193},
 primaryClass = {astro-ph.CO},
       adsurl = {https://ui.adsabs.harvard.edu/abs/2011arXiv1110.3193L}
}

@ARTICLE{Zasowski:2025gki,
  author = {{Zasowski}, Gail and {Jha}, Saurabh W. and {Chomiuk}, Laura and {Fan}, Xiaohui and {Hickox}, Ryan and {Huber}, Dan and {Kerins}, Eamonn and {Kobulnicky}, Chip and {Lauer}, Tod and {Sako}, Masao and {Shapley}, Alice and {Stephens}, Denise and {Weinberg}, David and {Williams}, Ben},
  title = "{Roman Observations Time Allocation Committee: Final Report and Recommendations}",
  journal = {arXiv e-prints},
  year = 2025,
  month = may,
  eid = {arXiv:2505.10574},
  pages = {arXiv:2505.10574},
  doi = {10.48550/arXiv.2505.10574},
  archivePrefix = {arXiv},
  eprint = {2505.10574},
  primaryClass = {astro-ph.IM},
  adsurl = {https://ui.adsabs.harvard.edu/abs/2025arXiv250510574O}
}

@ARTICLE{Euclid:2024yrr,
       author = {{Euclid Collaboration} and {Mellier}, Y. and {Abdurro'uf} and {Acevedo Barroso}, J.~A. and {Ach{\'u}carro}, A. and {Adamek}, J. and {Adam}, R. and {Addison}, G.~E. and {Aghanim}, N. and {Aguena}, M. and {Ajani}, V. and {Akrami}, Y. and {Al-Bahlawan}, A. and {Alavi}, A. and {Albuquerque}, I.~S. and {Alestas}, G. and {Alguero}, G. and {Allaoui}, A. and {Allen}, S.~W. and {Allevato}, V. and {Alonso-Tetilla}, A.~V. and {Altieri}, B. and {Alvarez-Candal}, A. and {Alvi}, S. and {Amara}, A. and {Amendola}, L. and {Amiaux}, J. and {Andika}, I.~T. and {Andreon}, S. and {Andrews}, A. and {Angora}, G. and {Angulo}, R.~E. and {Annibali}, F. and {Anselmi}, A. and {Anselmi}, S. and {Arcari}, S. and {Archidiacono}, M. and {Aric{\`o}}, G. and {Arnaud}, M. and {Arnouts}, S. and {Asgari}, M. and {Asorey}, J. and {Atayde}, L. and {Atek}, H. and {Atrio-Barandela}, F. and {Aubert}, M. and {Aubourg}, E. and {Auphan}, T. and {Auricchio}, N. and {Aussel}, B. and {Aussel}, H. and {Avelino}, P.~P. and {Avgoustidis}, A. and {Avila}, S. and {Awan}, S. and {Azzollini}, R. and {Baccigalupi}, C. and {Bachelet}, E. and {Bacon}, D. and {Baes}, M. and {Bagley}, M.~B. and {Bahr-Kalus}, B. and {Balaguera-Antolinez}, A. and {Balbinot}, E. and {Balcells}, M. and {Baldi}, M. and {Baldry}, I. and {Balestra}, A. and {Ballardini}, M. and {Ballester}, O. and {Balogh}, M. and {Ba{\~n}ados}, E. and {Barbier}, R. and {Bardelli}, S. and {Baron}, M. and {Barreiro}, T. and {Barrena}, R. and {Barriere}, J.-C. and {Barros}, B.~J. and {Barthelemy}, A. and {Bartolo}, N. and {Basset}, A. and {Battaglia}, P. and {Battisti}, A.~J. and {Baugh}, C.~M. and {Baumont}, L. and {Bazzanini}, L. and {Beaulieu}, J.-P. and {Beckmann}, V. and {Belikov}, A.~N. and {Bel}, J. and {Bellagamba}, F. and {Bella}, M. and {Bellini}, E. and {Benabed}, K. and {Bender}, R. and {Benevento}, G. and {Bennett}, C.~L. and {Benson}, K. and {Bergamini}, P. and {Bermejo-Climent}, J.~R. and {Bernardeau}, F. and {Bertacca}, D. and {Berthe}, M. and {Berthier}, J. and {Bethermin}, M. and {Beutler}, F. and {Bevillon}, C. and {Bhargava}, S. and {Bhatawdekar}, R. and {Bianchi}, D. and {Bisigello}, L. and {Biviano}, A. and {Blake}, R.~P. and {Blanchard}, A. and {Blazek}, J. and {Blot}, L. and {Bosco}, A. and {Bodendorf}, C. and {Boenke}, T. and {B{\"o}hringer}, H. and {Boldrini}, P. and {Bolzonella}, M. and {Bonchi}, A. and {Bonici}, M. and {Bonino}, D. and {Bonino}, L. and {Bonvin}, C. and {Bon}, W. and {Booth}, J.~T. and {Borgani}, S. and {Borlaff}, A.~S. and {Borsato}, E. and {Bose}, B. and {Botticella}, M.~T. and {Boucaud}, A. and {Bouche}, F. and {Boucher}, J.~S. and {Boutigny}, D. and {Bouvard}, T. and {Bouwens}, R. and {Bouy}, H. and {Bowler}, R.~A.~A. and {Bozza}, V. and {Bozzo}, E. and {Branchini}, E. and {Brando}, G. and {Brau-Nogue}, S. and {Brekke}, P. and {Bremer}, M.~N. and {Brescia}, M. and {Breton}, M.-A. and {Brinchmann}, J. and {Brinckmann}, T. and {Brockley-Blatt}, C. and {Brodwin}, M. and {Brouard}, L. and {Brown}, M.~L. and {Bruton}, S. and {Bucko}, J. and {Buddelmeijer}, H. and {Buenadicha}, G. and {Buitrago}, F. and {Burger}, P. and {Burigana}, C. and {Busillo}, V. and {Busonero}, D. and {Cabanac}, R. and {Cabayol-Garcia}, L. and {Cagliari}, M.~S. and {Caillat}, A. and {Caillat}, L. and {Calabrese}, M. and {Calabro}, A. and {Calderone}, G. and {Calura}, F. and {Camacho Quevedo}, B. and {Camera}, S. and {Campos}, L. and {Ca{\~n}as-Herrera}, G. and {Candini}, G.~P. and {Cantiello}, M. and {Capobianco}, V. and {Cappellaro}, E. and {Cappelluti}, N. and {Cappi}, A. and {Caputi}, K.~I. and {Cara}, C. and {Carbone}, C. and {Cardone}, V.~F. and {Carella}, E. and {Carlberg}, R.~G. and {Carle}, M. and {Carminati}, L. and {Caro}, F. and {Carrasco}, J.~M. and {Carretero}, J. and {Carrilho}, P. and {Carron Duque}, J. and {Carry}, B.},
        title = "{Euclid: I. Overview of the Euclid mission}",
      journal = {\aap},
         year = 2025,
        month = may,
       volume = {697},
          eid = {A1},
        pages = {A1},
          doi = {10.1051/0004-6361/202450810},
archivePrefix = {arXiv},
       eprint = {2405.13491},
 primaryClass = {astro-ph.CO},
       adsurl = {https://ui.adsabs.harvard.edu/abs/2025A&A...697A...1E}
}

@ARTICLE{roman2018,
       author = {{Akeson}, Rachel and {Armus}, Lee and {Bachelet}, Etienne and {Bailey}, Vanessa and {Bartusek}, Lisa and {Bellini}, Andrea and {Benford}, Dominic and {Bennett}, David and {Bhattacharya}, Aparna and {Bohlin}, Ralph and {Boyer}, Martha and {Bozza}, Valerio and {Bryden}, Geoffrey and {Calchi Novati}, Sebastiano and {Carpenter}, Kenneth and {Casertano}, Stefano and {Choi}, Ami and {Content}, David and {Dayal}, Pratika and {Dressler}, Alan and {Dor{\'e}}, Olivier and {Fall}, S. Michael and {Fan}, Xiaohui and {Fang}, Xiao and {Filippenko}, Alexei and {Finkelstein}, Steven and {Foley}, Ryan and {Furlanetto}, Steven and {Kalirai}, Jason and {Gaudi}, B. Scott and {Gilbert}, Karoline and {Girard}, Julien and {Grady}, Kevin and {Greene}, Jenny and {Guhathakurta}, Puragra and {Heinrich}, Chen and {Hemmati}, Shoubaneh and {Hendel}, David and {Henderson}, Calen and {Henning}, Thomas and {Hirata}, Christopher and {Ho}, Shirley and {Huff}, Eric and {Hutter}, Anne and {Jansen}, Rolf and {Jha}, Saurabh and {Johnson}, Samson and {Jones}, David and {Kasdin}, Jeremy and {Kelly}, Patrick and {Kirshner}, Robert and {Koekemoer}, Anton and {Kruk}, Jeffrey and {Lewis}, Nikole and {Macintosh}, Bruce and {Madau}, Piero and {Malhotra}, Sangeeta and {Mandel}, Kaisey and {Massara}, Elena and {Masters}, Daniel and {McEnery}, Julie and {McQuinn}, Kristen and {Melchior}, Peter and {Melton}, Mark and {Mennesson}, Bertrand and {Peeples}, Molly and {Penny}, Matthew and {Perlmutter}, Saul and {Pisani}, Alice and {Plazas}, Andr{\'e}s and {Poleski}, Radek and {Postman}, Marc and {Ranc}, Cl{\'e}ment and {Rauscher}, Bernard and {Rest}, Armin and {Roberge}, Aki and {Robertson}, Brant and {Rodney}, Steven and {Rhoads}, James and {Rhodes}, Jason and {Ryan}, Jr., Russell and {Sahu}, Kailash and {Sand}, David and {Scolnic}, Dan and {Seth}, Anil and {Shvartzvald}, Yossi and {Siellez}, Karelle and {Smith}, Arfon and {Spergel}, David and {Stassun}, Keivan and {Street}, Rachel and {Strolger}, Louis-Gregory and {Szalay}, Alexander and {Trauger}, John and {Troxel}, M.~A. and {Turnbull}, Margaret and {van der Marel}, Roeland and {von der Linden}, Anja and {Wang}, Yun and {Weinberg}, David and {Williams}, Benjamin and {Windhorst}, Rogier and {Wollack}, Edward and {Wu}, Hao-Yi and {Yee}, Jennifer and {Zimmerman}, Neil},
        title = "{The Wide Field Infrared Survey Telescope: 100 Hubbles for the 2020s}",
      journal = {arXiv e-prints},
         year = 2019,
        month = feb,
          eid = {arXiv:1902.05569},
        pages = {arXiv:1902.05569},
          doi = {10.48550/arXiv.1902.05569},
archivePrefix = {arXiv},
       eprint = {1902.05569},
 primaryClass = {astro-ph.IM},
       adsurl = {https://ui.adsabs.harvard.edu/abs/2019arXiv190205569A}
}

@phdthesis{ajaythesis,
  author      = {Gill, Ajay Singh},
  title       = {A Multiwavelength View of Galaxy Clusters and Environments: Instrumentation and Observations},
  school      = {University of Toronto},
  year        = {2024},
  type        = {PhD thesis},
  address     = {Toronto, Canada},
  url         = {https://utoronto.scholaris.ca/items/9b6ae5fe-04cd-44bf-be63-67e6ea6dd68f}
}

@ARTICLE{bf1,
       author = {{Antilogus}, P. and {Astier}, P. and {Doherty}, P. and {Guyonnet}, A. and {Regnault}, N.},
        title = "{The brighter-fatter effect and pixel correlations in CCD sensors}",
      journal = {Journal of Instrumentation},
         year = 2014,
        month = mar,
       volume = {9},
       number = {3},
          eid = {C03048},
        pages = {C03048},
          doi = {10.1088/1748-0221/9/03/C03048},
archivePrefix = {arXiv},
       eprint = {1402.0725},
 primaryClass = {astro-ph.IM},
       adsurl = {https://ui.adsabs.harvard.edu/abs/2014JInst...9C3048A}
}

@ARTICLE{Finner:2023nwl,
       author = {{Finner}, Kyle and {Lee}, Bomee and {Chary}, Ranga-Ram and {Jee}, M. James and {Hirata}, Christopher and {Congedo}, Giuseppe and {Taylor}, Peter and {HyeongHan}, Kim},
        title = "{Near-IR Weak-lensing (NIRWL) Measurements in the CANDELS Fields. I. Point-spread Function Modeling and Systematics}",
      journal = {\apj},
         year = 2023,
        month = nov,
       volume = {958},
       number = {1},
          eid = {33},
        pages = {33},
          doi = {10.3847/1538-4357/acfafd},
archivePrefix = {arXiv},
       eprint = {2301.07725},
 primaryClass = {astro-ph.GA},
       adsurl = {https://ui.adsabs.harvard.edu/abs/2023ApJ...958...33F}
}

@ARTICLE{Zhang:2022aer,
       author = {{Zhang}, Tianqing and {Almoubayyed}, Husni and {Mandelbaum}, Rachel and {Meyers}, Joshua E. and {Jarvis}, Mike and {Kannawadi}, Arun and {Schmitz}, Morgan A. and {Guinot}, Axel and {LSST Dark Energy Science Collaboration}},
        title = "{Impact of point spread function higher moments error on weak gravitational lensing - II. A comprehensive study}",
      journal = {\mnras},
         year = 2023,
        month = apr,
       volume = {520},
       number = {2},
        pages = {2328-2350},
          doi = {10.1093/mnras/stac3350},
archivePrefix = {arXiv},
       eprint = {2205.07892},
 primaryClass = {astro-ph.CO},
       adsurl = {https://ui.adsabs.harvard.edu/abs/2023MNRAS.520.2328Z}
}

@ARTICLE{DES:2025fvt,
       author = {{Schutt}, T. and {Jarvis}, M. and {Roodman}, A. and {Amon}, A. and {Becker}, M.~R. and {Gruendl}, R.~A. and {Yamamoto}, M. and {Bechtol}, K. and {Bernstein}, G.~M. and {Gatti}, M. and {Rykoff}, E.~S. and {Sheldon}, E. and {Troxel}, M.~A. and {Abbott}, T.~M.~C. and {Aguena}, M. and {Andrade-Oliveira}, F. and {Brooks}, D. and {Rosell}, A. Carnero and {Carretero}, J. and {Chang}, C. and {Choi}, A. and {Crocce}, M. and {da Costa}, L.~N. and {Davis}, T.~M. and {De Vicente}, J. and {Desai}, S. and {Diehl}, H.~T. and {Doel}, P. and {Fert{\'e}}, A. and {Frieman}, J. and {Garc{\'\i}a-Bellido}, J. and {Gaztanaga}, E. and {Gruen}, D. and {Gutierrez}, G. and {Hinton}, S.~R. and {Hollowood}, D.~L. and {Honscheid}, K. and {Kuehn}, K. and {Lahav}, O. and {Lee}, S. and {Lima}, M. and {Marshall}, J.~L. and {Mena-Fern{\'a}ndez}, J. and {Miquel}, R. and {Mohr}, J.~J. and {Myles}, J. and {Ogando}, R.~L.~C. and {Pieres}, A. and {Malag{\'o}n}, A.~A. Plazas and {Porredon}, A. and {Samuroff}, S. and {Sanchez}, E. and {Cid}, D. Sanchez and {Sevilla-Noarbe}, I. and {Smith}, M. and {Suchyta}, E. and {Tarle}, G. and {Vikram}, V. and {Walker}, A.~R. and {Weaverdyck}, N.},
        title = "{Dark Energy Survey Year 6 Results: Point-Spread Function Modeling}",
      journal = {The Open Journal of Astrophysics},
         year = 2025,
        month = mar,
       volume = {8},
          eid = {26},
        pages = {26},
          doi = {10.33232/001c.132299},
archivePrefix = {arXiv},
       eprint = {2501.05781},
 primaryClass = {astro-ph.CO},
       adsurl = {https://ui.adsabs.harvard.edu/abs/2025OJAp....8E..26S}
}

@software{johannes_sahlmann_2019_3516964,
  author       = {Johannes Sahlmann and
                  Shannon Osborne and
                  Colin Cox and
                  charlesrp and
                  David Law and
                  Marshall Perrin and
                  marthaboyer and
                  Joseph Hunkeler},
  title        = {{spacetelescope/pysiaf: Introduce check against
                   online version}},
  month        = oct,
  year         = 2019,
  publisher    = {Zenodo},
  version      = {v0.5.1},
  doi          = {10.5281/zenodo.3516964},
  url          = {https://doi.org/10.5281/zenodo.3516964}
}

@ARTICLE{Golovich:2017tke,
       author = {{Golovich}, Nathan and {Dawson}, William A. and {Wittman}, David M. and {Jee}, Myungkook J. and {Benson}, Bryant and {Lemaux}, Brian and {van Weeren}, Reinout J. and {Andrade-Santos}, Felipe and {Sobral}, David and {de Gasperin}, Francesco and {Bruggen}, Marcus and {Bradac}, Marusa and {Finner}, Kyle and {Peter}, Annika},
        title = "{Merging Cluster Collaboration: Optical and Spectroscopic Survey of a Radio-Selected Sample of Twenty Nine Merging Galaxy Clusters}",
      journal = {arXiv e-prints},
         year = 2017,
        month = nov,
          eid = {arXiv:1711.01347},
        pages = {arXiv:1711.01347},
          doi = {10.48550/arXiv.1711.01347},
archivePrefix = {arXiv},
       eprint = {1711.01347},
 primaryClass = {astro-ph.CO},
       adsurl = {https://ui.adsabs.harvard.edu/abs/2017arXiv171101347G}
}

@ARTICLE{susan2024,
author = {Susan F. Redmond and Steven J. Benton and Christopher J. Damaren and Spencer W. Everett and Aurelien A. Fraisse and Ajay S. Gill and John W. Hartley and David Harvey and Bradley Holder and Eric M. Huff and Mathilde Jauzac and William C. Jones and David Lagattuta and Jason S.-Y. Leung and Lun Li and Thuy Vy T. Luu and Richard Massey and Jacqueline E. McCleary and Johanna M. Nagy and C. Barth Netterfield and Emaad Paracha and Jason D. Rhodes and Andrew Robertson and L. Javier Romualdez and J{\"u}rgen Schmoll and Mohamed M. Shaaban and Ellen L. Sirks and Georgios N. Vassilakis and Andr{\'e} Z. Vitorelli},
title = {{To the stratosphere and beyond! Super-pressure balloon flight overview for the Super-pressure Balloon-borne Imaging Telescope (SuperBIT)}},
journal = {Proc. SPIE},
series = {Ground-based and Airborne Telescopes X},
volume = {13094},
eid = {130942P},
year = {2024},
doi = {10.1117/12.3020296},
URL = {https://doi.org/10.1117/12.3020296}
}

@ARTICLE{noisebias_1,
       author = {{Refregier}, Alexandre and {Kacprzak}, Tomasz and {Amara}, Adam and {Bridle}, Sarah and {Rowe}, Barnaby},
        title = "{Noise bias in weak lensing shape measurements}",
      journal = {\mnras},
         year = 2012,
        month = sep,
       volume = {425},
       number = {3},
        pages = {1951-1957},
          doi = {10.1111/j.1365-2966.2012.21483.x},
archivePrefix = {arXiv},
       eprint = {1203.5050},
 primaryClass = {astro-ph.CO},
       adsurl = {https://ui.adsabs.harvard.edu/abs/2012MNRAS.425.1951R}
}

@ARTICLE{noisebias_2,
       author = {{Melchior}, P. and {Viola}, M.},
        title = "{Means of confusion: how pixel noise affects shear estimates for weak gravitational lensing}",
      journal = {\mnras},
         year = 2012,
        month = aug,
       volume = {424},
       number = {4},
        pages = {2757-2769},
          doi = {10.1111/j.1365-2966.2012.21381.x},
archivePrefix = {arXiv},
       eprint = {1204.5147},
 primaryClass = {astro-ph.IM},
       adsurl = {https://ui.adsabs.harvard.edu/abs/2012MNRAS.424.2757M}
}

@software{gaiaxpy,
  author       = {{Gaia DPAC}},
  title        = {GaiaXPy},
  year         = {2025},
  version      = {2.1.4},
  doi          = {10.5281/zenodo.17943031},
  url          = {https://gaia-dpci.github.io/GaiaXPy-website/},
  note         = {Software}
}

@ARTICLE{gaiadr3_cv,
       author = {{Babusiaux}, C. and {Fabricius}, C. and {Khanna}, S. and {Muraveva}, T. and {Reyl{\'e}}, C. and {Spoto}, F. and {Vallenari}, A. and {Luri}, X. and {Arenou}, F. and {{\'A}lvarez}, M.~A. and {Anders}, F. and {Antoja}, T. and {Balbinot}, E. and {Barache}, C. and {Bauchet}, N. and {Bossini}, D. and {Busonero}, D. and {Cantat-Gaudin}, T. and {Carrasco}, J.~M. and {Dafonte}, C. and {Diakit{\'e}}, S. and {Figueras}, F. and {Garcia-Gutierrez}, A. and {Garofalo}, A. and {Helmi}, A. and {Jim{\'e}nez-Arranz}, {\'O}. and {Jordi}, C. and {Kervella}, P. and {Kostrzewa-Rutkowska}, Z. and {Leclerc}, N. and {Licata}, E. and {Manteiga}, M. and {Masip}, A. and {Mongui{\'o}}, M. and {Ramos}, P. and {Robichon}, N. and {Robin}, A.~C. and {Romero-G{\'o}mez}, M. and {S{\'a}ez}, A. and {Santove{\~n}a}, R. and {Spina}, L. and {Torralba Elipe}, G. and {Weiler}, M.},
        title = "{Gaia Data Release 3. Catalogue validation}",
      journal = {\aap},
         year = 2023,
        month = jun,
       volume = {674},
          eid = {A32},
        pages = {A32},
          doi = {10.1051/0004-6361/202243790},
archivePrefix = {arXiv},
       eprint = {2206.05989},
 primaryClass = {astro-ph.SR},
       adsurl = {https://ui.adsabs.harvard.edu/abs/2023A&A...674A..32B}
}
\bibliographystyle{aasjournalv7}



\end{document}